\documentclass[reqno,12pt]{amsart}
\usepackage{amsfonts}
\usepackage{amssymb}
\usepackage[pdftex]{graphicx}
\usepackage{caption}
\usepackage{subcaption}
\usepackage{float}
\usepackage{fullpage}
\usepackage{url}

\begin{document}

\newtheorem{thm}{Theorem}
\newtheorem{lem}{Lemma}
\newtheorem{prop}{Proposition}
\newtheorem{cor}{Corollary}
\numberwithin{equation}{section}

%my macros for LaTeX fixes

% refs:

%put `Eq.( )' around an equation ref in latex
\def\Eqref#1{Eq.~(\ref{#1})}
\def\Eqrefs#1#2{Eqs.~(\ref{#1}) and~(\ref{#2})}
\def\Eqsref#1#2{Eqs.~(\ref{#1}) to~(\ref{#2})}
\def\Sysref#1#2{Eqs. (\ref{#1})--(\ref{#2})}

%put `Fig.' before a figure ref in latex
\def\figref#1{Fig.~\ref{#1}}

%put `Sec. ' before a section ref in latex
\def\secref#1{Sec.~\ref{#1}}
\def\secrefs#1#2{Sec.~\ref{#1} and~\ref{#2}}

%put `App. ' before an appendix ref in latex
\def\appref#1{Appendix~\ref{#1}}

%put `Ref. ' before a bibitem ref in latex
\def\Ref#1{Ref.~\cite{#1}}
\def\Refs#1{Refs.~\cite{#1}}

%use footnote style for a bibitem ref in latex
\def\Cite#1{${\mathstrut}^{\cite{#1}}$}

% abbrevs for latex commands:
%equations
\def\EQ{\begin{equation}}
\def\endEQ{\end{equation}}

%my macros

\def\fewquad{\qquad\qquad}
\def\severalquad{\qquad\fewquad}
\def\manyquad{\qquad\severalquad}
\def\manymanyquad{\manyquad\manyquad}

\def\sech{\;{\rm sech}}
\def\sgn{{\rm sgn}}
\def\Re{{\rm Re}\,}
\def\Im{{\rm Im}\,}
\def\cubrt#1{\textstyle\sqrt[3]{#1}}

\def\k{\mathrm{k}}
\def\w{\mathrm{w}}

\def\wtil{\widetilde}

\def\ie/{i.e.}
\def\eg/{e.g.}
\def\etc/{etc.}
\def\const{{\rm const.}}

\allowdisplaybreaks[4]

%\begin{frontmatter}

\title{Oscillatory solitons of U(1)-invariant mKdV equations I: Envelope speed and temporal frequency}

\author{
Stephen C. Anco$^1$,
Abdus Sattar Mia$^{1,2}$,
Mark R. Willoughby$^{1}$
\\\\\lowercase{\scshape{
${}^1$
Department of Mathematics\\
Brock University\\
St. Catharines, ON Canada \\
${}^2$
Department of Mathematics and Statistics\\
University of Saskatchewan\\
Saskatoon, SK Canada \\
}}
}

\begin{abstract}
Harmonically modulated complex solitary waves 
which are a generalized type of envelope soliton 
(herein called {\em oscillatory solitons}) 
are studied for the two $U(1)$-invariant integrable generalizations of 
the modified Korteweg-de Vries equation, 
given by the Hirota equation and the Sasa-Satsuma equation. 
A bilinear formulation of these two equations is used to derive
the oscillatory $1$-soliton and $2$-soliton solutions,
which are then written out in a physical form parameterized in terms of
their speed, modulation frequency, and phase. 
Depending on the modulation frequency, 
the speeds of oscillatory waves ($1$-solitons) 
can be positive, negative, or zero, 
in contrast to the strictly positive speed of ordinary solitons. 
When the speed is zero, an oscillatory wave is a time-periodic standing wave.
Properties of the amplitude and phase of oscillatory $1$-solitons are derived.
Oscillatory $2$-solitons are graphically illustrated to describe 
collisions between two oscillatory $1$-solitons
in the case when the speeds are distinct. 
In the special case of equal speeds, 
oscillatory $2$-solitons are shown to reduce to 
harmonically modulated breather waves. 
\end{abstract}

%\begin{keyword}
%\MSC
%\end{keyword}
\keywords{mKdV equation, Hirota equation, Sasa-Satsuma equation, solitary wave, envelope soliton, oscillatory soliton, breather, overtake collision, head-on collision}
\subjclass[2010]{}
%PACS 05.45.Yv,  02.30.Ik
%\thanks{}

\maketitle

\section{ Introduction }

The modified Korteweg-de Vries (mKdV) equation 
\EQ
u_t +a u^2 u_x +b u_{xxx} =0
\label{mkdveq}
\endEQ
(where $a$ and $b$ are arbitrary positive constants)
is an integrable evolution equation which arises in many physical applications,
such as acoustic waves in anharmonic lattices \cite{mkdv1}
and Alfven waves in collision-free plasmas \cite{mkdv2}. 
Its well-known integrability properties consist of 
multi-soliton solutions, 
a Lax pair, a bi-Hamiltonian structure, an infinite hierarchy of 
symmetries and conservation laws, and a bilinear formulation. 
Soliton solutions of the mKdV equation are solitary waves 
\EQ
u(t,x) = \epsilon\sqrt{\frac{6c}{a}} \sech\Big(\sqrt{\frac{c}{b}}(x-ct)\Big)
\label{realsoliton}
\endEQ
whose shape, wave speed $c>0$, and up/down orientation $\epsilon=\pm 1$
are preserved after undergoing collisions. 
These waves have a mathematical characterization as 
stable travelling wave solutions that are 
single-peaked, unidirectional, and decaying for large $|x|$. 
Collisions of two or more mKdV solitary waves are described by 
multi-soliton solutions that reduce to 
a linear superposition of distinct solitary waves 
in the asymptotic past and future. 
All mKdV soliton solutions carry mass, momentum, energy, 
as well as Galilean energy associated with the motion of center of momentum, 
which are constants of motion arising from conservation laws 
for the mKdV equation \eqref{mkdveq}. 
Remarkably, the only net effect of a collision is 
to shift the asymptotic positions of the solitary waves such that 
the center of momentum moves at a constant speed throughout the collision. 

There are exactly two integrable complex generalizations 
\cite{complexmkdveqs1,complexmkdveqs2} of the mKdV equation \eqref{mkdveq}. 
One is the Hirota equation \cite{Hir1973}
\begin{equation}
u_t +a |u|^2 u_x +b u_{xxx} =0 , 
\label{hmkdveq}
\end{equation}
and the other is the Sasa-Satsuma equation \cite{SasSat1991}
\begin{equation}
u_t +\tfrac{1}{4} a( u\bar u_x +3u_x\bar u )u +b u_{xxx} =0 . 
\label{ssmkdveq}
\end{equation}
These two equations share the same scaling symmetry
\EQ
t\rightarrow \lambda^3 t,\quad
x\rightarrow \lambda x,\quad
u\rightarrow \lambda^{-1} u 
\label{scalingsymm}
\endEQ
admitted by the mKdV equation \eqref{mkdveq},
and possess an additional $U(1)$ phase symmetry 
\EQ
u\rightarrow \exp(i\phi) u .
\label{phasesymm}
\endEQ
Both the Hirota equation \eqref{hmkdveq} and the Sasa-Satsuma equation \eqref{ssmkdveq} 
are interesting physically and mathematically. 
In particular, 
under a Galilean transformation 
$t\rightarrow t$, $x\rightarrow x-vt$
combined with a phase-modulation transformation 
$u\rightarrow \exp(i(kx+\omega t))u$, 
each equation takes the form of 
a 3rd order generalization of the nonlinear Schrodinger equation, 
describing short wave pulses in optical fibers 
\cite{Pot,CavCreCru}
and deep water waves
\cite{Sed,Slu}. 
Both equations have integrability properties similar to 
those of the mKdV equation,
and their solitary wave solutions have the form of mKdV solitons 
up to a phase factor
\EQ
u(t,x) = \exp(i\phi) f_{\rm mKdV}(x-ct)
\label{mkdv1soliton}
\endEQ
where $c>0$ is the speed and $-\pi\leq \phi\leq \pi$ is the phase angle,
and where $f_{\rm mKdV}$ is the envelope function
\EQ\label{mkdvenvel}
f_{\rm mKdV}(x-ct)= 
\sqrt{\frac{6c}{a}} \sech\Big(\sqrt{\frac{c}{b}}(x-ct)\Big) .
\endEQ
For each equation \eqref{hmkdveq} and \eqref{ssmkdveq}, 
collisions of two or more solitary waves are described by 
multi-soliton solutions with the main feature that the net effect 
on the solitary waves is a shift in their asymptotic positions, 
while their asymptotic phases 
stay unchanged in the case of the Hirota equation \eqref{hmkdveq}
but undergo a shift in the case of the Sasa-Satsuma equation \eqref{ssmkdveq}. 
The actual nonlinear interaction of these solitary waves during a collision 
exhibits interesting features which depend on the speed ratios 
and relative phase angles of the waves,
as studied in recent work \cite{AncNgaWil}. 
(See the animations at 
http://lie.math.brocku.ca/\url{~}sanco/solitons/mkdv\_solitons.php)

Most interestingly, 
both the Hirota equation \eqref{hmkdveq}
and the Sasa-Satsuma equation \eqref{ssmkdveq}
possess a more general type of soliton solution \cite{GilHieNimOht}
\EQ
u(t,x) = \exp(i\phi) \exp(i(\kappa x +\omega t)) f(kx+wt)
\label{envel1soliton}
\endEQ
which has the form of a solitary wave $\exp(i\phi)f(kx+wt)$, 
with speed $c=-w/k$ and phase angle $\phi$, 
modulated by a harmonic wave $\exp(i(\kappa x +\omega t))$,
with frequency $\omega/(2\pi)$ and wave length $2\pi/\kappa$,
satisfying the algebraic relations
\EQ\label{wkrels}
w=-b k(k^2-3\kappa^2),
\quad 
\omega= -b \kappa(3k^2-\kappa^2),
\quad
\kappa \neq 0 .
\endEQ
The envelope function $f$ in this solution differs from $f_{\rm mKdV}$,
specifically 
\EQ\label{henvel}
f_{\rm H}(kx+wt)= 
\sqrt{\frac{6b}{a}} |k| \sech\big(kx+wt\big) 
\endEQ
in the case of the Hirota equation \eqref{hmkdveq},
and
\EQ\label{ssenvel}
f_{\rm SS}(kx+wt)= 
\sqrt{\frac{6b}{a}} 
\frac{|k|\big(k^2+\kappa^2 +(\kappa+ik)(\kappa/2)\exp\big(2(kx+wt)\big)\big)}
{(k^2+\kappa^2)\cosh\big(kx+wt\big) + (\kappa^2/8)\exp\big(3(kx+wt)\big)}
\endEQ
in the case of the Sasa-Satsuma equation \eqref{ssmkdveq}. 
If $\kappa=0$ (and hence $\omega=0$) then these envelope functions 
\eqref{henvel} and \eqref{ssenvel} reduce to $f_{\rm mKdV}$,
whereby the soliton solution \eqref{envel1soliton} reduces to 
the solitary wave \eqref{mkdv1soliton}. 
In contrast to an ordinary soliton \eqref{mkdv1soliton}, 
the envelope speed
\EQ
c=-w/k=b(k^2-3\kappa^2)
\endEQ
can be positive, negative, or zero, 
depending on whether $|\kappa|$ is less than, greater than, or equal to 
$|k|/\sqrt{3}$, respectively. 
Consequently, 
harmonically modulated solitons can have three different types of collisions:
(1) {\em right-overtake} --- 
where a faster right-moving soliton overtakes a slower right-moving soliton 
or a stationary soliton;
(2) {\em left-overtake} --- 
where a faster left-moving soliton overtakes a slower left-moving soliton
or a stationary soliton;
(3) {\em head-on} --- 
where a right-moving soliton collides with a left-moving soliton. 
All of these collisions can be expected to exhibit 
highly interesting new features compared to collisions of ordinary solitons.
However, very little seems to be known about the explicit behaviour of 
colliding harmonically modulated soliton solutions
in the literature on the Hirota equation \eqref{hmkdveq} 
and the Sasa-Satsuma equation \eqref{ssmkdveq},
although general formulas yielding the harmonically modulated multi-soliton solutions 
for both equations have been known for some time \cite{Hir1973,SasSat1991,TaoHe2012}. 

The present paper and a sequel paper will be devoted to studying
the basic properties of harmonically modulated complex solitons 
and their nonlinear interactions. 
One new aspect of the analysis is that 
it will introduce a direct physical parameterization for these solitons, 
which will greatly help in understanding their interaction properties. 

In section~\ref{bilinear}, 
we use a bilinear formulation of general $U(1)$-invariant complex mKdV equations
to derive explicit expressions 
for the harmonically modulated 2-soliton solutions 
\begin{equation}\label{envel2soliton}
\begin{aligned}
u(t,x) = & 
\exp(i\phi_1)\exp(i(\kappa_1 x+\omega_1 t)) f_1(k_1x+w_1t,k_2x+w_2t)
\\&\qquad
+ \exp(i\phi_2)\exp(i(\kappa_2 x+\omega_2 t)) f_2(k_1x+w_1t,k_2x+w_2t)
\end{aligned}
\end{equation}
of the Hirota equation \eqref{hmkdveq} and the Sasa-Satsuma equation \eqref{ssmkdveq}.
The envelope functions $f_1$ and $f_2$ turn out to be complex functions,
which implies that the solitary wave envelopes are given by $|f_1|$ and $|f_2|$
while their modulation comes from the overall phases
$\kappa_1 x+\omega_1 +\arg(f_1)$ and $\kappa_2 x+\omega_2 +\arg(f_2)$ . 
Consequently, 
for the purpose of analytically and graphically understanding these solutions,
it is mathematically and physically preferable to rewrite the expressions
\eqref{envel2soliton} in a different parameterization given by 
the speed of the envelopes and the frequency of the modulations. 

In section~\ref{kinematics}, 
we first express the harmonically modulated 1-soliton solutions \eqref{envel1soliton}
in the direct physical parameterization 
\EQ
u(t,x) = \exp(i\phi)\exp(i\nu t) \tilde f(x-ct)
\label{oscil1soliton}
\endEQ
involving only the envelope speed $c=-w/k$, 
a temporal modulation frequency $\nu=\omega+c\kappa$,
and the phase angle $\phi$. 
We show that $c$ and $\nu$ obey a simple kinematic relation 
which gives a direct way to classify the cases for which 
$c$ is positive, negative, or zero, depending only on $\nu$. 
In particular, when the envelope speed is $c=0$, 
these solutions describe time-periodic standing waves
\EQ
u(t,x) = \exp(i\phi)\exp(i\nu t) \tilde f(x) .
\label{standing}
\endEQ
We next express the harmonically modulated 2-soliton solutions \eqref{envel2soliton} 
for the Hirota equation \eqref{hmkdveq} and the Sasa-Satsuma equation \eqref{ssmkdveq}
in an analogous physical parameterization, given by 
\EQ
u(t,x) = \exp(i\phi_1)\exp(i\nu_1 t) \tilde f_1(x-c_1 t,x-c_2 t) + \exp(i\phi_2)\exp(i\nu_2 t) \tilde f_2(x-c_1 t,x-c_2 t) 
\label{oscil2soliton}
\endEQ
if the envelope speeds $c_1=-w_1/k_1$ and $c_2=-w_2/k_2$ are distinct, 
or 
\EQ
u(t,x) = \exp(i\phi_1)\exp(i\nu_1 t) \tilde f_1(x-ct,(\nu_1-\nu_2)t) + \exp(i\phi_2)\exp(i\nu_2 t) \tilde f_2(x-ct,(\nu_1-\nu_2)t) 
\label{specialoscil2soliton}
\endEQ
with $c=-w_1/k_1=-w_2/k_2$ if the envelope speeds are equal. 

We will call solitary wave solutions of the form \eqref{oscil1soliton} 
an {\em oscillatory soliton},
and solutions of the form \eqref{oscil2soliton} 
an {\em oscillatory $2$-soliton}. 
The special case \eqref{specialoscil2soliton} describes 
a solitary wave solution that has speed $c$ 
and involves two temporal frequencies $\nu_1$ and $\nu_2$. 
When these frequencies are related by $\nu_1+\nu_2=0$, 
the resulting wave is a {\em breather}, 
which has the equivalent general form 
\EQ
u(t,x) = \exp(i\phi_0)\tilde f(x-ct,\nu t+\phi)
\label{breather}
\endEQ
involving the frequency $\nu=\nu_1=-\nu_2$ 
and the phase angles $\phi=(\phi_1-\phi_2)/2$ and $\phi_0=(\phi_1+\phi_2)/2$. 
When the frequencies are independent, $\nu_1+\nu_2\neq 0$, 
we will write the solitary wave solution \eqref{specialoscil2soliton} 
in a similar physical form 
\EQ
u(t,x) = \exp(i(\nu_0 t+\phi_0)) \tilde f(x-ct,\nu t+\phi)
\label{oscilbreather}
\endEQ
which we call an {\em oscillatory breather},
with an envelope frequency $\nu=(\nu_1-\nu_2)/2$ 
and phase angle $\phi=(\phi_1-\phi_2)/2$, 
and a temporal modulation frequency $\nu_0=(\nu_1+\nu_2)/2\neq 0$
and phase angle $\phi_0=(\phi_1+\phi_2)/2$. 

We emphasize that this oscillatory form for
the harmonically modulated 1-soliton and 2-soliton solutions 
presented in section~\ref{kinematics} 
is more mathematically clear and more physically meaningful 
than the standard form \eqref{envel1soliton} and \eqref{envel2soliton}
found in the literature. 
In particular, the frequency and speed parameters used 
in the oscillatory form \eqref{oscil1soliton} and \eqref{oscil2soliton}
have a direct physical meaning which is conceptually simpler 
for studying collision properties of harmonically modulated solitary waves. 

Our main results are obtained in section~\ref{properties}. 
We first derive the basic properties of the amplitude $|u|$ and phase $\arg(u)$
for the oscillatory $1$-solitons \eqref{oscil1soliton} of
the Hirota equation \eqref{hmkdveq} and the Sasa-Satsuma equation \eqref{ssmkdveq}.
Next we graphically illustrate 
that the oscillatory $2$-solitons \eqref{oscil2soliton} of these two equations 
describe collisions of oscillatory $1$-solitons with distinct speeds $c_1\neq c_2$, 
and that the oscillatory breathers \eqref{oscilbreather} of the equations 
describe solitary waves whose amplitude displays time-periodic oscillations 
with frequency $\nu$ and speed $c$. 

Finally, in section~\ref{conclude}
we make some concluding remarks. 

All computations in the paper have been carried out by use of Maple. 
Hereafter, by scaling variables $t,x,u$, we will put 
\EQ\label{scaling}
a=24, 
\quad
b=1
\endEQ 
for convenience.

\section{Derivation of harmonically modulated soliton solutions}
\label{bilinear} 

Consider a general $U(1)$-invariant complex mKdV equation
\begin{equation}
u_t +( \alpha u\bar u_x + \beta u_x\bar u )u + \gamma u_{xxx} =0
\label{genmkdveq}
\end{equation}
where $\alpha,\beta,\gamma$ are constants. 
A bilinear formulation of this equation can be obtained 
by the following steps \cite{Hirbook,GilHieNimOht}. 
First, we convert equation \eqref{genmkdveq} into a rational form 
through the standard transformation 
\begin{equation}
u = G/F, \quad
\bar u =\bar G/F
\end{equation}
where $F(t,x)$ is a real function and $G(t,x)$ is a complex function. 
Second, we express all derivatives of $F$, $G$ and $\bar G$ in terms of
Hirota's bilinear operator defined by 
\begin{align}
& D(f,g)=g Df-f Dg \\
& D^2(f,g)=gD^2 f + fD^2 g-2(Df)Dg \\
& D^3(f,g)=gD^3 f -3(Dg)D^2 f +3(Df)D^2 g -f D^3 g
\end{align}
where $D$ denotes a total derivative. 
Last, we split the resulting rational equation 
\begin{equation}
0= F^2(\gamma D_x^3(G,F) + D_t(G,F)) - \alpha GF D_x(G,\bar G) -(3\gamma D_x^2(F,F)-(\alpha+\beta)G\bar G)D_x(G,F)
\end{equation}
into a system of bilinear equations 
\begin{subequations}\label{genbilinsys}
\begin{align}
& D_t(G,F) + \gamma D_x^3(G,F) =GH 
\label{genbilineq1}\\
& 3\gamma D_x^2(F,F) -(\alpha+\beta) G\bar G =\lambda GH
\label{genbilineq2}\\
& \alpha D_x(G,\bar G)+\lambda D_x(G,F) = FH
\label{genbilineq3}
\end{align}
\end{subequations}
where $H(t,x)$ is an auxiliary function
and $\lambda$ is a complex constant. 

When the original equation \eqref{genmkdveq} is integrable, 
$N$-soliton solutions can be derived via the Hirota ansatz 
\begin{align}
& \lambda = 0 
\label{separable}\\
& G= \text{ complex polynomial of odd degree in } e^{\Theta_i},e^{\bar\Theta_j} 
\label{Gansatz}\\
& F= \text{ real polynomial of even degree in } e^{\Theta_i},e^{\bar\Theta_j} 
\label{Fansatz}\\
& H= \text{ complex polynomial of even degree in } e^{\Theta_i},e^{\bar\Theta_j} 
\label{Hansatz}
\end{align}
with 
\begin{equation}
\Theta_i = \k_i x + \w_i t, \quad
\bar\Theta_i =\bar\k_i x +\bar\w_i t,
\quad
i=1,\ldots, N
\end{equation}
where $\k_i$ and $\w_i$ are complex constants when harmonically modulated solitons are sought 
or real constants if ordinary solitons are sought instead. 
Substitution of this ansatz \eqref{separable}--\eqref{Hansatz} 
into the bilinear system \eqref{genbilinsys}
yields a system of algebraic equations given by 
polynomials in $e^{\Theta_i},e^{\bar\Theta_j}$, 
whose monomial coefficients must separately vanish. 
This system can be solved degree by degree,
starting at the respective degrees $1$, $2$, $2$ 
in equations \eqref{genbilineq1}, \eqref{genbilineq2}, \eqref{genbilineq3}, 
and stopping at some degree such that the corresponding highest-degree coefficients 
in the polynomials $G,F,H$ are found to vanish. 
(This termination generally will not occur unless the original equation is integrable.)

\subsection{Harmonically modulated Hirota solitons}
The bilinear system \eqref{genbilinsys} 
combined with the ansatz equation \eqref{separable}
applied to the Hirota equation 
\begin{equation}\label{hmkdveqscaled}
u_t +24 |u|^2 u_x + u_{xxx} =0
\end{equation}
gives
\begin{subequations}\label{hbilinsys}
\begin{align}
& H=0 
\\
& D_t(G,F) + D_x^3(G,F) = 0 
\\
& D_x^2(F,F) -8 G\bar G =0 .
\end{align}
\end{subequations}
To set up the soliton ansatz \eqref{Gansatz}--\eqref{Fansatz}, 
let all monomial terms of fixed degree $n\geq0$ in $G,F$ be denoted by
$G^{(n)},F^{(n)}$, so thus the ansatz is written as 
\begin{equation}\label{hsolitonansatz}
G= G^{(1)} +G^{(3)} +\cdots,
\quad
F= 1+ F^{(2)} + F^{(4)}+\cdots
\end{equation}
(where $F^{(0)}$ has been normalized to $1$ without loss of generality). 
Then it is straightforward to split the bilinear system \eqref{hbilinsys}
into a hierarchy of equations indexed by degree:
\begin{subequations}\label{hdegs}
\begin{align}
\label{hdeg1}
& D_t(G^{(1)},1) + D_x^3(G^{(1)},1) = 0 
\\
\label{hdeg2}
& D_x^2(F^{(2)},1) - 4 G^{(1)}\bar G^{(1)} =0
\\
\label{hdeg3}
& D_t(G^{(3)},1) + D_x^3(G^{(3)},1) = - D_t(G^{(1)},F^{(2)}) - D_x^3(G^{(1)},F^{(2)}) 
\\
\label{hdeg4}
& D_x^2(F^{(4)},1) - 4 (G^{(3)}\bar G^{(1)} + G^{(1)}\bar G^{(3)}) =-\tfrac{1}{2} D_x^2(F^{(2)},F^{(2)})
\\
&\text{\etc/}
\nonumber
\end{align}
\end{subequations}

The hierarchy of bilinear equations \eqref{hdegs} is solved in Appendix~\ref{A}
to obtain the harmonically modulated $1$-soliton and $2$-soliton solutions
for the Hirota equation \eqref{hmkdveqscaled}. 

\begin{prop}\label{prop:hrationalcosh1soliton}
The harmonically modulated $1$-soliton \eqref{envel1soliton}
for the Hirota equation \eqref{hmkdveqscaled}
has an explicit rational $\cosh$ form 
given by the envelope function 
(up to space translations on $x$ and phase shifts on $\phi$) 
\begin{equation}\label{hrationalcosh1soliton}
f_{\rm H}(\theta)= \frac{|k|}{2\cosh(\theta)},
\quad
\theta = k(x -(k^2-3\kappa^2)t)
\end{equation}
which is invariant under reflection $k\rightarrow -k$. 
\end{prop}

\begin{prop}\label{prop:henvel2soliton}
When 
\begin{equation}\label{henvel2solitoncond}
w_1/k_1\neq w_2/k_2,
\end{equation}
the harmonically modulated $2$-soliton \eqref{envel2soliton}
for the Hirota equation \eqref{hmkdveqscaled}
has an explicit rational $\cosh$ form given by 
the envelope functions (up to space-time translations on $t,x$) 
\begin{equation}\label{henvelfuncts}
f_{1\rm H}(\theta_1,\theta_2) =
X_{1\rm H}(\theta_1,\theta_2)/Y_{\rm H}(\theta_1,\theta_2),
\quad
f_{2\rm H}(\theta_1,\theta_2) =
X_{2\rm H}(\theta_1,\theta_2)/Y_{\rm H}(\theta_1,\theta_2)
\end{equation}
with 
\begin{equation}\label{hX1X2}
X_{1\rm H}(\theta_1,\theta_2) 
= |k_1|\sqrt{\Gamma} \cosh(\theta_2 +i\gamma_2),
\quad
X_{2\rm H}(\theta_1,\theta_2) 
= |k_2|\sqrt{\Gamma} \cosh(\theta_1 +i\gamma_1)
\end{equation}
\begin{equation}\label{hY}
Y_{\rm H}(\theta_1,\theta_2) 
= \cosh(\theta_1-\theta_2) +\Gamma\cosh(\theta_1+\theta_2)
-|\Gamma-1| \cos(\mu_1\theta_1-\mu_2\theta_2 +\phi_1-\phi_2)
\end{equation}
where $w_1,w_2,\omega_1,\omega_2$ are given by equation \eqref{w1w2rels}
and $\mu_1,\mu_2$ are given by equation \eqref{mus}, 
and where
\begin{gather}
\Gamma=\frac{(k_1-k_2)^2+(\kappa_1-\kappa_2)^2}{(k_1+k_2)^2+(\kappa_1-\kappa_2)^2} , 
\label{h2solitonGamma}\\
\begin{aligned}
& \gamma_1 = \arg\big( (k_1+k_2)(k_1-k_2)-(\kappa_1-\kappa_2)^2 + i2k_1(\kappa_1-\kappa_2) \big) , 
\\
& \gamma_2 = \arg\big( (k_1+k_2)(k_2-k_1)-(\kappa_1-\kappa_2)^2 -i2k_2(\kappa_1-\kappa_2) \big) , 
\end{aligned}
\label{h2solitongammas}\\
\begin{aligned}
& \theta_1 = k_1x +w_1t = k_1(x -(k_1^2-3\kappa_1^2)t) , 
\\
& \theta_2 = k_2x +w_2t = k_2(x -(k_2^2-3\kappa_2^2)t) . 
\end{aligned}
\label{henvelthetas}
\end{gather}
\end{prop}

The kinematic condition \eqref{henvel2solitoncond} will be seen later 
to have the interpretation that the harmonically modulated $2$-soliton 
describes a collision of harmonically modulated $1$-solitons with distinct speeds
$c_1=-w_1/k_1\neq c_2=-w_2/k_2$. 
These explicit expressions \eqref{henvelfuncts}--\eqref{henvelthetas}
for the harmonically modulated Hirota $2$-soliton \eqref{envel2soliton}
have not previously appeared in the literature. 
They reduce to the ordinary $2$-soliton solution \cite{Hir1972,AncNgaWil} 
for the Hirota equation 
in the case $\kappa_1=\kappa_2=0$. 

We remark that in the case of equal speeds $c_1=c_2$, 
Proposition~\ref{prop:henvel2soliton} remains valid 
if $\mu_1\theta_1-\mu_2\theta_2$ is replaced by $\vartheta_1-\vartheta_2$
through equation \eqref{Imthetaid},
and if $k_1x$ and $k_2x$ are respectively replaced by $k_1x+\chi$ and $k_2x-\chi$
through the equation $k_1(x_0+a_1)+w_1t_0= -k_2(x_0+a_2)-w_2t_0=-\chi$, 
where $\chi=(a_2-a_1)k_1k_2/(k_1+k_2)$ is a shift 
which cannot be absorbed by a space-time translation \eqref{spacetimetrans}. 

Finally, we examine the properties of the Hirota envelope functions
\eqref{henvelfuncts}--\eqref{hY} under reflections 
\begin{equation}\label{reflk1k2}
k_1 \rightarrow \mp k_1, 
\quad
k_2 \rightarrow \pm k_2
\end{equation}
which will be important when expressing the Hirota $2$-soliton solution 
in oscillatory form. 
We first note that 
$\theta_1\rightarrow \mp \theta_1$, $\theta_2\rightarrow \pm \theta_2$
and that $w_1\rightarrow \mp w_1$, $w_2\rightarrow \pm w_2$
while $\omega_1\rightarrow \omega_1$, $\omega_2\rightarrow \omega_2$
from equation \eqref{w1w2rels}. 
We then have 
\begin{gather}
\mu_1\rightarrow -\mu_1, 
\quad
\mu_2\rightarrow -\mu_2
\label{reflmus}
\\
\Gamma\rightarrow 1/\Gamma,
\quad
\gamma_1\rightarrow -\gamma_1,
\quad
\gamma_2\rightarrow -\gamma_2
\end{gather}
from equations \eqref{mus}, \eqref{h2solitonGamma}, \eqref{h2solitongammas},
and thus 
\begin{equation}
X_{1\rm H}\rightarrow (1/\Gamma) X_{1\rm H} ,
\quad
X_{2\rm H}\rightarrow (1/\Gamma) X_{2\rm H} ,
\quad
Y_{\rm H}\rightarrow (1/\Gamma)Y_{\rm H} . 
\end{equation}
These transformations establish the following reflection property. 

\begin{lem}\label{lem:hrefl}
The Hirota envelope functions \eqref{henvelfuncts} are invariant under 
reflections \eqref{reflk1k2}.
\end{lem}

\subsection{Harmonically modulated Sasa-Satsuma solitons}
For the Sasa-Satsuma equation 
\begin{equation}\label{ssmkdveqscaled}
u_t +6( u\bar u_x +3u_x\bar u )u + u_{xxx} =0
\end{equation}
the bilinear system \eqref{genbilinsys} 
combined with the ansatz equation \eqref{separable} gives
\begin{subequations}\label{ssbilinsys}
\begin{align}
& D_t(G,F) + D_x^3(G,F) = GH
\\
& D_x^2(F,F) -8 G\bar G =0
\\
& 6D_x(G,\bar G) =FH
\end{align}
\end{subequations}
which is more complicated than in the case of the Hirota equation.
To set up the soliton ansatz \eqref{Gansatz}--\eqref{Fansatz}, 
let all monomial terms of fixed degree $n\geq0$ in $G,F,H$ be denoted by
$G^{(n)},F^{(n)},H^{(n)}$. 
The ansatz is thus written as 
\begin{equation}\label{sssolitonansatz}
G= G^{(1)} +G^{(3)} +\cdots,
\quad
F= 1+ F^{(2)} + F^{(4)}+\cdots,
\quad
H= H^{(2)} + H^{(4)}+\cdots
\end{equation}
(where $F^{(0)}$ has been normalized to $1$ without loss of generality). 
Then the bilinear system \eqref{ssbilinsys} splits into 
a hierarchy of equations indexed by degree:
\begin{subequations}\label{ssdegs}
\begin{align}
\label{ssdeg1}
& D_t(G^{(1)},1) + D_x^3(G^{(1)},1) = 0 
\\
\label{ssdeg2}
& D_x^2(F^{(2)},1) - 4 G^{(1)}\bar G^{(1)} =0
\\
\label{ssH2}
& H^{(2)} = 6D_x(G^{(1)},\bar G^{(1)})
\\
\label{ssdeg3}
& D_t(G^{(3)},1) + D_x^3(G^{(3)},1) = - D_t(G^{(1)},F^{(2)}) - D_x^3(G^{(1)},F^{(2)}) +G^{(1)} H^{(2)}
\\
\label{ssdeg4}
& D_x^2(F^{(4)},1) - 4 (G^{(3)}\bar G^{(1)} + G^{(1)}\bar G^{(3)}) = -\tfrac{1}{2} D_x^2(F^{(2)},F^{(2)})
\\
\label{ssH4}
& H^{(4)} = 6D_x(G^{(1)},\bar G^{(3)}) + 6D_x(G^{(3)},\bar G^{(1)}) -F^{(2)} H^{(2)}
\\
\label{ssdeg5}
& \begin{aligned}
D_t(G^{(5)},1) + D_x^3(G^{(5)},1) =& - D_t(G^{(3)},F^{(2)}) - D_x^3(G^{(3)},F^{(2)}) 
+G^{(3)}H^{(2)} 
\\&\qquad 
- D_t(G^{(1)},F^{(4)}) - D_x^3(G^{(1)},F^{(4)}) +G^{(1)} H^{(4)} 
\end{aligned}
\\
\label{ssdeg6}
& D_x^2(F^{(6)},1) - 4 (G^{(5)}\bar G^{(1)} + G^{(1)}\bar G^{(5)}) = 4 G^{(3)}\bar G^{(3)} - D_x^2(F^{(4)},F^{(2)})
\\
&\text{\etc/}
\nonumber
\end{align}
\end{subequations}
Note that $H^{(2)}$, $H^{(4)}$, \etc/ can be successively eliminated through
equations \eqref{ssH2}, \eqref{ssH4}, and so on. 

The hierarchy of bilinear equations \eqref{ssdegs} is solved in Appendix~\ref{B}
to obtain the harmonically modulated $1$-soliton and $2$-soliton solutions
for the Sasa-Satsuma equation \eqref{ssmkdveqscaled}. 

\begin{prop}\label{prop:ssrationalcosh1soliton}
The harmonically modulated $1$-soliton \eqref{envel1soliton} 
for the Sasa-Satsuma equation \eqref{ssmkdveqscaled}
has an explicit rational $\cosh$ form when $\kappa\neq0$ 
given by the envelope function 
(up to space translations on $x$ and phase shifts on $\phi$) 
\begin{equation}\label{ssrationalcosh1soliton}
f_{\rm SS}(\theta)= 
\frac{|k|(2|\kappa|)^{1/2}(k^2+\kappa^2)^{1/4} \cosh(\theta +i\lambda/2)}{|\kappa|\cosh(2\theta) +(k^2+\kappa^2)^{1/2}},
\quad
\theta = k(x -(k^2-3\kappa^2)t)
\end{equation}
where $\lambda$ is given by equation \eqref{sslam}.
This function is invariant under reflections $k\rightarrow -k$. 
\end{prop}

\begin{prop}\label{prop:ssenvel2soliton}
When 
\begin{equation}\label{ssenvel2solitoncond}
w_1/k_1\neq w_2/k_2,
\quad
\kappa_1\neq 0,
\quad
\kappa_2\neq 0
\end{equation}
the harmonically modulated $2$-soliton \eqref{envel2soliton} 
for the Sasa-Satsuma equation \eqref{ssmkdveqscaled}
has an explicit rational $\cosh$ form 
given by the envelope functions
(up to space-time translations on $t,x$) 
\begin{equation}\label{ssenvelfuncts}
f_{1\rm SS}(\theta_1,\theta_2) =
X_{1\rm SS}(\theta_1,\theta_2)/Y_{\rm SS}(\theta_1,\theta_2),
\quad
f_{2\rm SS}(\theta_1,\theta_2) =
X_{2\rm SS}(\theta_1,\theta_2)/Y_{\rm SS}(\theta_1,\theta_2)
\end{equation}
with 
\begin{equation}\label{ssenvelX1}
\begin{aligned}
X_{1\rm SS}(\theta_1,\theta_2) = & 
|k_1|(k_1^2+\kappa_1^2)^{1/4} |2\kappa_1|^{1/2}
\times
\\&\bigg(
|\kappa_2|\Big( \sqrt{\Delta\Gamma} \cosh(\theta_1+2\theta_2+i(\alpha_2+\gamma_2))
+\frac{1}{\sqrt{\Delta\Gamma}}\cosh(\theta_1-2\theta_2+i(\upsilon_2-\gamma_2)) \Big)
\\&\qquad\qquad
+(k_2^2+\kappa_2^2)^{1/2}\Big( 
-8k_2^2 |\kappa_2|\frac{1}{\sqrt{\Omega\Upsilon}} \cosh(\theta_1+i(\rho+\rho_2+\varpi_1+\gamma_1))
\\&\qquad\qquad
+\frac{\sqrt{\Gamma}}{\sqrt{\Delta}} \cosh(\theta_1+i(\alpha_2-\gamma_2))
+ \frac{\sqrt{\Delta}}{\sqrt{\Gamma}} \cosh(\theta_1+i(\upsilon_2+\gamma_2)) \Big)
\bigg)
\\&
-k_1^2 |k_2|(k_2^2+\kappa_2^2)^{1/4} |8\kappa_2|^{1/2} \times
\\&\bigg(
\frac{\sqrt{\Omega}}{\sqrt{\Upsilon}} 
\exp(i(\mu_1\theta_1-\mu_2\theta_2+\phi_1-\phi_2)) \cosh(\theta_2-i(\rho+\rho_1+\varpi_2-\gamma_2)) 
\bigg) , 
\end{aligned}
\end{equation}
\begin{equation}
\begin{aligned}\label{ssenvelX2}
X_{2\rm SS}(\theta_1,\theta_2) = & 
|k_2|(k_2^2+\kappa_2^2)^{1/4} |2\kappa_2|^{1/2}
\times
\\&\bigg(
|\kappa_1|\Big( \sqrt{\Delta\Gamma} \cosh(\theta_2+2\theta_1+i(\alpha_1+\gamma_1))
+\frac{1}{\sqrt{\Delta\Gamma}}\cosh(\theta_2-2\theta_1+i(\upsilon_1-\gamma_1)) \Big)
\\&\qquad\qquad
+(k_1^2+\kappa_1^2)^{1/2}\Big( 
-8k_1^2 |\kappa_1| \frac{1}{\sqrt{\Omega\Upsilon}} 
\cosh(\theta_2+i(\rho+\rho_1+\varpi_2+\gamma_2)) 
\\&\qquad\qquad
+\frac{\sqrt{\Gamma}}{\sqrt{\Delta}} \cosh(\theta_2+i(\alpha_1-\gamma_1))
+ \frac{\sqrt{\Delta}}{\sqrt{\Gamma}} \cosh(\theta_2+i(\upsilon_1+\gamma_1)) \Big)
\bigg)
\\&
-k_2^2 |k_1|(k_1^2+\kappa_1^2)^{1/4} |8\kappa_1|^{1/2} \times
\\&\bigg(
\frac{\sqrt{\Omega}}{\sqrt{\Upsilon}} 
\exp(i(\mu_2\theta_2-\mu_1\theta_1+\phi_2-\phi_1)) \cosh(\theta_1-i(\rho+\rho_2+\varpi_1-\gamma_1)) 
\bigg) , 
\end{aligned}
\end{equation}
\begin{equation}\label{ssenvelY}
\begin{aligned}
Y_{\rm SS}(\theta_1,\theta_2)= &
(k_1^2+\kappa_1^2)^{1/2}(k_2^2+\kappa_2^2)^{1/2}
\Big(\frac{\Gamma}{\Delta} +\frac{\Delta}{\Gamma} 
+64k_1^2 k_2^2\kappa_1\kappa_2 \frac{1}{\Omega\Upsilon}\Big)
\\&
+|\kappa_1\kappa_2|\Big( \Delta\Gamma \cosh(2(\theta_1+\theta_2))
+ \frac{1}{\Delta\Gamma} \cosh(2(\theta_1-\theta_2)) \Big)
\\&
+2|\kappa_2|(k_1^2+\kappa_1^2)^{1/2} \cosh(2\theta_2) 
+ 2|\kappa_1|(k_2^2+\kappa_2^2)^{1/2} \cosh(2\theta_1)
\\&
+4k_1^2 k_2^2\frac{\Omega}{\Upsilon}
\cos(2(\mu_1\theta_1-\mu_2\theta_2)+2(\phi_1-\phi_2)+\rho_2-\rho_1)
\\&
-16 |k_1k_2||\kappa_1\kappa_2|^{1/2} (k_1^2+\kappa_1^2)^{1/4}(k_2^2+\kappa_2^2)^{1/4}
\Re\Big( \exp(i(\mu_1\theta_1-\mu_2\theta_2+\phi_1-\phi_2))
\times
\\&\qquad
\Big( \frac{\sqrt{\Gamma}}{\sqrt{\Upsilon}} \cosh(\theta_1+\theta_2+i(\varpi_1-\varpi_2))
+ \frac{1}{\sqrt{\Gamma\Upsilon}} \cosh(\theta_1-\theta_2+i(\varpi_1+\varpi_2)) \Big) \Big) , 
\end{aligned}
\end{equation}
where $w_1,w_2,\omega_1,\omega_2$ are given by equation \eqref{w1w2rels}
and $\mu_1,\mu_2$ are given by equation \eqref{mus}, 
and where
\begin{gather}
\Omega = \sqrt{\big((k_1+k_2)^2+(\kappa_1+\kappa_2)^2\big)\big((k_1 -k_2)^2+(\kappa_1 +\kappa_2)^2\big)} , 
\label{ss2solitonOmega}
\\
\Upsilon = \big((k_1-k_2)^2+(\kappa_1 -\kappa_2)^2\big)\big((k_1+k_2)^2+(\kappa_1-\kappa_2)^2\big) ,
\label{ss2solitonUpsilon}
\\
\Delta = \sqrt{\frac{(k_1-k_2)^2+(\kappa_1+\kappa_2)^2}{(k_1 +k_2)^2+(\kappa_1 +\kappa_2)^2}} ,
\label{ss2solitonDelta}
\\
\Gamma = \frac{(k_1-k_2)^2+(\kappa_1-\kappa_2)^2}{(k_1 +k_2)^2+(\kappa_1 -\kappa_2)^2} ,
\label{ss2solitonGamma}
\\
\alpha_1 = (\lambda_2+\delta_1)/2 , 
\quad
\alpha_2 = (\lambda_1+\delta_2)/2 ,
\label{ss2solitonalphas}
\\
\upsilon_1 = (\lambda_2-\delta_1)/2 , 
\quad
\upsilon_2 = (\lambda_1-\delta_2)/2 ,
\label{ss2solitonupsilons}
\\
\varpi_1=(\lambda_1 -\delta_1)/2 , 
\quad
\varpi_2=(\lambda_2 -\delta_2)/2 ,
\label{ss2solitonvarpis}
\\
\begin{aligned}
\gamma_1=\arg( k_1^2-k_2^2 -(\kappa_1-\kappa_2)^2 +i2k_1(\kappa_1-\kappa_2) ) ,
\\
\gamma_2=\arg( k_2^2-k_1^2 -(\kappa_1-\kappa_2)^2 -i2k_2(\kappa_1-\kappa_2) ) ,
\end{aligned}
\label{ss2solitongammas}
\\
\begin{aligned}
\delta_1=\arg( k_2^2-k_1^2 +(\kappa_1+\kappa_2)^2 +i2k_1(\kappa_1+\kappa_2) ) ,
\\
\delta_2=\arg( k_1^2-k_2^2 +(\kappa_1+\kappa_2)^2+i2k_2(\kappa_1+\kappa_2) ) ,
\end{aligned}
\label{ss2solitondeltas}
\\
\lambda_1 = \arg( \kappa_1(\kappa_1+ik_1) ) ,
\quad
\lambda_2 = \arg( \kappa_2(\kappa_2+ik_2) ) ,
\label{ss2solitonlambdas}
\\
\begin{aligned}
& \theta_1 = k_1x +w_1t = k_1(x -(k_1^2-3\kappa_1^2)t) ,
\\
& \theta_2 = k_2x +w_2t = k_2(x -(k_2^2-3\kappa_2^2)t) .
\end{aligned}
\label{ssenvelthetas}
\end{gather}
\end{prop}

It will be useful to write out the half-angle expressions for 
$\lambda_1/2$, $\lambda_2/2$, $\delta_1/2$, $\delta_2/2$ 
appearing in the envelope functions \eqref{ssenvelX1}--\eqref{ssenvelY}. 
From the angle expressions \eqref{ss2solitonlambdas} and \eqref{ss2solitondeltas}, 
we obtain by a direct calculation 
\begin{align}
\lambda_1/2 = \arg\Big(\sqrt{\sqrt{1+k_1^2/\kappa_1^2} +1} +i\varepsilon_1\sqrt{\sqrt{1+k_1^2/\kappa_1^2} -1}\Big) ,
\label{ss2solitonhalflambda1}
\\
\lambda_2/2 = \arg\Big(\sqrt{\sqrt{1+k_2^2/\kappa_2^2} +1} +i\varepsilon_2\sqrt{\sqrt{1+k_2^2/\kappa_2^2} -1}\Big) ,
\label{ss2solitonhalflambda2}
\end{align}
where 
\begin{equation}\label{sgns}
\varepsilon_1=\sgn(\kappa_1) ,
\quad
\varepsilon_2=\sgn(\kappa_2) ,
\end{equation}
and
\begin{align}
\begin{aligned}
\delta_1/2=& 
\arg\Big( (\epsilon^2\epsilon_- + (1-\epsilon^2\epsilon_-)\sgn(\kappa_1+\kappa_2))
 \sqrt{k_2^2-k_1^2 +(\kappa_1+\kappa_2)^2+\Omega} 
\\&\qquad
+i \varepsilon_1( 1+\epsilon^2\epsilon_-(\sgn(\kappa_1+\kappa_2)-1)) \sqrt{k_2^2-k_1^2 +(\kappa_1+\kappa_2)^2-\Omega} \Big) ,
\end{aligned}
\label{ss2solitonhalfdelta1}
\\
\begin{aligned}
\delta_2/2=&
\arg\Big( (\epsilon^2\epsilon_+ + (1-\epsilon^2\epsilon_+)\sgn(\kappa_1+\kappa_2))\sqrt{k_1^2-k_2^2 +(\kappa_1+\kappa_2)^2+\Omega} 
\\&\qquad 
+i\varepsilon_2( 1+\epsilon^2\epsilon_+(\sgn(\kappa_1+\kappa_2)-1))\sqrt{k_1^2-k_2^2 +(\kappa_1+\kappa_2)^2-\Omega} \Big) ,
\end{aligned}
\label{ss2solitonhalfdelta2}
\end{align}
where 
\begin{equation}\label{sgnss}
\epsilon_\pm = (1\pm \epsilon)/2,  
\quad
\epsilon=\sgn(|k_1|-|k_2|) 
=\begin{cases}
1, & |k_1|>|k_2|\\
-1, & |k_1|<|k_2|\\
0, & |k_1|=|k_2|
\end{cases} .
\end{equation}
Then, combining expressions \eqref{ss2solitonhalfdelta1} and \eqref{ss2solitonhalfdelta2}, we have 
\begin{equation}
(\delta_1\pm\delta_2)/2= \arg\Big( \varepsilon (\kappa_1+\kappa_2 +i(k_1\pm k_2)) \Big)
\end{equation}
where 
\begin{equation}\label{sssgn}
\varepsilon=\epsilon^2 +(1-\epsilon^2)\sgn(\kappa_1+\kappa_2)
=\begin{cases}
1, & |k_1|\neq|k_2|\\
\sgn(\kappa_1+\kappa_2), & |k_1|=|\k_2|
\end{cases} .
\end{equation} 
Additionally, we note 
\begin{equation}\label{ssrhos}
\exp(i\rho_1)=\varepsilon_1 ,
\qquad
\exp(i\rho_2)=\varepsilon_2 ,
\qquad
\exp(i\rho)=\varepsilon .
\end{equation} 

These explicit expressions \eqref{ssenvelfuncts}--\eqref{ssenvelthetas}
and \eqref{ss2solitonhalflambda1}--\eqref{ssrhos}
for the harmonically modulated Sasa-Satsuma $2$-soliton \eqref{envel2soliton}
have not previously appeared in the literature. 
The kinematic condition \eqref{ssenvel2solitoncond} will be seen later 
to imply that this solution describes a collision of 
harmonically modulated $1$-solitons with distinct speeds
$c_1=-w_1/k_1\neq c_2=-w_2/k_2$. 

In the case of equal speeds $c_1=c_2$, 
we remark that Proposition~\ref{prop:ssenvel2soliton} remains valid 
if $\mu_1\theta_1-\mu_2\theta_2$ is replaced by $\vartheta_1-\vartheta_2$
through equation \eqref{Imthetaid},
and if $k_1x$ and $k_2x$ are respectively replaced by $k_1x+\chi$ and $k_2x-\chi$
through the equation $k_1(x_0+a_1)+w_1t_0= -k_2(x_0+a_2)-w_2t_0=-\chi$, 
where $\chi=(a_2-a_1)k_1k_2/(k_1+k_2)$ is a shift 
which cannot be absorbed by a space-time translation \eqref{spacetimetrans}. 
The solution in this particular case has previously appeared in an equivalent 
envelope form in \Ref{Mih1993,Mih1994}.

For expressing the Sasa-Satsuma $2$-soliton in oscillatory form, 
the properties of the envelope functions \eqref{ssenvelfuncts}--\eqref{ssenvelY}
under reflections \eqref{reflk1k2} will be important. 
As in the Hirota case, we have 
\begin{gather}
\theta_1\rightarrow \mp \theta_1 ,
\quad
\theta_2\rightarrow \pm \theta_2 ,
\\
\mu_1\rightarrow -\mu_1 , 
\quad
\mu_2\rightarrow -\mu_2 .
\end{gather}
Also, from equations \eqref{ss2solitonOmega}--\eqref{ss2solitonlambdas}, 
we have 
\begin{gather}
\Omega \rightarrow \Omega ,
\quad
\Upsilon\rightarrow \Upsilon ,
\\
\Gamma\rightarrow 1/\Gamma ,
\quad
\Delta\rightarrow 1/\Delta ,
\\
\lambda_1\rightarrow \mp\lambda_1 ,
\quad
\lambda_2\rightarrow \pm\lambda_2 ,
\\
\gamma_1\rightarrow \mp\gamma_1 ,
\quad
\gamma_2\rightarrow \pm\gamma_2 ,
\\
\delta_1\rightarrow \mp\delta_1 ,
\quad
\delta_2\rightarrow \pm\delta_2 .
\end{gather}
These transformations yield
\begin{equation}
X_{1\rm SS}\rightarrow X_{1\rm SS},
\quad
X_{2\rm SS}\rightarrow X_{2\rm SS},
\quad
Y_{\rm SS}\rightarrow Y_{\rm SS} .
\end{equation}
which establishes the following reflection property. 

\begin{lem}\label{lem:ssrefl}
The Sasa-Satsuma envelope functions \eqref{ssenvelfuncts} are invariant under 
reflections \eqref{reflk1k2}.
\end{lem}

\section{Oscillatory parameterization}
\label{kinematics} 

The harmonically modulated $1$-soliton solutions shown in 
Proposition~\ref{prop:hrationalcosh1soliton} 
for the Hirota equation \eqref{hmkdveqscaled}
and Proposition~\ref{prop:ssrationalcosh1soliton} 
for the Sasa-Satsuma equation \eqref{ssmkdveqscaled}
will now be expressed in the more physical oscillatory form \eqref{oscil1soliton}.

We write 
\begin{equation}\label{oscilrels}
kx+wt = k(x-ct) ,
\quad
\kappa x+\omega t = \kappa (x-ct)+\nu t ,
\end{equation}
where
\begin{equation}
c= -w/k,
\quad 
\nu = \omega -w\kappa/k . 
\end{equation}
From relations \eqref{wkrels}, \eqref{scaling} for $w$ and $\omega$, 
we get 
\begin{gather}
c= k^2-3\kappa^2 ,
\label{speed}
\\
\nu = -2\kappa(k^2+\kappa^2) ,
\label{freq}
\end{gather}
with 
\begin{equation}
\kappa\neq 0 . 
\end{equation}
By combining these equations \eqref{speed} and \eqref{freq}, 
we have a cubic equation that determines $\kappa$, 
\begin{equation}\label{kappaeq}
8\kappa^3+2\kappa c+\nu=0 
\end{equation}
and an elementary quadratic equation that determines $|k|$, 
\begin{equation}\label{keq}
k^2=3\kappa^2+c  . 
\end{equation}
The discriminant of the cubic equation \eqref{kappaeq} is 
\begin{equation}
\mathit{\Delta} = -\frac{64}{81}(\tilde{c}^3+\tilde{\nu}^2)
\end{equation}
where
\begin{equation}\label{cnu}
\tilde{c}=c/3 ,
\quad
\tilde{\nu}=\nu/2 \neq 0 . 
\end{equation}
There are three cases to consider. 

First, if $\mathit{\Delta}<0$, \ie/ $\tilde{c}^3>-\tilde{\nu}^2$, 
then equation \eqref{kappaeq} has only one real root, 
\begin{equation}\label{kapparoot}
\kappa=\frac{\beta_{-} - \beta_{+}}{2}
\end{equation}
where 
\begin{equation}
\beta_{\pm} =
\cubrt{\sqrt{\tilde{c}^3 +\tilde{\nu}^2} \pm \tilde{\nu}}
%\sqrt[3]{\mathstrut}\big(\sqrt{\tilde{c}^3 +\tilde{\nu}^2} \pm \tilde{\nu}\big)
\end{equation}
which is defined as the real cube root. 
Equation \eqref{keq} then becomes
\begin{equation}
k^2 = 3(\kappa^2+\tilde{c}) =\frac{3(\beta_{-} + \beta_{+})^2}{4} 
\end{equation}
so thus 
\begin{equation}\label{kroot}
|k|=\frac{\sqrt{3}(\beta_{-} + \beta_{+})}{2}
\end{equation}
where $\beta_{+} > -\beta_{-} >0$ when $\tilde\nu>0$
and $\beta_{-} > -\beta_{+} >0$ when $\tilde\nu<0$. 

Next, if $\mathit{\Delta}=0$, \ie/ $\tilde{c}^3=-\tilde{\nu}^2$, 
then equation \eqref{kappaeq} has three real roots, two of which are repeated, 
\begin{equation}
\kappa= \beta_{-}, 
\quad
\kappa=\beta_{+}/2
\quad\text{(repeated)}    
\end{equation}
where $\beta_{\pm} = \pm \cubrt{\tilde{\nu}}\neq 0$. 
The single root coincides with the previous real root \eqref{kapparoot},
while the repeated roots violate equation \eqref{keq} because 
\begin{equation}
0\leq k^2 = 3(\kappa^2+\tilde{c})
=3( (\beta_{+}/2)^2 - \beta_{+}^2 ) =-(3\beta_{+}/2)^2 <0 . 
\end{equation}

Last, if $\mathit{\Delta}>0$, \ie/ $\tilde{c}^3<-\tilde{\nu}^2$,
then equation \eqref{kappaeq} has three distinct real roots, 
\begin{equation}
\begin{gathered}
\kappa =
|\beta_{+}|\cos(\psi),
\\
\psi = \arg\beta_{+}, 
\quad 
\psi=\arg\beta_{+}+2\pi/3,
\quad
\psi=\arg\beta_{+}-2\pi/3
\end{gathered}
\end{equation}
where $|\beta_{\pm}| = \sqrt{-\tilde{c}}\neq 0$
and $\tan(\arg\beta_{\pm}) = \pm \sqrt{|\tilde{c}^3 +\tilde{\nu}^2|}/\tilde{\nu} \neq 0$.
But all three roots violate equation \eqref{keq}, 
\begin{equation}
0\leq k^2 = 3(\kappa^2+\tilde{c})=3|\beta_{+}|^2\big(\cos^2(\psi)-1\big) < 0
\end{equation}
due to $|\cos(\psi)|\neq 1$ which is the condition for no roots to be repeated.
 
Hence we have established the following main identities.

\begin{lem}\label{lem:oscilform}
(i) Let $w=-k(k^2-3\kappa^2)$, $\omega= -\kappa(3k^2-\kappa^2)$, 
and $\kappa \neq 0$. 
Then 
\begin{equation}
|w|=-c|k|,
\quad
\omega = c\kappa +\nu
\end{equation}
is an identity, where $\kappa$ and $|k|$ are given by 
equations \eqref{cnu}, \eqref{kapparoot}, \eqref{kroot}
in terms of $c$ and $\nu$. 
(ii) Let the function $f(kx+wt)=f(k(x-ct))$ be invariant 
under reflection $k\rightarrow -k$. 
Then the harmonically modulated function $\exp(i(\kappa x +\omega t))f(kx+wt)$ 
is reflection invariant, 
in which case it can be expressed in the equivalent form 
\begin{equation}\label{enveloscilrels}
\exp(i(\kappa x +\omega t))f(kx+wt) 
= \exp(i\nu t)\tilde f(x-ct),
\quad
\tilde f(\xi) = \exp(i\kappa\xi)f(k\xi) 
\end{equation}
in terms of 
\begin{gather}
k = \frac{\sqrt{3}}{2}\Big( \cubrt{\sqrt{(c/3)^3 +(\nu/2)^2} - \nu/2} 
+ \cubrt{\sqrt{(c/3)^3 +(\nu/2)^2} + \nu/2}\; \Big) , 
\label{krel}
\\
\kappa = \frac{1}{2}\Big( \cubrt{\sqrt{(c/3)^3 +(\nu/2)^2} - \nu/2} 
- \cubrt{\sqrt{(c/3)^3 +(\nu/2)^2} + \nu/2}\; \Big) , 
\label{kapparel}
\end{gather}
where
\begin{equation}\label{kinrel}
(c/3)^3 +(\nu/2)^2 \geq 0 . 
\end{equation}
\end{lem}

Applying Lemma~\ref{lem:oscilform} to the Hirota and Sasa-Satsuma 
harmonically modulated $1$-solitons, 
which are given by the reflection-invariant envelope functions 
\eqref{hrationalcosh1soliton} and \eqref{ssrationalcosh1soliton}, 
we obtain the following result. 

\begin{thm}\label{thm:1soliton}
The equivalent oscillatory form \eqref{oscil1soliton} of 
a harmonically modulated $1$-soliton \eqref{envel1soliton} is parameterized by 
a phase angle $\phi$, a temporal frequency $\nu$, and a speed $c$,
which satisfy the kinematic relation \eqref{kinrel},
where $k,\kappa$ are given in terms of $c,\nu$ by equations \eqref{krel}, \eqref{kapparel}. 
For the Hirota equation \eqref{hmkdveqscaled}
and the Sasa-Satsuma equation \eqref{ssmkdveqscaled}, 
the oscillatory form of the harmonically modulated $1$-soliton solutions 
\eqref{hrationalcosh1soliton} and \eqref{ssrationalcosh1soliton}
expressed using a travelling wave coordinate $\xi=x-ct$ 
is given by 
\begin{equation}\label{1soliton}
u(t,x) = \exp(i\phi)\exp(i\nu t) \tilde f(\xi) 
\end{equation}
in terms of the respective functions
\begin{gather}
\tilde f_{\rm H}(\xi) = 
\frac{k\exp(i\kappa\xi)}{2\cosh(k\xi)} , 
\label{hoscilfunct}
\\
\tilde f_{\rm SS}(\xi) = 
\frac{k(2|\kappa|)^{1/2}(k^2+\kappa^2)^{1/4}\exp(i\kappa\xi) \cosh(k\xi +i\lambda/2)}{|\kappa|\cosh(2k\xi) +(k^2+\kappa^2)^{1/2}} ,
\quad
\kappa\neq 0 , 
\label{ssoscilfunct}
\end{gather}
where $\lambda$ is given by equation \eqref{sslam}.
When $c=0$ (and $\kappa\neq0$), 
these $1$-soliton solutions are standing waves 
(\ie/ harmonically modulated stationary solitons). 
\end{thm}

We emphasize that, due to the invariance of the
Hirota and Sasa-Satsuma envelope functions 
\eqref{hrationalcosh1soliton} and \eqref{ssrationalcosh1soliton}
under reflection $k\rightarrow -k$, 
the parameters $(\kappa,k)$ and $(\kappa,-k)$ 
correspond to the same harmonically-modulated 1-soliton. 
Thus, the change of parameterization from $(\kappa,\pm k)$ to $(\nu,c)$
in Theorem~\ref{thm:1soliton} is one-to-one,
such that the parameter range 
\begin{equation}
0\leq |\kappa| <\infty,
\quad
0< k <\infty
\end{equation}
corresponds to the kinematic range 
\begin{equation}
0\leq |\nu| <\infty,
\quad
-(3/\cubrt{4})(\cubrt{|\nu|})^2 < c <\infty 
\end{equation}
through the relations \eqref{speed}--\eqref{freq}
and \eqref{krel}--\eqref{kapparel}.

From relations \eqref{kapparel} and \eqref{krel}, 
we note that $\kappa=0$ iff $\nu=0$ and that $k=0$ iff $(c/3)^3 +(\nu/2)^2 =0$. 
Consequently, Theorem~\ref{thm:1soliton} implies the following result. 

\begin{cor}\label{cor:1soliton}
For the Hirota and Sasa-Satsuma equations, 
a harmonically modulated $1$-soliton solution 
\begin{equation}\label{enveloscilwave}
u(t,x) = \exp(i\phi)\exp(i(\kappa x +\omega t)) f(kx+wt) 
=\exp(i\phi)\exp(i\nu t) \tilde f(x-ct)
\end{equation}
is distinguished from an ordinary $1$-soliton solution 
\begin{equation}\label{travelwave}
u(t,x) = \exp(i\phi) f(x-ct)
\end{equation}
by the kinematic conditions
\begin{equation}
\nu \neq 0 , 
\quad
(c/3)^3 +(\nu/2)^2 >0 , 
\end{equation}
or equivalently 
\begin{equation}\label{1solitonkinrel}
c > -(3/\cubrt{4})(\cubrt{\nu})^2 \neq 0 . 
\end{equation}
\end{cor}

Next, the harmonically modulated $2$-soliton solutions shown 
in Proposition~\ref{prop:henvel2soliton} 
for the Hirota equation \eqref{hmkdveqscaled}
and Proposition~\ref{prop:ssenvel2soliton} 
for the Sasa-Satsuma equation \eqref{ssmkdveqscaled}
will be expressed in the analogous oscillatory form \eqref{oscil2soliton}.

From Lemmas~\ref{lem:hrefl} and~\ref{lem:ssrefl}, 
the envelope functions \eqref{henvelfuncts} and \eqref{ssenvelfuncts}
in these $2$-soliton solutions are invariant under reflections \eqref{reflk1k2},
so thus the parameters 
$(\kappa_1,\kappa_2,\pm k_1,k_2)$ and $(\kappa_1,\kappa_2,k_1,\pm k_2)$ 
correspond to the same solution. 
This invariance property allows Lemma~\ref{lem:oscilform} 
to be applied as follows. 
We define 
\begin{gather}
k_1 = \sqrt{3}(\beta_{1-} + \beta_{1+})/2 ,
\quad
\kappa_1 = (\beta_{1-} - \beta_{1+})/2 ,
\label{rels1}
\\
k_2 = \sqrt{3}(\beta_{2-} + \beta_{2+})/2 ,
\quad
\kappa_2 = (\beta_{2-} - \beta_{2+})/2 ,
\label{rels2}
\end{gather}
in terms of 
\begin{equation}\label{betapms}
\begin{aligned}
&
\beta_{1\pm} = \cubrt{\sqrt{(c_1/3)^3 +(\nu_1/2)^2} \pm \nu_1/2}, 
\\
&
\beta_{2\pm} = \cubrt{\sqrt{(c_2/3)^3 +(\nu_2/2)^2} \pm \nu_2/2} , 
\end{aligned}
\end{equation}
where
\begin{equation}\label{kinrels}
(c_1/3)^3 +(\nu_1/2)^2 \geq 0 ,
\quad
(c_2/3)^3 +(\nu_2/2)^2 \geq 0 .
\end{equation}
We then have the identities
\begin{equation}
\begin{aligned}
& c_1= -w_1/k_1 ,
\quad 
\nu_1 = \omega_1 -w_1\kappa_1/k_1 ,
\\
& c_2= -w_2/k_2,
\quad 
\nu_2 = \omega_2 -w_2\kappa_2/k_2 ,
\end{aligned}
\end{equation}
holding from the algebraic relations \eqref{w1w2rels}. 
Through these expressions, the identity \eqref{Imthetaid}--\eqref{mus}
becomes
\begin{equation}
(\kappa_1-\kappa_2)x+(\omega_1-\omega_2)t
= (\kappa_1-\mu)(x-c_1t) - (\kappa_2-\mu)(x-c_2t) ,
\quad
c_1\neq c_2
\end{equation}
where 
\begin{equation}
\mu = (\nu_1-\nu_2)/(c_1-c_2) .
\label{mu}
\end{equation}

We emphasize that this change of parameterization 
from $(\kappa_1,\pm k_1)$ and $(\kappa_2,\pm k_2)$ to $(\nu_1,c_1)$ and $(\nu_2,c_2)$ 
is one-to-one, 
such that the parameter range 
\begin{equation}
0\leq |\kappa_1| <\infty, 
\quad
0\leq |\kappa_2| <\infty, 
\quad
0< k_1 <\infty ,
\quad
0< k_2 <\infty 
\end{equation}
corresponds to the kinematic range 
\begin{equation}
\begin{gathered}
0\leq |\nu_1| <\infty,
\\
0\leq |\nu_2| <\infty,
\\
-(3/\cubrt{4})(\cubrt{|\nu_1|})^2 < c_1 <\infty,
\\
-(3/\cubrt{4})(\cubrt{|\nu_2|})^2 < c_2 <\infty. 
\end{gathered}
\end{equation}
This leads to the following main result. 

\begin{thm}\label{thm:2soliton}
The equivalent oscillatory form \eqref{oscil2soliton} of 
a harmonically modulated $2$-soliton \eqref{envel2soliton}
has parameters $\phi_1,\phi_2,\nu_1,\nu_2,c_1,c_2$,
satisfying the kinematic relations \eqref{kinrels},
where $k_1,\kappa_1$ are given in terms of $c_1,\nu_1$ by equation \eqref{rels1},
and $k_2,\kappa_2$ are given in terms of $c_2,\nu_2$ by equation \eqref{rels2}.
As expressed using travelling wave coordinates 
$\xi_1=x-c_1 t$ and $\xi_2=x-c_2 t$ when $c_1\neq c_2$, 
the oscillatory form for the $2$-soliton solutions 
\eqref{henvelfuncts}--\eqref{henvelthetas} 
for the Hirota equation \eqref{hmkdveqscaled}
and \eqref{ssenvelfuncts}--\eqref{ssenvelthetas}
for the Sasa-Satsuma equation \eqref{ssmkdveqscaled}
is given by 
\begin{equation}\label{2soliton}
u(t,x) = \exp(i\phi_1)\exp(i\nu_1 t) \tilde f_1(\xi_1,\xi_2) + \exp(i\phi_2)\exp(i\nu_2 t) \tilde f_2(\xi_1,\xi_2)
\end{equation}
where 
\begin{equation}\label{f2soliton}
\tilde f_1(\xi_1,\xi_2) = \tilde X_1(\xi_1,\xi_2)/\tilde Y(\xi_1,\xi_2) ,
\quad
\tilde f_2(\xi_1,\xi_2) = \tilde X_2(\xi_1,\xi_2)/\tilde Y(\xi_1,\xi_2)
\end{equation}
are the respective functions:
\begin{align}
& \tilde X_{1\rm H}(\xi_1,\xi_2)
= k_1\exp(i\kappa_1\xi_1)  \cosh(k_2\xi_2 +i\gamma_2) , 
\label{hoscilX1}\\
%&\nonumber\\
& \tilde X_{2\rm H}(\xi_1,\xi_2)
= k_2\exp(i\kappa_2\xi_2)  \cosh(k_1\xi_1 +i\gamma_1) , 
\label{hoscilX2}\\
%&\nonumber\\
&\begin{aligned}
\tilde Y_{\rm H}(\xi_1,\xi_2)
= & \sqrt{\Gamma}\cosh(k_1\xi_1+k_2\xi_2) +\frac{1}{\sqrt{\Gamma}}\cosh(k_1\xi_1-k_2\xi_2) 
\\&\qquad
-\frac{4k_1k_2}{\sqrt{\Upsilon}} \cos(\kappa_1\xi_1-\kappa_2\xi_2+\mu(\xi_2-\xi_1)+\phi_1-\phi_2)
\end{aligned}
\label{hoscilY}
\end{align}
in the Hirota case;
and 
\begin{align}
&\begin{aligned}
\tilde X_{1\rm SS}(\xi_1,\xi_2) = 
\exp(i\kappa_1\xi_1) \bigg( 
& 
k_1(k_1^2+\kappa_1^2)^{1/4} |2\kappa_1|^{1/2}
\bigg(
|\kappa_2|\Big( \sqrt{\Delta\Gamma} \cosh(k_1\xi_1+2k_2\xi_2+i(\alpha_2+\gamma_2))
\\&
+\frac{1}{\sqrt{\Delta\Gamma}}\cosh(k_1\xi_1-2k_2\xi_2+i(\upsilon_2-\gamma_2)) \Big)
\\&
+(k_2^2+\kappa_2^2)^{1/2}\Big( 
-8k_2^2\kappa_2 \varepsilon \frac{1}{\sqrt{\Omega\Upsilon}} \cosh(k_1\xi_1+i(\varpi_1+\gamma_1))
\\&
+\sqrt{\frac{\Gamma}{\Delta}} \cosh(k_1\xi_1+i(\alpha_2-\gamma_2))
+ \sqrt{\frac{\Delta}{\Gamma}} \cosh(k_1\xi_1+i(\upsilon_2+\gamma_2)) \Big)
\bigg)
\\&
+ k_1k_2 \varepsilon \sqrt{\frac{\Omega}{\Upsilon}} \bigg(
k_2 (k_1^2+\kappa_1^2)^{1/4} |8\kappa_2|^{1/2} \varepsilon_2 \cosh(k_1\xi_1+i(\varpi_1-\gamma_1)) 
\\&\qquad
- k_1(k_2^2+\kappa_2^2)^{1/4} |32\kappa_1|^{1/2} \varepsilon_1
\Re\Big( \cosh(k_2\xi_2+i(\gamma_2-\varpi_2)) 
\times
\\&\qquad\qquad
\exp(i(\kappa_1\xi_1-\kappa_2\xi_2+\mu(\xi_2-\xi_1)+\phi_1-\phi_2)) \Big) \bigg) 
\bigg) ,
\end{aligned}
\label{ssoscilX1}\\
&\nonumber\\
&\nonumber\\
&\begin{aligned}
\tilde X_{2\rm SS}(\xi_1,\xi_2) = 
\exp(i\kappa_2\xi_2) \bigg( & 
k_2(k_2^2+\kappa_2^2)^{1/4} |2\kappa_2|^{1/2}
\bigg(
|\kappa_1|\Big( \sqrt{\Delta\Gamma} \cosh(k_2\xi_2+2k_1\xi_1+i(\alpha_1+\gamma_1))
\\& 
+\frac{1}{\sqrt{\Delta\Gamma}}\cosh(k_2\xi_2-2k_1\xi_1+i(\upsilon_1-\gamma_1)) \Big)
\\&
+(k_1^2+\kappa_1^2)^{1/2}\Big( 
-8k_1^2\kappa_1 \varepsilon \frac{1}{\sqrt{\Omega\Upsilon}} 
\cosh(k_2\xi_2+i(\varpi_2+\gamma_2)) 
\\&
+\sqrt{\frac{\Gamma}{\Delta}} \cosh(k_2\xi_2+i(\alpha_1-\gamma_1))
+ \sqrt{\frac{\Delta}{\Gamma}} \cosh(k_2\xi_2+i(\upsilon_1+\gamma_1)) \Big)
\bigg)
\\&
+ k_1k_2 \varepsilon \sqrt{\frac{\Omega}{\Upsilon}} \bigg(
k_1(k_2^2+\kappa_2^2)^{1/4} |8\kappa_1|^{1/2} \varepsilon_1 \cosh(k_2\xi_2+i(\varpi_2-\gamma_2)) 
\\&\qquad
- k_2(k_1^2+\kappa_1^2)^{1/4} |32\kappa_2|^{1/2} \varepsilon_2
\Re\Big( \cosh(k_1\xi_1+i(\gamma_1-\varpi_1)) 
\times
\\&\qquad\qquad
\exp(i(\kappa_2\xi_2-\kappa_1\xi_1+\mu(\xi_1-\xi_2)+\phi_2-\phi_1)) \Big) \bigg) 
\bigg) ,
\end{aligned}
\label{ssoscilX2}\\
&\nonumber\\
&\nonumber\\
&\begin{aligned}
\tilde Y_{\rm SS}(\xi_1,\xi_2)= &
|\kappa_1\kappa_2|\Big( \Delta\Gamma \cosh(2(k_1\xi_1+k_2\xi_2))
+ \frac{1}{\Delta\Gamma} \cosh(2(k_1\xi_1-k_2\xi_2)) \Big)
\\&
+2(k_1^2+\kappa_1^2)^{1/2}|\kappa_2| \cosh(2k_2\xi_2) 
+ 2(k_2^2+\kappa_2^2)^{1/2}|\kappa_1| \cosh(2k_1\xi_1)
\\&
+4k_1^2 k_2^2\varepsilon_1\varepsilon_2\frac{\Omega}{\Upsilon}
\cos(2(\kappa_1\xi_1-\kappa_2\xi_2+\mu(\xi_2-\xi_1))+2(\phi_1-\phi_2))
\\&
+(k_1^2+\kappa_1^2)^{1/2}(k_2^2+\kappa_2^2)^{1/2}
\bigg(\frac{\Gamma}{\Delta} +\frac{\Delta}{\Gamma} 
+64k_1^2 k_2^2\kappa_1\kappa_2 \frac{1}{\Omega\Upsilon}
\\&\qquad
-16 k_1k_2 |\kappa_1\kappa_2|^{1/2} 
\Re\Big( \exp(i(\kappa_1\xi_1-\kappa_2\xi_2+\mu(\xi_2-\xi_1)+\phi_1-\phi_2))
\times
\\&\qquad\qquad
\Big( \sqrt{\frac{\Gamma}{\Upsilon}} \cosh(k_1\xi_1+k_2\xi_2+i(\varpi_1-\varpi_2))
\\&\qquad\qquad\qquad
+ \frac{1}{\sqrt{\Gamma\Upsilon}} \cosh(k_1\xi_1-k_2\xi_2+i(\varpi_1+\varpi_2)) \Big) \Big)
\bigg) 
\end{aligned}
\label{ssoscilY}
\end{align}
in the Sasa-Satsuma case when $\kappa_1\neq0$ and $\kappa_2\neq0$. 
In both cases, 
$\mu$ is given by equation \eqref{mu}, 
$\Omega$, $\Upsilon$, $\Delta$, $\Gamma$
are given by equations \eqref{ss2solitonOmega}--\eqref{ss2solitonGamma}, 
$\alpha_1,\alpha_2$, $\upsilon_1,\upsilon_2$, $\varpi_1$, $\varpi_2$,
$\gamma_1,\gamma_2$
are given by equations \eqref{ss2solitonalphas}--\eqref{ss2solitongammas},
and $\lambda_1,\lambda_2$, $\delta_1,\delta_2$, 
$\varepsilon_1$, $\varepsilon_2$, $\varepsilon$ 
are given by equations \eqref{ss2solitonhalflambda1}--\eqref{sssgn}.
\end{thm}

Similarly to the $1$-soliton case, 
Theorem~\ref{thm:2soliton} implies the following result. 

\begin{cor}\label{cor:2soliton}
For the Hirota and Sasa-Satsuma equations, 
a harmonically modulated $2$-soliton solution 
\begin{equation}\label{enveloscil2soliton}
\begin{aligned}
u(t,x) & =  \exp(i\phi_1)\exp(i(\kappa_1 x +\omega_1 t)) f_1(k_1x+w_1t,k_2x+w_2t)
\\& 
\qquad 
+ \exp(i\phi_2)\exp(i(\kappa_2 x +\omega_2 t)) f_2(k_1x+w_1t,k_2x+w_2t) 
\\& 
=\exp(i\phi_1)\exp(i\nu_1 t) \tilde f_1(x-c_1t,x-c_2t)
\\& 
\qquad 
+ \exp(i\phi_2)\exp(i\nu_2 t) \tilde f_2(x-c_1t,x-c_2t) ,
\quad c_1\neq c_2
\end{aligned}
\end{equation}
is distinguished from an ordinary $2$-soliton solution 
\begin{equation}\label{ordinary2soliton}
u(t,x) =\exp(i\phi_1) f(x-c_1t,x-c_2t) + \exp(i\phi_2) f(x-c_1t,x-c_2t), 
\quad c_1\neq c_2
\end{equation}
by the kinematic conditions
\begin{gather}
\nu_1 \neq 0 ,
\quad
(c_1/3)^3 +(\nu_1/2)^2 >0 , 
\\
\nu_2 \neq 0 , 
\quad
(c_2/3)^3 +(\nu_2/2)^2 >0 ,
\end{gather}
or equivalently 
\begin{equation}\label{kinconds}
c_1 > -(3/\cubrt{4})(\cubrt{\nu_1})^2 \neq 0 , 
\quad
c_2 > -(3/\cubrt{4})(\cubrt{\nu_2})^2 \neq 0 .
\end{equation}
\end{cor}

\subsection{Breathers}

From the remarks made after Proposition~\ref{prop:henvel2soliton} 
for the Hirota equation \eqref{hmkdveqscaled}
and Proposition~\ref{prop:ssenvel2soliton} 
for the Sasa-Satsuma equation \eqref{ssmkdveqscaled},
we will now state a counterpart of Theorem~\ref{thm:2soliton}
for harmonically modulated breather solutions. 

Let 
\begin{equation}\label{betapmoscilbr}
\begin{aligned}
& \beta_{1\pm} = \cubrt{\sqrt{(c/3)^3 +((\nu_0+\nu)/2)^2} \pm (\nu_0+\nu)/2} 
\\
& \beta_{2\pm} = \cubrt{\sqrt{(c/3)^3 +((\nu_0-\nu)/2)^2} \pm (\nu_0-\nu)/2} 
\end{aligned}
\end{equation}
with 
\begin{equation}\label{oscilbreatherkinrel}
(c/3)^3 +((|\nu_0|+|\nu|)/2)^2 \geq 0 . 
\end{equation}

\begin{thm}\label{thm:oscilbreather}
The equivalent oscillatory form \eqref{oscilbreather} of 
a harmonically modulated breather \eqref{specialoscil2soliton}
has parameters $\chi,\phi_0,\nu_0,c,\phi,\nu\neq0$, 
satisfying the kinematic relation \eqref{oscilbreatherkinrel}. 
As expressed using a travelling wave coordinate $\xi=x-c t$
and an oscillation coordinate $\tau=\nu t+\phi$, 
the oscillatory form for the breather solutions 
for the Hirota equation \eqref{hmkdveqscaled}
and the Sasa-Satsuma equation \eqref{ssmkdveqscaled}
is given by 
\begin{equation}\label{oscilbreathersoln}
\begin{aligned}
& u(t,x) = \exp(i(\nu_0 t+\phi_0)) \tilde f(\xi,\tau)
\\
& \tilde f(\xi,\tau) = 
\exp(i\tau) \tilde X_1(\xi,\tau)/\tilde Y(\xi,\tau) 
+ \exp(-i\tau)\tilde X_2(\xi,\tau)/\tilde Y(\xi,\tau)
\end{aligned}
\end{equation}
in terms of the respective functions:
\begin{align}
&\tilde X_{1\rm H}(\xi,\tau)
= k_1 \exp(i\kappa_1\xi) \cosh(k_2\xi -\chi+i\gamma_2) , 
\label{hoscilbrX1}\\
& \tilde X_{2\rm H}(\xi,\tau)
= k_2 \exp(i\kappa_2\xi) \cosh(k_1\xi +\chi+i\gamma_1) , 
\label{hoscilbrX2}\\
&\begin{aligned}
\tilde Y_{\rm H}(\xi,\tau)
= & \sqrt{\Gamma}\cosh((k_1+k_2)\xi) +\frac{1}{\sqrt{\Gamma}}\cosh((k_1-k_2)\xi+2\chi) 
%\\&\qquad
-\frac{4k_1k_2}{\sqrt{\Upsilon}} \cos((\kappa_1-\kappa_2)\xi+2\tau)
\end{aligned}
\label{hoscilbrY}
\end{align}
in the Hirota case;
and 
\begin{align}
&\begin{aligned}
\tilde X_{1\rm SS}(\xi,\tau) = 
\exp(i\kappa_1\xi) \bigg( 
& 
k_1(k_1^2+\kappa_1^2)^{1/4} |2\kappa_1|^{1/2}
\bigg(
|\kappa_2|\Big( \sqrt{\Delta\Gamma} \cosh((k_1+2k_2)\xi-\chi+i(\alpha_2+\gamma_2))
\\&
+\frac{1}{\sqrt{\Delta\Gamma}}\cosh((k_1-2k_2)\xi+3\chi+i(\upsilon_2-\gamma_2)) \Big)
\\&
+(k_2^2+\kappa_2^2)^{1/2}\Big( 
-8k_2^2\kappa_2 \epsilon \frac{1}{\sqrt{\Omega\Upsilon}} \cosh(k_1\xi+\chi+i(\varpi_1+\gamma_1))
\\&
+\sqrt{\frac{\Gamma}{\Delta}} \cosh(k_1\xi+\chi+i(\alpha_2-\gamma_2))
+ \sqrt{\frac{\Delta}{\Gamma}} \cosh(k_1\xi+\chi+i(\upsilon_2+\gamma_2)) \Big)
\bigg)
\\&
+ k_1k_2 \varepsilon \sqrt{\frac{\Omega}{\Upsilon}} \bigg(
k_2 (k_1^2+\kappa_1^2)^{1/4} |8\kappa_2|^{1/2} \varepsilon_2 \cosh(k_1\xi+\chi+i(\varpi_1-\gamma_1)) 
\\&\qquad
- k_1(k_2^2+\kappa_2^2)^{1/4} |32\kappa_1|^{1/2} \varepsilon_1
\Re\Big( \cosh(k_2\xi-\chi+i(\gamma_2-\varpi_2)) 
\times
\\&\qquad\qquad
\exp(i((\kappa_1-\kappa_2)\xi +2\tau)) \Big) \bigg) 
\bigg) , 
\end{aligned}
\label{ssoscilbrX1}\\
&\begin{aligned}
\tilde X_{2\rm SS}(\xi,\tau) = 
\exp(i\kappa_2\xi) \bigg( & 
k_2(k_2^2+\kappa_2^2)^{1/4} |2\kappa_2|^{1/2}
\bigg(
|\kappa_1|\Big( \sqrt{\Delta\Gamma} \cosh((k_2+2k_1)\xi+\chi+i(\alpha_1+\gamma_1))
\\& 
+\frac{1}{\sqrt{\Delta\Gamma}}\cosh((k_2-2k_1)\xi-3\chi+i(\upsilon_1-\gamma_1)) \Big)
\\&
+(k_1^2+\kappa_1^2)^{1/2}\Big( 
-8k_1^2\kappa_1 \varepsilon \frac{1}{\sqrt{\Omega\Upsilon}} 
\cosh(k_2\xi-\chi+i(\varpi_2+\gamma_2)) 
\\&
+\sqrt{\frac{\Gamma}{\Delta}} \cosh(k_2\xi-\chi+i(\alpha_1-\gamma_1))
+ \sqrt{\frac{\Delta}{\Gamma}} \cosh(k_2\xi-\chi+i(\upsilon_1+\gamma_1)) \Big)
\bigg)
\\&
+ k_1k_2 \varepsilon \sqrt{\frac{\Omega}{\Upsilon}} \bigg(
k_1(k_2^2+\kappa_2^2)^{1/4} |8\kappa_1|^{1/2} \varepsilon_1 \cosh(k_2\xi-\chi+i(\varpi_2-\gamma_2)) 
\\&\qquad
- k_2(k_1^2+\kappa_1^2)^{1/4} |32\kappa_2|^{1/2} \varepsilon_2
\Re\Big( \cosh(k_1\xi+\chi+i(\gamma_1-\varpi_1)) 
\times
\\&\qquad\qquad
\exp(i((\kappa_2-\kappa_1)\xi+2\tau)) \Big) \bigg) 
\bigg) , 
\end{aligned}
\label{ssoscilbrX2}\\
&\begin{aligned}
\tilde Y_{\rm SS}(\xi,\tau)= &
|\kappa_1\kappa_2|\Big( \Delta\Gamma \cosh(2(k_1+k_2)\xi)
+ \frac{1}{\Delta\Gamma} \cosh(2((k_1-k_2)\xi+2\chi)) \Big)
\\&
+2(k_1^2+\kappa_1^2)^{1/2}|\kappa_2| \cosh(2(k_2\xi-\chi)) 
+ 2(k_2^2+\kappa_2^2)^{1/2}|\kappa_1| \cosh(2(k_1\xi+\chi))
\\&
+4k_1^2 k_2^2\varepsilon_1\varepsilon_2\frac{\Omega}{\Upsilon}
\cos(2(\kappa_1-\kappa_2)\xi+4\tau)
\\&
+(k_1^2+\kappa_1^2)^{1/2}(k_2^2+\kappa_2^2)^{1/2}
\bigg(\frac{\Gamma}{\Delta} +\frac{\Delta}{\Gamma} 
+64k_1^2 k_2^2\kappa_1\kappa_2 \frac{1}{\Omega\Upsilon}
\\&\qquad
-16 k_1k_2 |\kappa_1\kappa_2|^{1/2} 
\Re\Big( \exp(i((\kappa_1-\kappa_2)\xi+2\tau))
\times
\\&\qquad\qquad
\Big( \sqrt{\frac{\Gamma}{\Upsilon}} \cosh((k_1+k_2)\xi+i(\varpi_1-\varpi_2))
\\&\qquad\qquad\qquad
+ \frac{1}{\sqrt{\Gamma\Upsilon}} \cosh((k_1-k_2)\xi+2\chi+i(\varpi_1+\varpi_2)) \Big) \Big)
\bigg) 
\end{aligned}
\label{ssoscilbrY}
\end{align}
in the Sasa-Satsuma case when $\kappa_1\neq0$ and $\kappa_2\neq0$. 
In both cases, 
$k_1$, $k_2$, $\kappa_1$, $\kappa_2$ are given in terms of 
$c$, $\nu\neq0$, $\nu_0$ by equations \eqref{rels1}--\eqref{rels2}, 
$\Omega$, $\Upsilon$, $\Delta$, $\Gamma$
are given by equations \eqref{ss2solitonOmega}--\eqref{ss2solitonGamma}, 
$\alpha_1,\alpha_2$, $\upsilon_1,\upsilon_2$, $\varpi_1$, $\varpi_2$,
$\gamma_1,\gamma_2$
are given by equations \eqref{ss2solitonalphas}--\eqref{ss2solitongammas},
and $\lambda_1,\lambda_2$, $\delta_1,\delta_2$, 
$\varepsilon_1$, $\varepsilon_2$, $\varepsilon$
are given by equations \eqref{ss2solitonhalflambda1}--\eqref{sssgn}.
\end{thm}

In the special case $\nu_0=0$, 
the oscillatory breathers \eqref{oscilbreathersoln} reduce to ordinary breathers \eqref{breather}, 
which are given by much simpler expressions. 

\begin{prop}\label{prop:breather}
For the Hirota equation \eqref{hmkdveqscaled}
and the Sasa-Satsuma equation \eqref{ssmkdveqscaled},
the ordinary breather solutions 
expressed using a travelling wave coordinate $\xi=x-c t$
and an oscillation coordinate $\tau=\nu t+\phi$
are given by 
\begin{equation}\label{breathersoln}
\begin{aligned}
& u(t,x) = \exp(i\phi_0) f(\xi,\tau)
\\
& f(\xi,\tau) = 
\exp(i\tau) X_1(\xi,\tau)/Y(\xi,\tau) 
+ \exp(-i\tau) X_2(\xi,\tau)/Y(\xi,\tau)
\end{aligned}
\end{equation}
in terms of the respective functions
\begin{align}
& X_{1\rm H}(\xi,\tau)
= k|\kappa|\sqrt{k^2+\kappa^2}\exp(i\kappa\xi) \sinh(k\xi -\chi-i\gamma) , 
\label{hbrX1}\\
& X_{2\rm H}(\xi,\tau)
= k|\kappa|\sqrt{k^2+\kappa^2}\exp(-i\kappa\xi) \sinh(k\xi +\chi+i\gamma) , 
\label{hbrX2}\\
&\begin{aligned}
Y_{\rm H}(\xi,\tau)
= & \kappa^2\cosh(2k\xi) +(k^2+\kappa^2)\cosh(2\chi) +k^2\cos(2(\kappa\xi+\tau))
\end{aligned}
\label{hbrY}
\end{align}
in the Hirota case, 
and in the Sasa-Satsuma case 
\begin{align}
&\begin{aligned}
X_{1\rm SS}(\xi,\tau) 
=& (k/\Sigma)\exp(i\kappa\xi) \bigg(
|\kappa|\sqrt{k^2+\kappa^2}\sinh(k\xi -\chi-i\gamma) 
\\&\qquad
-\frac{((k^2+\kappa^2)\Lambda)^{3/2}}{|\kappa|(1+\Sigma)^{1/2}}\exp(\chi+i(\lambda+\gamma/2))
\cosh(k\xi +\varpi+i(\lambda+\gamma/2)) 
\bigg) , 
\end{aligned}
\label{ssbrX1}\\
&\begin{aligned}
X_{2\rm SS}(\xi,\tau) 
=& (k/\Sigma)\exp(-i\kappa\xi) \bigg(
|\kappa|\sqrt{k^2+\kappa^2}\sinh(k\xi +\chi+i\gamma) 
\\&\qquad
-\frac{((k^2+\kappa^2)\Lambda)^{3/2}}{|\kappa|(1+\Sigma)^{1/2}}\exp(-\chi+i(\lambda-\gamma/2))
\cosh(k\xi +\varpi+i(\lambda-\gamma/2))
\bigg) , 
\end{aligned}
\label{ssbrX2}\\
&\begin{aligned}
Y_{\rm H}(\xi,\tau)
= & \kappa^2\Sigma\cosh(2k\xi) +(k^2+\kappa^2)\cosh(2\chi) +k^2\cos(2(\kappa\xi+\tau)) , 
\end{aligned}
\label{ssbrY}
\end{align}
where
\begin{gather}
\Lambda = \sinh(2|\chi|) ,
\quad
\Sigma = \sqrt{1+(1+k^2/\kappa^2)\Lambda^2} ,
\quad
\varpi= \tfrac{1}{4}\ln\Big(\frac{\Sigma+1}{\Sigma-1}\Big)
\\
\gamma=\arg( k+i\kappa ) ,
\quad
\lambda = \arg(1+i\sgn(\kappa\chi)) 
\end{gather}
and where $k$, $\kappa\neq0$ are given in terms of 
$c$, $\nu\neq0$ by equations \eqref{krel} and \eqref{kapparel}, 
such that the kinematic relation $(c/3)^3 +(\nu/2)^2 > 0$ holds. 
\end{prop}

When $\chi=\phi_0=0$,
we remark that both the Hirota and Sasa-Satsuma breather solutions 
in Proposition~\ref{prop:breather}
reduce to the well-known mKdV breather solution \cite{Wad}
\begin{equation}
u(t,x) = k|\kappa|\sqrt{k^2+\kappa^2}
\frac{\sinh(k\xi)\cos(\gamma)\cos(\kappa\xi+\tau)+\cosh(k\xi)\sin(\gamma)\sin(\kappa\xi+\tau)}{\kappa^2\cosh(k\xi)^2 +k^2\cos(\kappa\xi+\tau)^2} .
\end{equation}

\section{Properties of oscillatory soliton solutions}
\label{properties}

We begin by discussing some basic properties of 
the oscillatory $1$-soliton solutions from Theorem~\ref{thm:1soliton} 
for the Hirota equation \eqref{hmkdveqscaled}
and the Sasa-Satsuma equation \eqref{ssmkdveqscaled}.

\subsection{Oscillatory $1$-solitons}
An oscillatory wave \eqref{enveloscilwave} 
has amplitude $|u|=|\tilde f(\xi)|$ 
where $\xi=x-ct$ is a moving coordinate centered at $x=ct$. 
Hence the spatial shape of $|u|$ is determined by the properties of
the function $|\tilde f(\xi)|$. 
In both the Hirota and Sasa-Satsuma oscillatory $1$-soliton solutions, 
these functions $|\tilde f(\xi)|$ share two main properties,
as seen from expressions \eqref{hoscilfunct} and \eqref{ssoscilfunct}. 
First, for large $|\xi|$,
both functions exhibit exponential decay 
$|\tilde f(\xi)| \sim O(\exp(-k|\xi|))$. 
Second, both functions exhibit reflection-conjugation invariance 
$\tilde f(-\xi) = \overline{\tilde f(\xi)}$
(where a bar denotes complex conjugation),
implying that $|\tilde f(\xi)|$ is an even function of $\xi$
and thus $\Re(\tilde f'(0))=0$. 

In the case of the Hirota function $|\tilde f_{\rm H}(\xi)|$,
from expression \eqref{hoscilfunct} 
we find that $\Re(\tilde f_{\rm H}'(\xi))\neq 0$ when $\xi\neq 0$. 
Hence the function $|\tilde f_{\rm H}(\xi)|$ has a peak at $\xi=0$. 
In contrast, in the case of the Sasa-Satsuma function $|\tilde f_{\rm SS}(\xi)|$,
from expression \eqref{ssoscilfunct} 
we find that $\Re(\tilde f_{\rm SS}'(\xi))=0$ has roots $\xi\neq 0$ 
when (and only when) 
$\cosh(2k\xi)= (k^2-\kappa^2)/(|\kappa|\sqrt{k^2+\kappa^2})$,
which requires the condition 
$(k^2-\kappa^2)/(|\kappa|\sqrt{k^2+\kappa^2}) \geq 1$ on $k,\kappa$. 
This condition is equivalent to $k^2\geq 3\kappa^2$. 
Hence in this case the function $|\tilde f_{\rm SS}(\xi)|$ 
has a pair of peaks centered symmetrically around $\xi=0$. 

\begin{figure}[H]
\centering
\begin{subfigure}[t]{.4\textwidth}
\includegraphics[width=\textwidth]{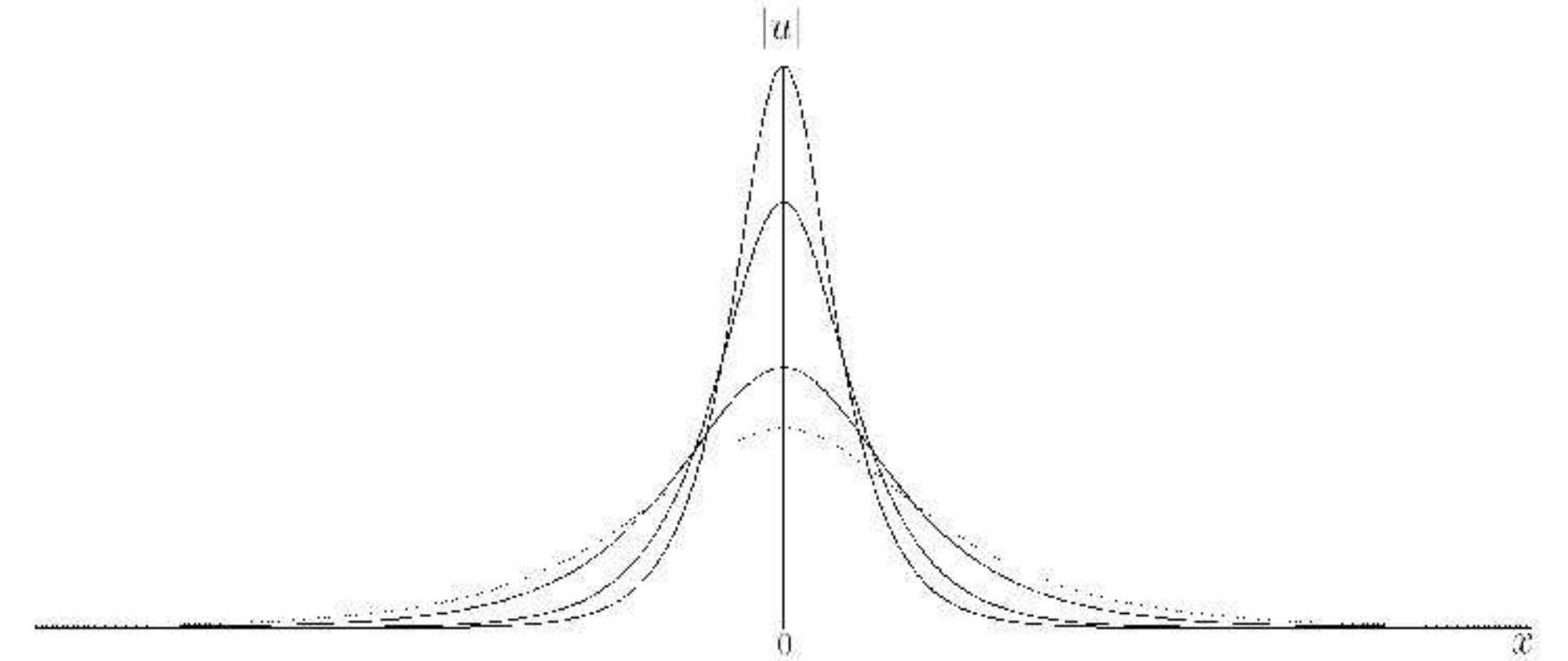} 
\captionof{figure}{amplitude}
\label{oscil_hirota_overlay_amplitude_c=4_nu=0_15_100_250} 
\end{subfigure}%
\begin{subfigure}[t]{.4\textwidth}
\includegraphics[width=\textwidth]{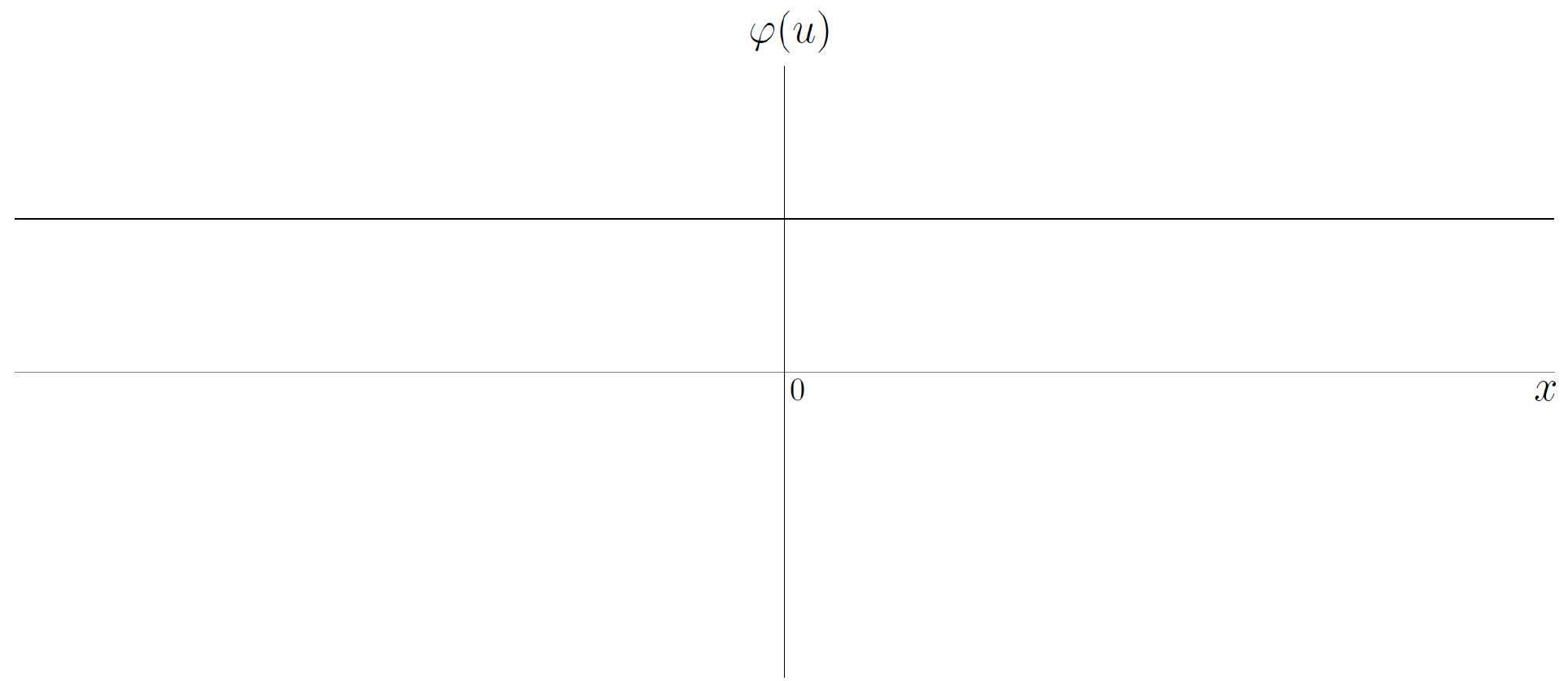}
\captionof{figure}{envelope phase}
\label{oscil_hirota_overlay_phase_c=4_phi=halfpi} 
\end{subfigure}
\caption{Hirota oscillatory $1$-solitons
with $c=4$, $|\nu|=15,100,250$ and $\nu=0$ (dotted line), $\phi=\pi/2$}
\end{figure}

\begin{figure}[H]
\centering
\begin{subfigure}[t]{.4\textwidth}
\includegraphics[width=\textwidth]{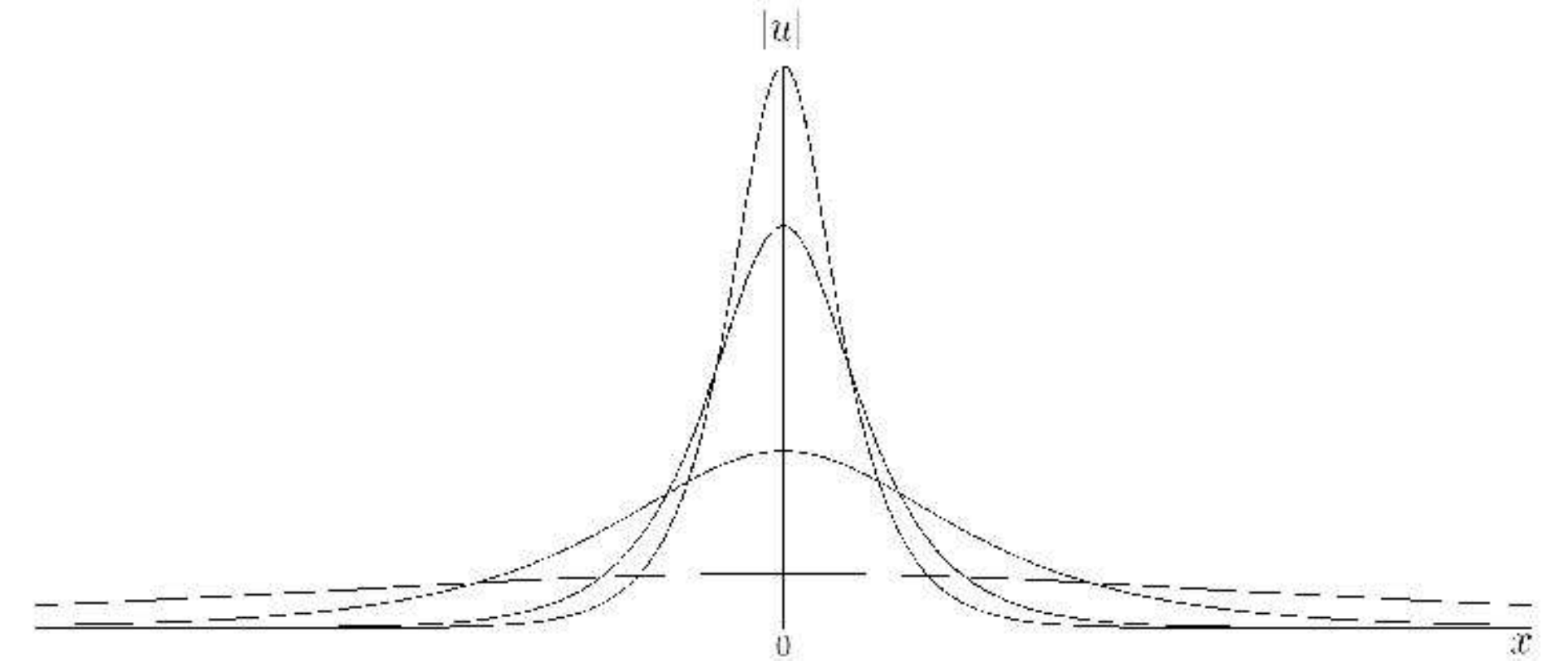} 
\captionof{figure}{amplitude}
\label{oscil_hirota_overlay_amplitude_c=-4_nu=4_15_100_250} 
\end{subfigure}%
\begin{subfigure}[t]{.4\textwidth}
\includegraphics[width=\textwidth]{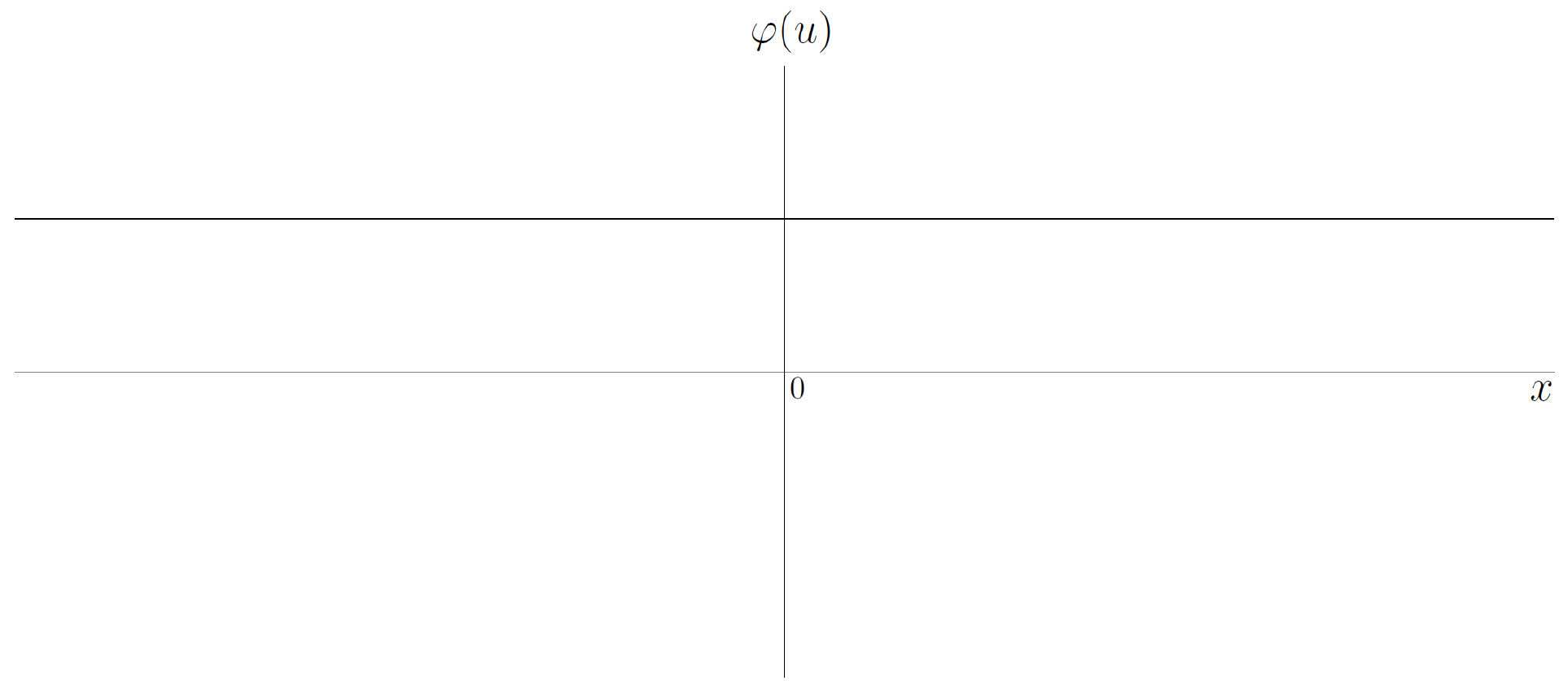}
\captionof{figure}{envelope phase}
\label{oscil_hirota_overlay_phase_c=-4_phi=halfpi} 
\end{subfigure}
\caption{Hirota oscillatory $1$-solitons 
with $c=-4$, $|\nu|=4,15,100,250$, $\phi=\pi/2$}
\end{figure}

\begin{figure}[H]
\centering
\begin{subfigure}[t]{.4\textwidth}
\includegraphics[width=\textwidth]{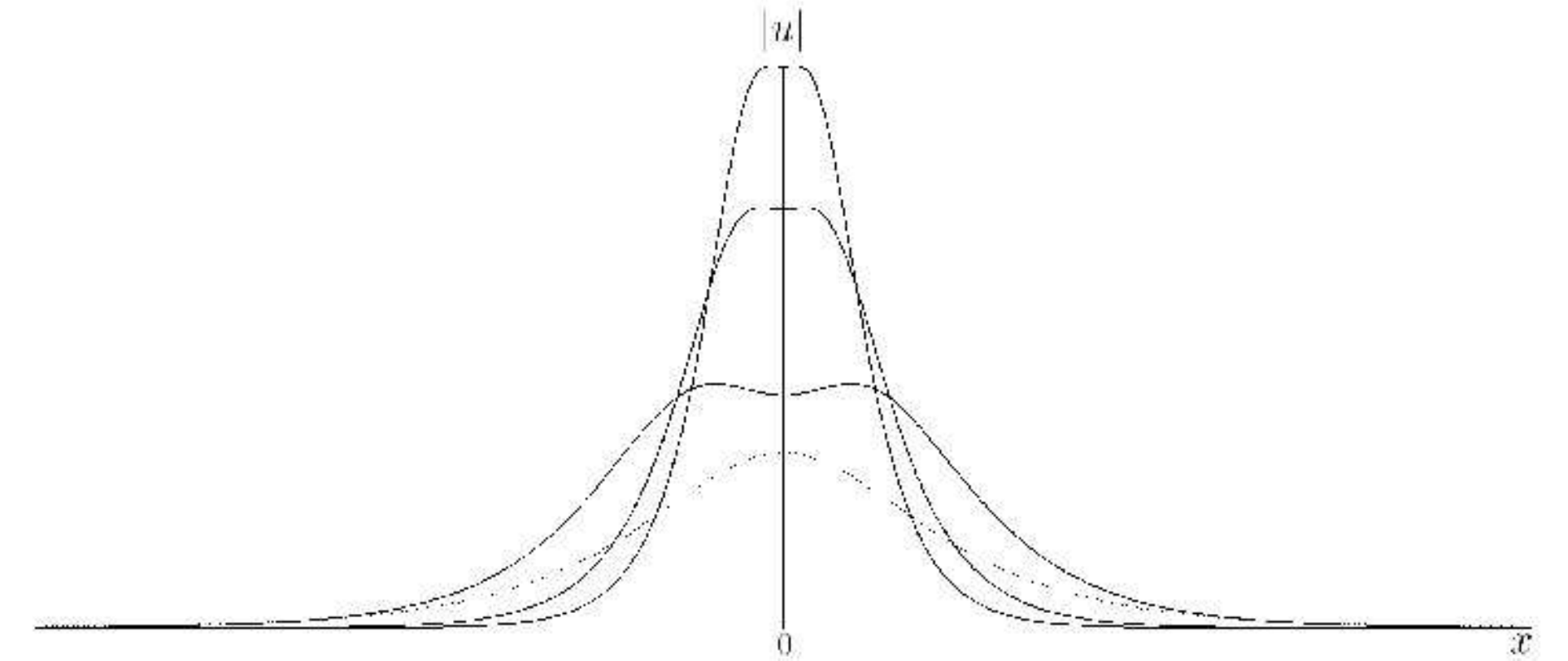} 
\captionof{figure}{amplitude}
\label{oscil_ss_overlay_amplitude_c=4_nu=0_15_100_250} 
\end{subfigure}%
\begin{subfigure}[t]{.4\textwidth}
\includegraphics[width=\textwidth]{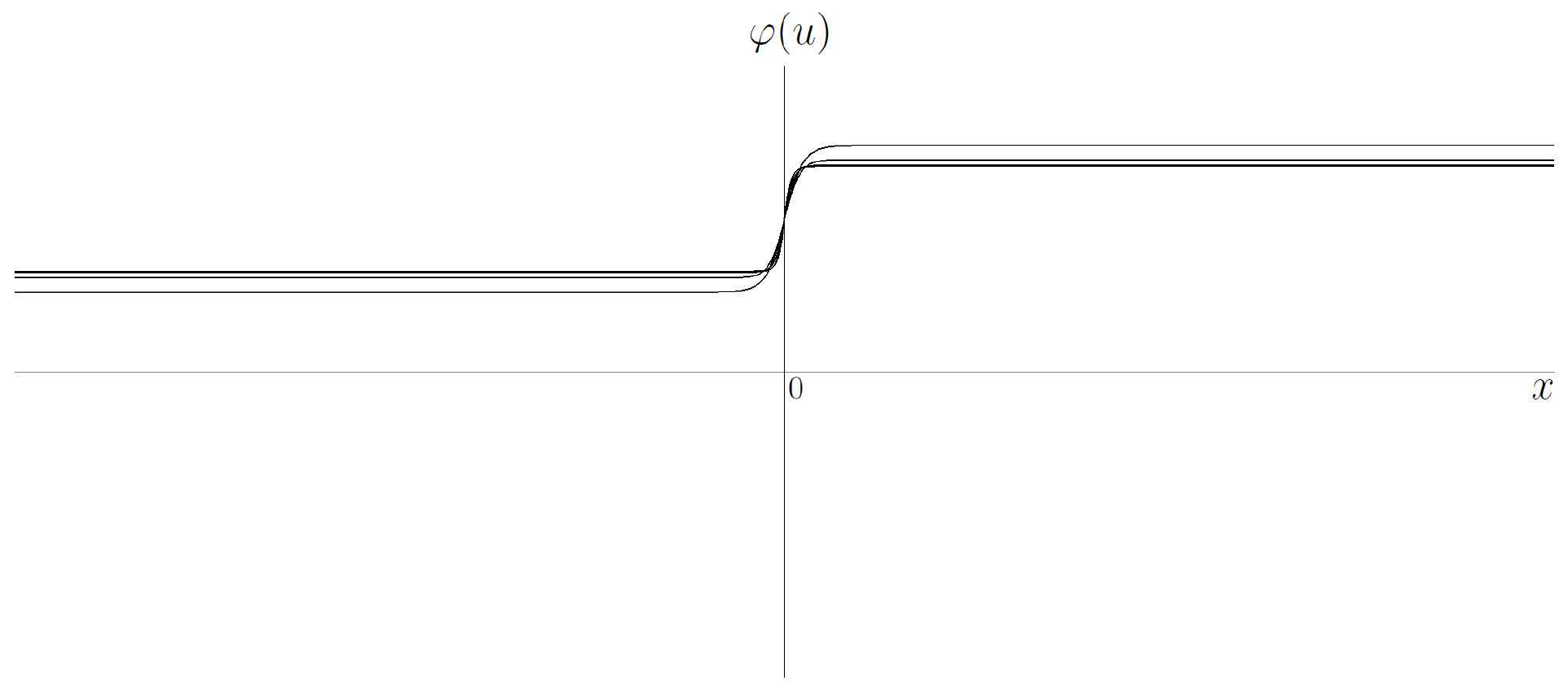}
\captionof{figure}{envelope phase}
\label{oscil_ss_overlay_phase_c=4_nu=0_15_100_250_phi=halfpi}
\end{subfigure}
\caption{Sasa-Satsuma oscillatory $1$-solitons
with $c=4$, $|\nu|=15,100,250$ and $\nu=1$ (dotted line), $\phi=\pi/2$}
\end{figure}

\begin{figure}[H]
\centering
\begin{subfigure}[t]{.4\textwidth}
\includegraphics[width=\textwidth]{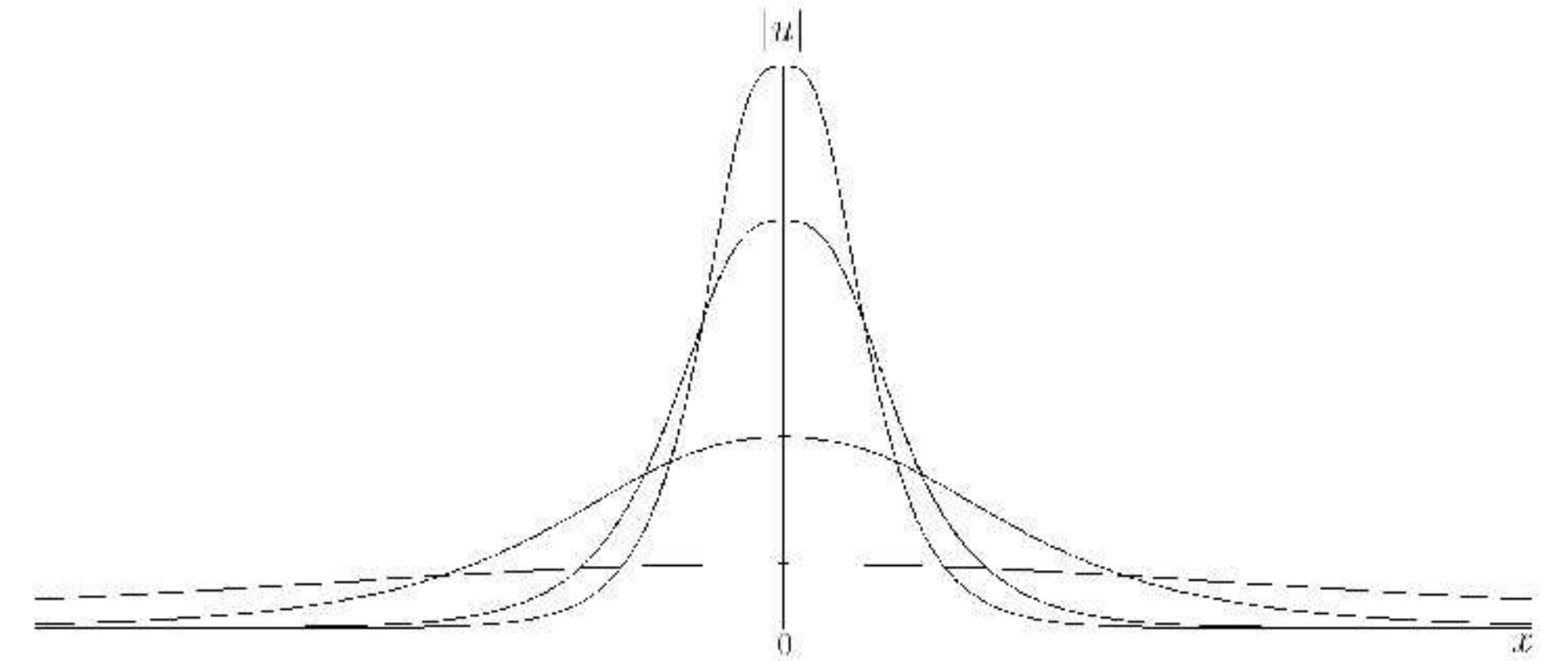} 
\captionof{figure}{amplitude}
\label{oscil_ss_overlay_amplitude_c=-4_nu=4_15_100_250} 
\end{subfigure}%
\begin{subfigure}[t]{.4\textwidth}
\includegraphics[width=\textwidth]{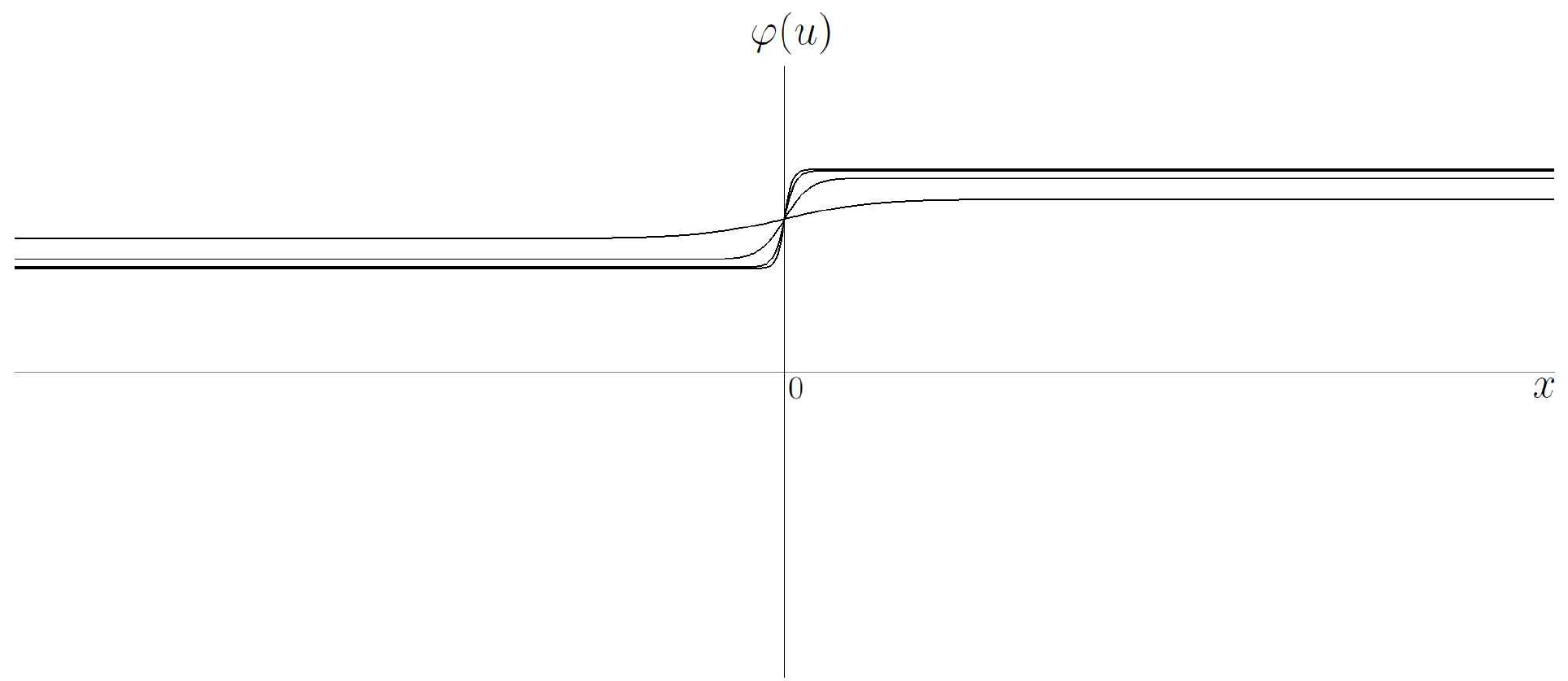}
\captionof{figure}{envelope phase}
\label{oscil_ss_overlay_phase_c=-4_nu=4_15_100_250_phi=halfpi}
\end{subfigure}
\caption{Sasa-Satsuma oscillatory $1$-solitons
with $c=-4$, $|\nu|=4,15,100,250$, $\phi=\pi/2$}
\end{figure}

From these properties, 
we obtain the following two results about the spatial shape of $|u|$, 
stated in terms of the notation 
$\beta_{\pm} = \cubrt{\sqrt{(c/3)^3 +(\nu/2)^2} \pm \nu/2}$. 
Cases $c>0$, $c<0$, $c=0$ are illustrated in 
\figref{oscil_hirota_overlay_amplitude_c=4_nu=0_15_100_250}, 
\figref{oscil_hirota_overlay_amplitude_c=-4_nu=4_15_100_250}, 
%\figref{standing_hirota_overlay_amplitude_nu=1_15_100_250}
for the Hirota oscillatory $1$-soliton, 
and in 
\figref{oscil_ss_overlay_amplitude_c=4_nu=0_15_100_250}, 
\figref{oscil_ss_overlay_amplitude_c=-4_nu=4_15_100_250},
%\figref{standing_ss_overlay_amplitude_nu=1_15_100_250}
for the Sasa-Satsuma oscillatory $1$-soliton. 

\begin{prop}\label{prop:1solitonheightwidth}
For both the Hirota oscillatory $1$-soliton 
\eqref{1soliton}, \eqref{hoscilfunct}, 
and the Sasa-Satsuma oscillatory $1$-soliton 
\eqref{1soliton}, \eqref{ssoscilfunct}, 
the amplitude $|u|$ is an even function of $x-ct$
and decays exponentially for 
$|x-ct|\gg 1/k = 2/(\sqrt{3}(\beta_{-} + \beta_{+}))$. 
The Hirota oscillatory $1$-soliton 
and the Sasa-Satsuma oscillatory $1$-soliton for $c\leq 0$
each have a single peak centered at $x=ct$, 
with the height of the respective peaks given by 
\begin{equation}
|u|\big|_{x=ct} = \frac{k}{2} 
= \frac{\sqrt{3}}{4}(\beta_{-} + \beta_{+})
\end{equation}
and 
\begin{equation}
|u|\big|_{x=ct} = \frac{k|\kappa|^{1/2}}{(|\kappa| +(k^2+\kappa^2)^{1/2})^{1/2}}
= \frac{\sqrt{3}(\beta_{-} + \beta_{+})|\nu+(\beta_{-} - \beta_{+})c|^{1/4}}{2(2|\nu|^{1/2}+ |\nu+(\beta_{-} -\beta_{+})c|^{1/2})^{1/2}} . 
\end{equation}
For $c>0$, 
the Sasa-Satsuma oscillatory $1$-soliton instead has a symmetrical pair of peaks
at $x=ct\pm x_0$, where 
\begin{equation}
x_0= \frac{k^2-\kappa^2+k(k^2+\kappa^2)^{1/2}}{|\kappa|(k^2+\kappa^2)^{1/2}}
=\frac{|\beta_{+} -\beta_{-}|c +|\beta_{+}^2 -\beta_{-}^2|(3c)^{1/2}+|\nu|}{|\nu|^{1/2}|\beta_{+} -\alpha_1|^{3/2}} 
>0 , 
\end{equation}
with the height of the peaks given by 
\begin{equation}
|u|\big|_{x=ct\pm x_0} = \frac{1}{2}(k^2+\kappa^2)^{1/2} 
= \frac{|\nu|^{1/2}}{2|\beta_{+} -\beta_{-}|} . 
\end{equation}
\end{prop}

In general, 
an oscillatory wave \eqref{enveloscilwave}
can be factorized into a harmonic wave part $\exp(i(\kappa x +\omega t))$ 
and a travelling wave part $f(kx+wt)$ 
through equation \eqref{enveloscilrels}. 
The envelope function $f$ in this factorization will be real-valued 
when (and only when) it satisfies $f=|f|$ or $\arg(f)=0$,
corresponding to $\arg(u)=\phi+\kappa x +\omega t$ being linear in $x,t$
(in which case $u$ is a harmonically modulated travelling wave). 
This condition is equivalent to the relation 
$\arg(\tilde f(\xi)) = \kappa \xi$
given in terms of the function $\tilde f(\xi)$ in the oscillatory wave \eqref{enveloscilwave}. 
It is thereby convenient to define 
\begin{equation}\label{nonlinphase}
\varphi(u) = \arg(u) -\kappa x -(\nu-\kappa c) t
\mod 2\pi . 
\end{equation}
Since an oscillatory wave \eqref{enveloscilwave} 
has $\arg(u)=\phi+\nu t +\arg(\tilde f(\xi))$, 
we then see that the condition $\arg(\tilde f(\xi)) = \kappa \xi$ 
can be formulated simply as $\varphi(u) =\phi$ is constant. 
Therefore, the property $\varphi(u) \neq\const$ distinguishes 
a general oscillatory wave from a harmonically modulated travelling wave
with a real envelope function. 
Correspondingly, 
it is natural to define 
\begin{equation}\label{winding}
\ell = \frac{1}{2\pi}\int_{-\infty}^{\infty} \varphi(u)_x\; dx 
= \frac{\varphi(u)}{2\pi}\Big|_{x=-\infty}^{x=\infty}
\end{equation}
which can be regarded as measuring 
the net winding contributed by the envelope function of 
an oscillatory wave \eqref{enveloscilwave}. 
(Note that the net winding will be $\ell=0$
when the envelope function is real-valued.)
We will thus refer to the expressions \eqref{nonlinphase} and \eqref{winding} 
as the {\em envelope phase} and {\em envelope winding number}, respectively.

It is straightforward to derive the following phase properties of 
the Hirota and Sasa-Satsuma oscillatory $1$-soliton solutions. 

\begin{prop}\label{prop:1solitonphase}
For both the Hirota oscillatory $1$-soliton 
\eqref{1soliton}, \eqref{hoscilfunct}, 
and the Sasa-Satsuma oscillatory $1$-soliton 
\eqref{1soliton}, \eqref{ssoscilfunct}, 
the envelope phase $\varphi(u)$ is an even function of $x-ct$
and equals $\phi$ at $x=ct$. 
Away from $x=ct$, 
the envelope phase has the features 
\begin{gather}
\varphi(u) = \phi, 
\quad
\pm(x-ct)>0
\\
\ell = 0
\end{gather}
for the Hirota $1$-soliton,
and 
\begin{gather}
\varphi(u) \sim \phi \pm \lambda/2, 
\quad
\pm(x-ct) \gg 1/k = 2/(\sqrt{3}(\beta_{-} + \beta_{+}))
\\
\ell = \lambda/(2\pi)
\end{gather}
for the Sasa-Satsuma $1$-soliton,
where $\lambda$ is given by equation \eqref{sslam}.
\end{prop}

Cases $c>0$, $c<0$, $c=0$ are shown in 
\figref{oscil_hirota_overlay_phase_c=4_phi=halfpi},
\figref{oscil_hirota_overlay_phase_c=-4_phi=halfpi},
%\figref{standing_hirota_overlay_phase_phi=halfpi},
for the Hirota oscillatory $1$-soliton, 
and in 
\figref{oscil_ss_overlay_phase_c=4_nu=0_15_100_250_phi=halfpi}, 
\figref{oscil_ss_overlay_phase_c=-4_nu=4_15_100_250_phi=halfpi},
%\figref{standing_ss_overlay_phase_nu=1_15_100_250}
for the Sasa-Satsuma oscillatory $1$-soliton. 



%\begin{figure}[H]
%\centering
%\begin{subfigure}[t]{.5\textwidth}
%\includegraphics[width=\textwidth]{standing_hirota_overlay_amplitude_nu=1_15_100_250-compressed.pdf}
%\captionof{figure}{amplitude}
%\label{standing_hirota_overlay_amplitude_nu=1_15_100_250}
%\end{subfigure}%
%\begin{subfigure}[t]{.5\textwidth}
%\includegraphics[width=\textwidth]{oscil_hirota_phase_phi=halfpi-compressed.pdf}
%\captionof{figure}{envelope phase}
%\label{standing_hirota_overlay_phase_phi=halfpi}
%\end{subfigure}
%\caption{Hirota standing wave ($c=0$) with $|\nu|=1,15,100,250$}
%\end{figure}

%\begin{figure}[H]
%\centering
%\begin{subfigure}[t]{.5\textwidth}
%\includegraphics[width=\textwidth]{standing_ss_overlay_amplitude_nu=1_15_100_250-compressed.pdf}
%\captionof{figure}{amplitude}
%\label{standing_ss_overlay_amplitude_nu=1_15_100_250}
%\end{subfigure}%
%\begin{subfigure}[t]{.5\textwidth}
%\includegraphics[width=\textwidth,bb=0 0 731 325]{standing_ss_overlay_phase_nu=1_15_100_250-compressed.pdf}
%\captionof{figure}{envelope phase}
%\label{standing_ss_overlay_phase_nu=1_15_100_250}
%\end{subfigure}
%\caption{Sasa-Satsuma standing wave ($c=0$) with $|\nu|=1,15,100,250$}
%\end{figure}

\subsection{Oscillatory $2$-solitons}
We now illustrate some properties of 
the oscillatory $2$-soliton solutions from Theorem~\ref{thm:2soliton},
shown graphically by the amplitude $|u|$ and the phase gradient $\arg(u)_x$.

An oscillatory $2$-soliton \eqref{enveloscil2soliton} 
is parameterized by phases $\phi_1,\phi_2$, frequencies $\nu_1,\nu_2$, 
and speeds $c_1\neq c_2$, 
satisfying the kinematic relations \eqref{kinrels}. 
These solitons are symmetric under simultaneously interchanging 
$c_1 \longleftrightarrow c_2$, $\nu_1 \longleftrightarrow \nu_2$, 
$\phi_1 \longleftrightarrow \phi_2$, 
and thus we can assume $c_1>c_2$ without loss of generality. 

\begin{figure}[H]
\begin{subfigure}[t]{0.45\textwidth}
\includegraphics[width=\textwidth]{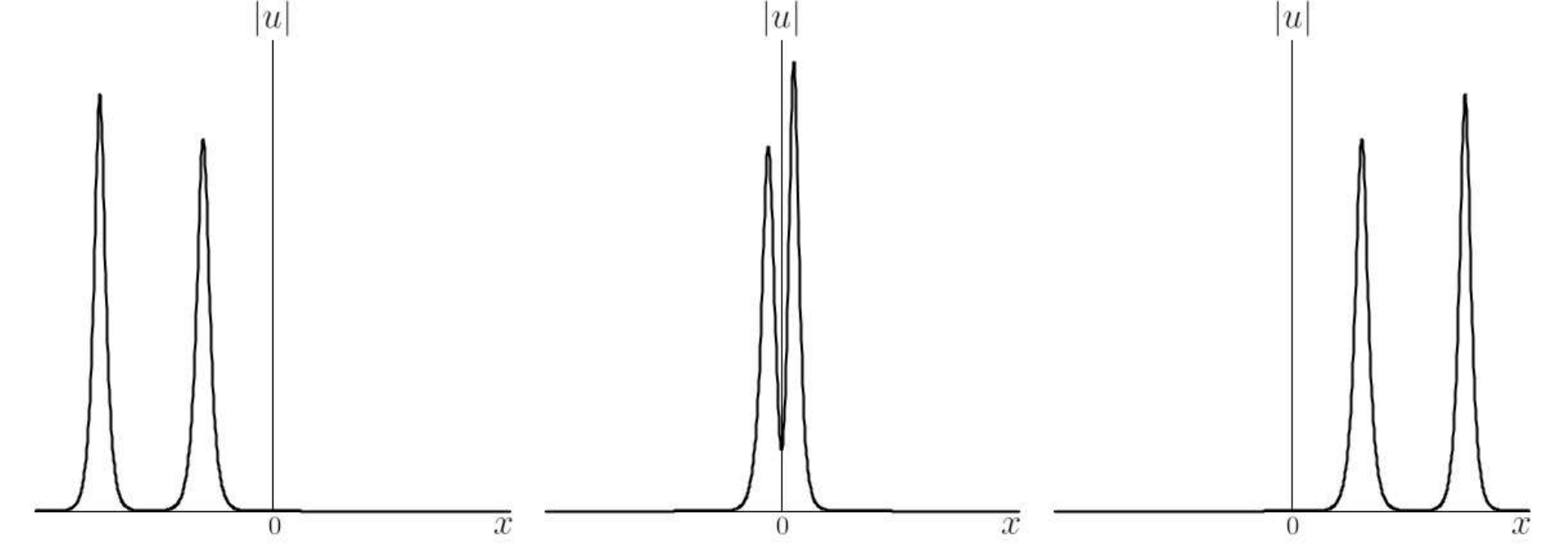}
\captionof{figure}{amplitude}
\end{subfigure}%
\begin{subfigure}[t]{0.45\textwidth}
\includegraphics[width=\textwidth]{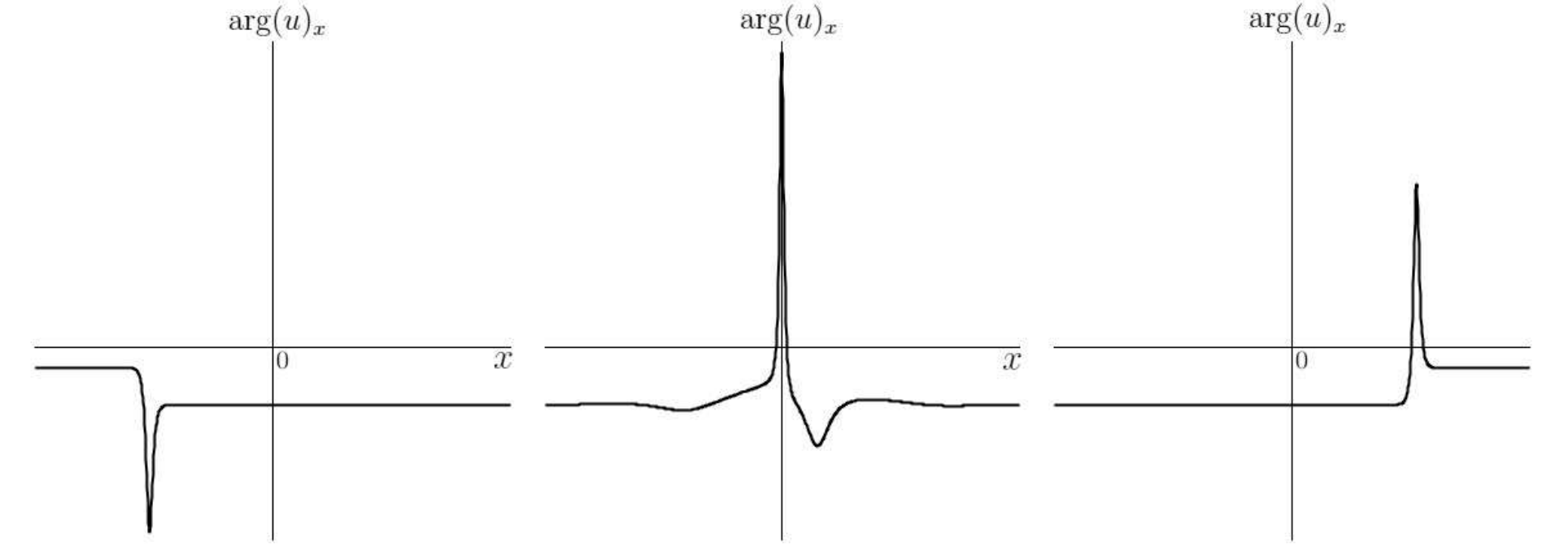}
\captionof{figure}{phase gradient}
\end{subfigure}
\caption{Hirota oscillatory $2$-soliton right-overtake 
with $c_1=4$, $c_2=2$, $\nu_1=2$, $\nu_2=5$, $\phi_1=0$, $\phi_2=\pi/2$
 ($t_0=-4$)}
\label{right-overtake_hirota_c1=4_c2=2_nu1=2_nu2=5_phi1=0_phi2=halfpi}
\end{figure}

\begin{figure}[H]
\begin{subfigure}[t]{0.45\textwidth}
\includegraphics[width=\textwidth]{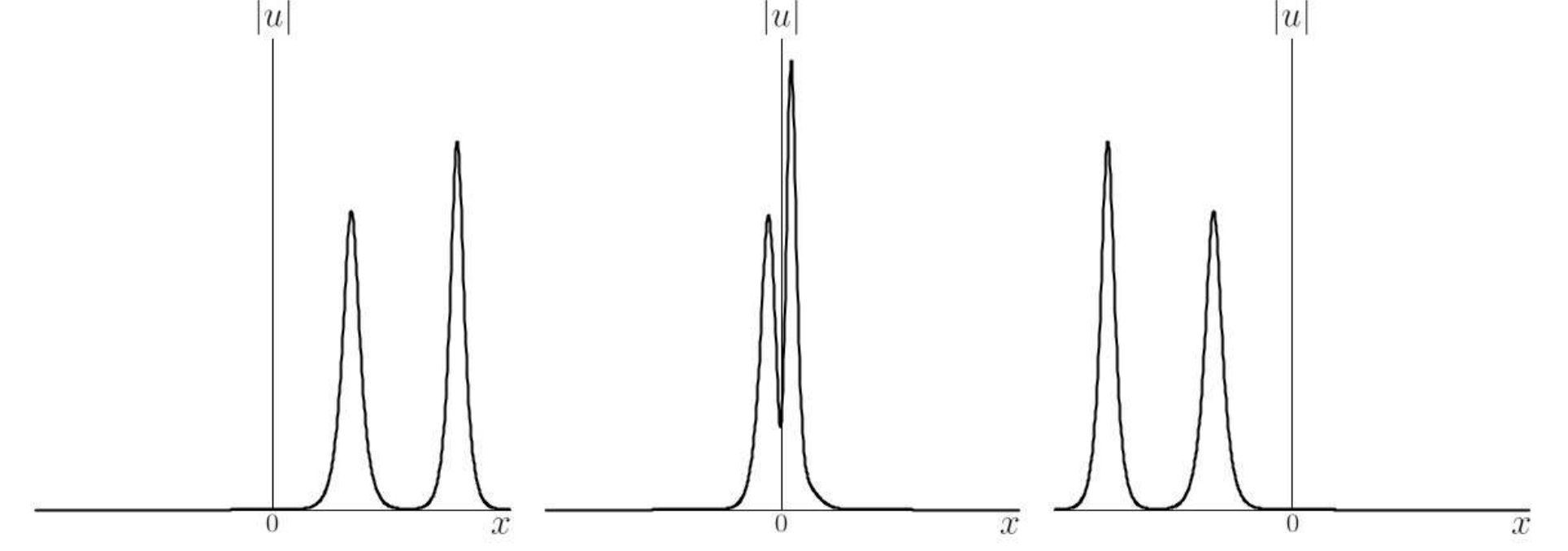}
\captionof{figure}{amplitude}
\end{subfigure}%
\begin{subfigure}[t]{0.45\textwidth}
\includegraphics[width=\textwidth]{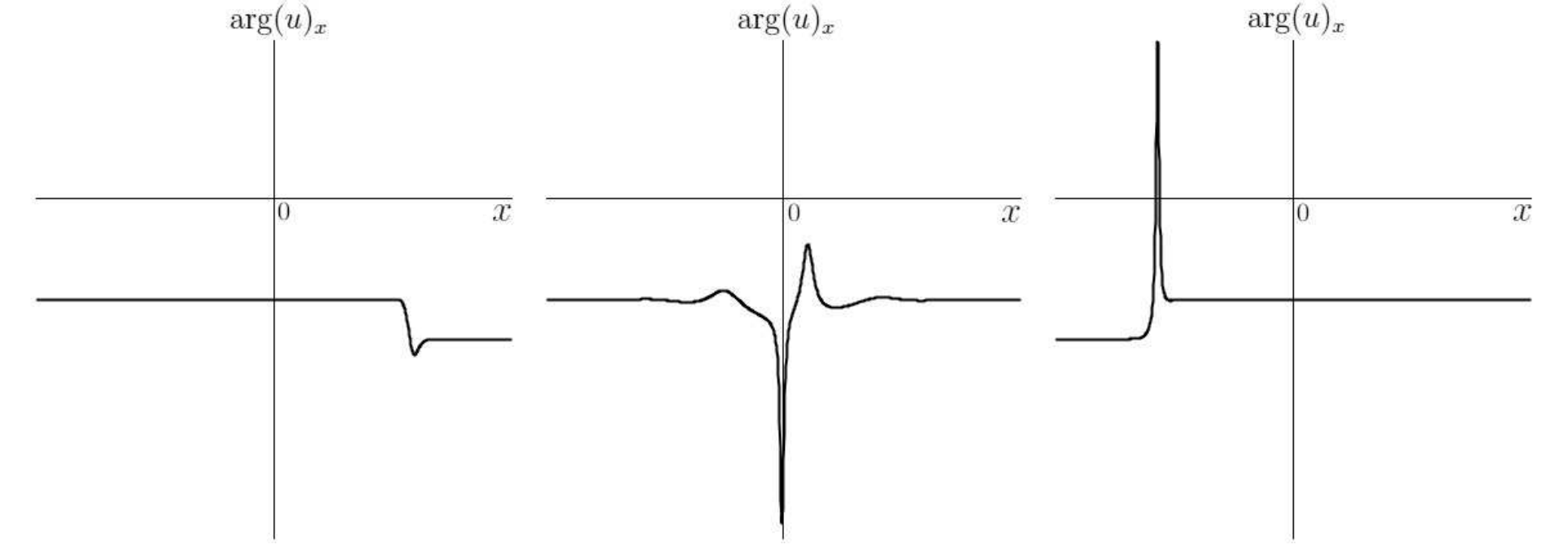}
\captionof{figure}{phase gradient}
\end{subfigure}
\caption{Hirota oscillatory $2$-soliton left-overtake 
with $c_1=-2$, $c_2=-4$, $\nu_1=2$, $\nu_2=5$, $\phi_1=0$, $\phi_2=\pi/2$ ($t_0=-10$)}
\label{left-overtake_hirota_c1=-2_c2=-4_nu1=2_nu2=5_phi1=0_phi2=halfpi}
\end{figure}

\begin{figure}[H]
\begin{subfigure}[t]{0.45\textwidth}
\includegraphics[width=\textwidth]{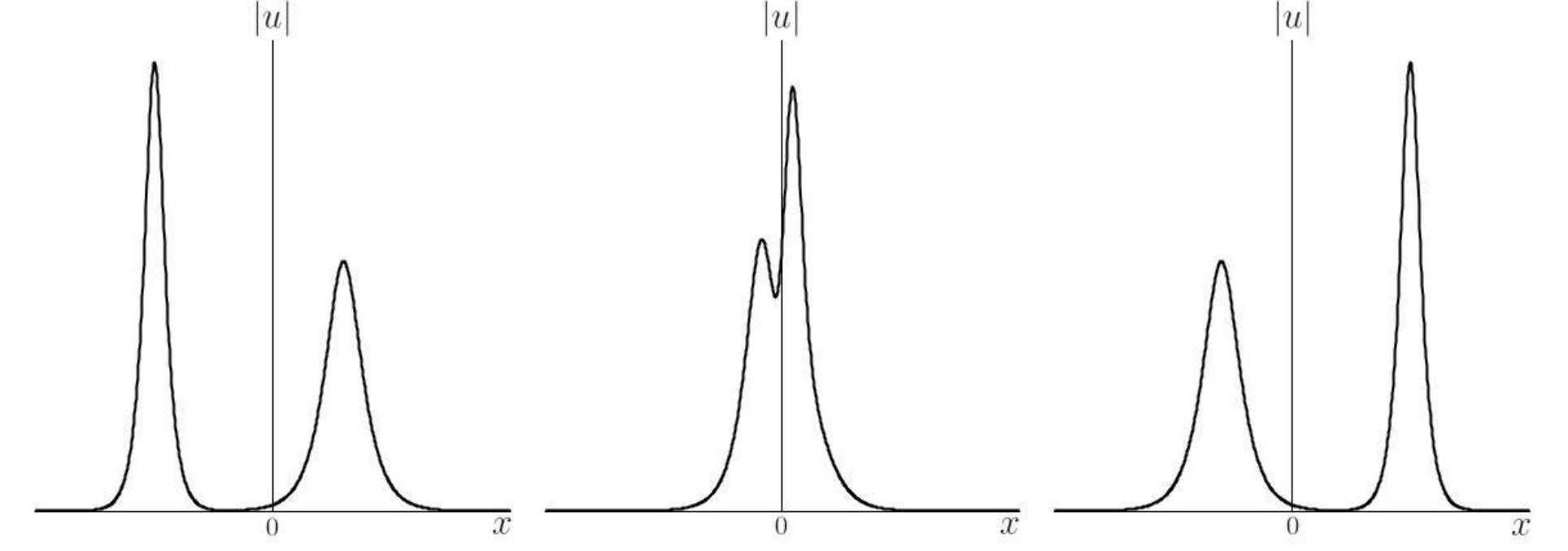}
\captionof{figure}{amplitude}
\end{subfigure}%
\begin{subfigure}[t]{0.45\textwidth}
\includegraphics[width=\textwidth]{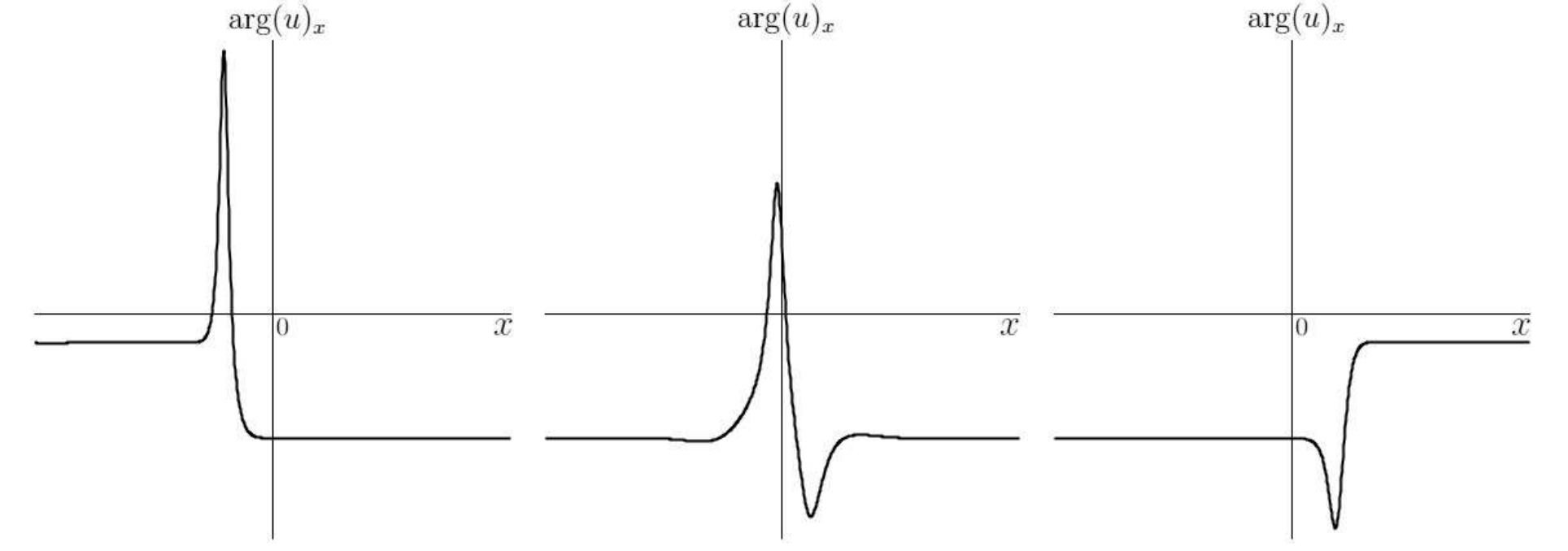}
\captionof{figure}{phase gradient}
\end{subfigure}
\caption{Hirota oscillatory $2$-soliton head-on
with $c_1=4$, $c_2=-2$, $\nu_1=2$, $\nu_2=5$, $\phi_1=0$, $\phi_2=\pi/2$ ($t_0=-1.5$)}
\label{headon_hirota_c1=4_c2=-2_nu1=2_nu2=5_phi1=0_phi2=halfpi}
\end{figure}

\begin{figure}[H]
\begin{subfigure}[t]{0.45\textwidth}
\includegraphics[width=\textwidth]{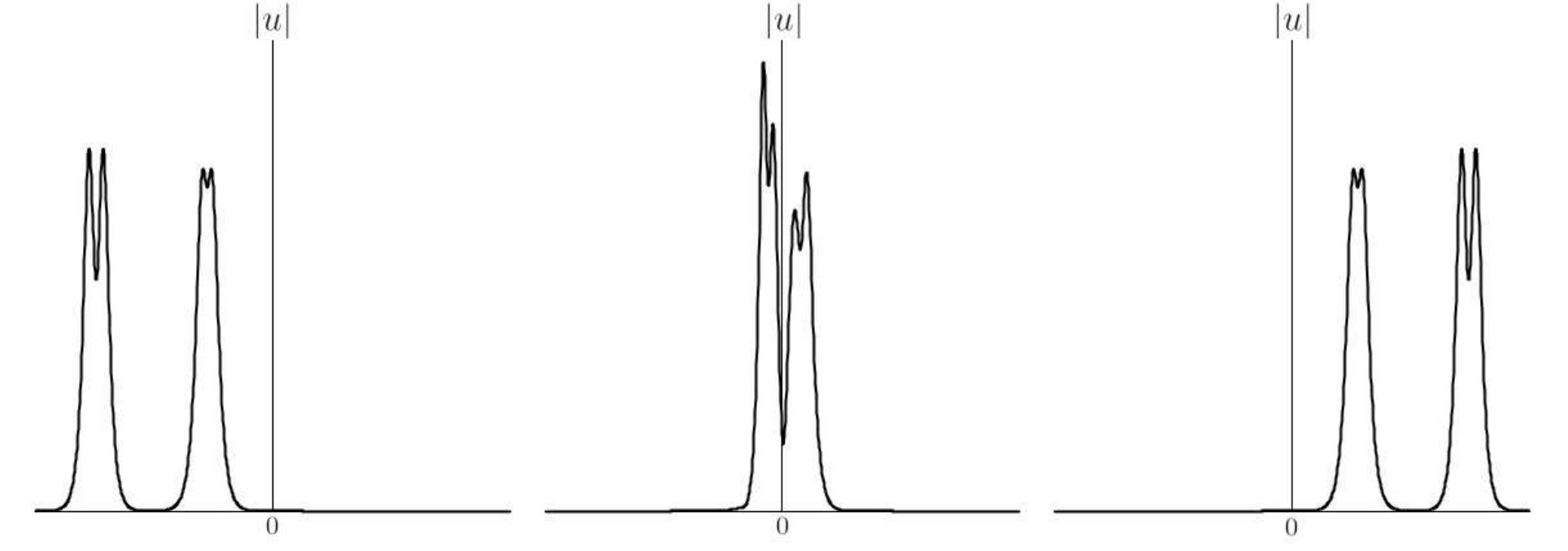}
\captionof{figure}{amplitude}
\end{subfigure}%
\begin{subfigure}[t]{0.45\textwidth}
\includegraphics[width=\textwidth]{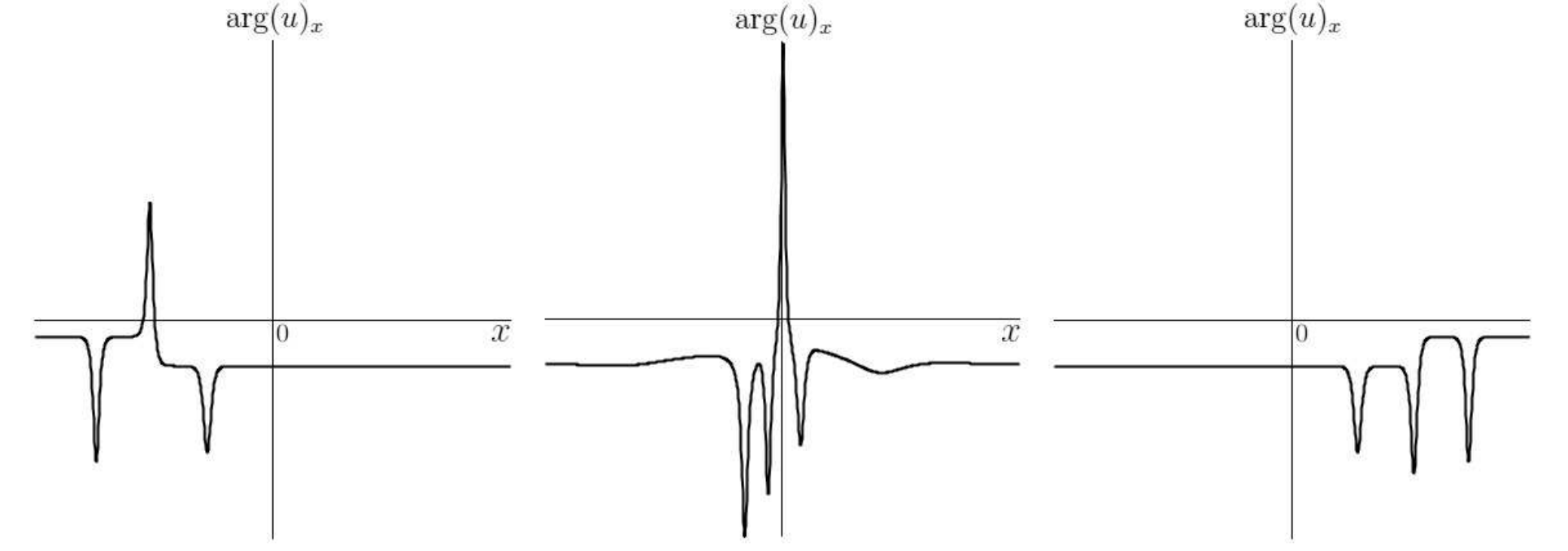}
\captionof{figure}{phase gradient}
\end{subfigure}
\caption{Sasa-Satsuma oscillatory $2$-soliton right-overtake 
with $c_1=4$, $c_2=2$, $\nu_1=2$, $\nu_2=5$, $\phi_1=0$, $\phi_2=\pi/2$ ($t_0=-4$)}
\label{right-overtake_ss_c1=4_c2=2_nu1=2_nu2=5_phi1=0_phi2=halfpi}
\end{figure}

\begin{figure}[H]
\begin{subfigure}[t]{0.45\textwidth}
\includegraphics[width=\textwidth]{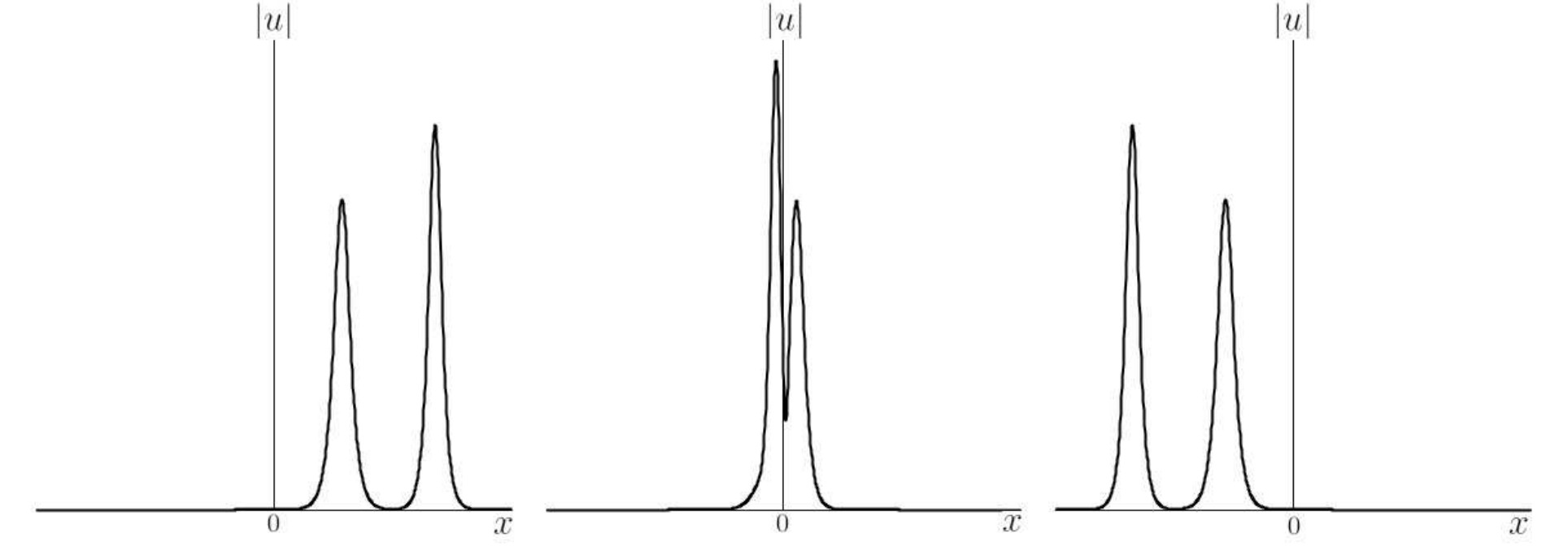}
\captionof{figure}{amplitude}
\end{subfigure}%
\begin{subfigure}[t]{0.45\textwidth}
\includegraphics[width=\textwidth]{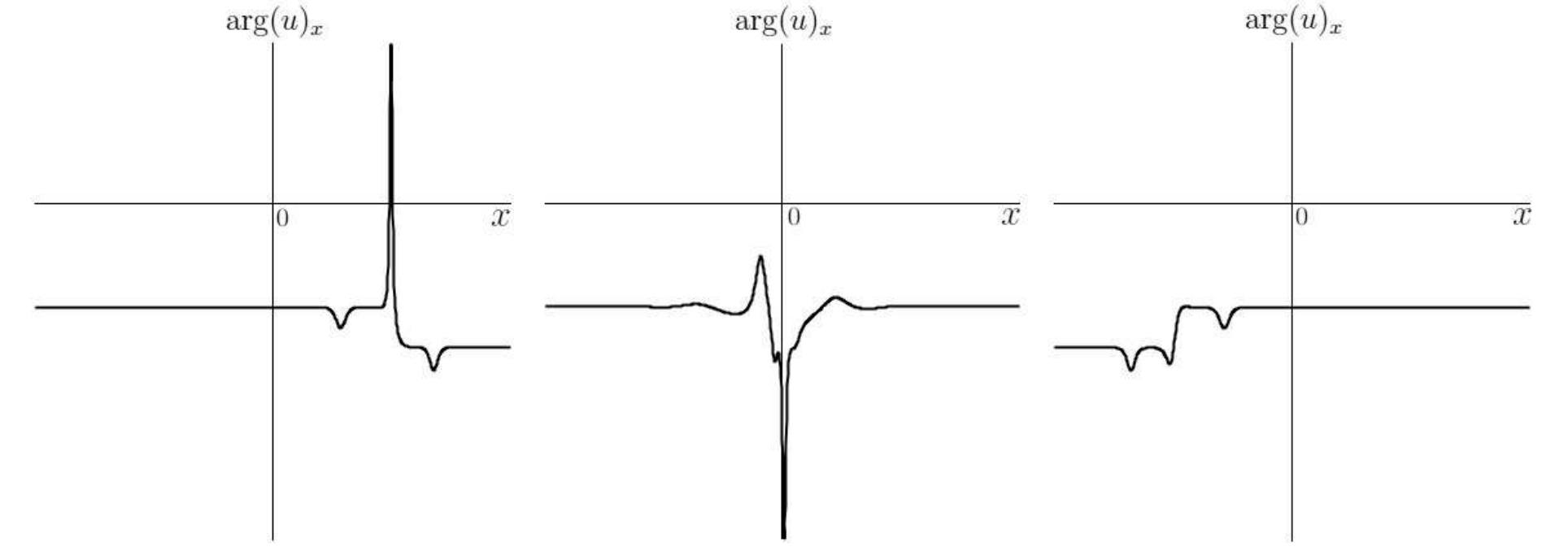}
\captionof{figure}{phase gradient}
\end{subfigure}
\caption{Sasa-Satsuma oscillatory $2$-soliton left-overtake 
with $c_1=-2$, $c_2=-4$, $\nu_1=2$, $\nu_2=5$, $\phi_1=0$, $\phi_2=\pi/2$ ($t_0=-10$)}
\label{left-overtake_ss_c1=-2_c2=-4_nu1=2_nu2=5_phi1=0_phi2=halfpi}
\end{figure}

\begin{figure}[H]
\begin{subfigure}[t]{0.45\textwidth}
\includegraphics[width=\textwidth]{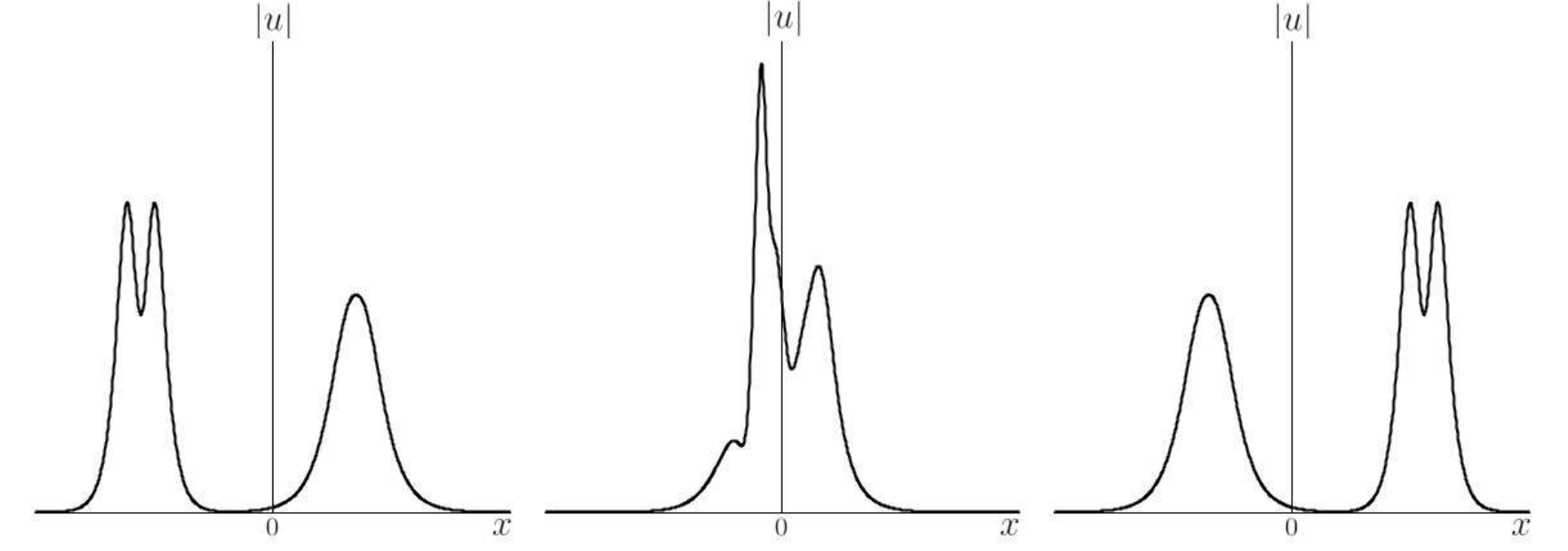}
\captionof{figure}{amplitude}
\end{subfigure}%
\begin{subfigure}[t]{0.45\textwidth}
\includegraphics[width=\textwidth]{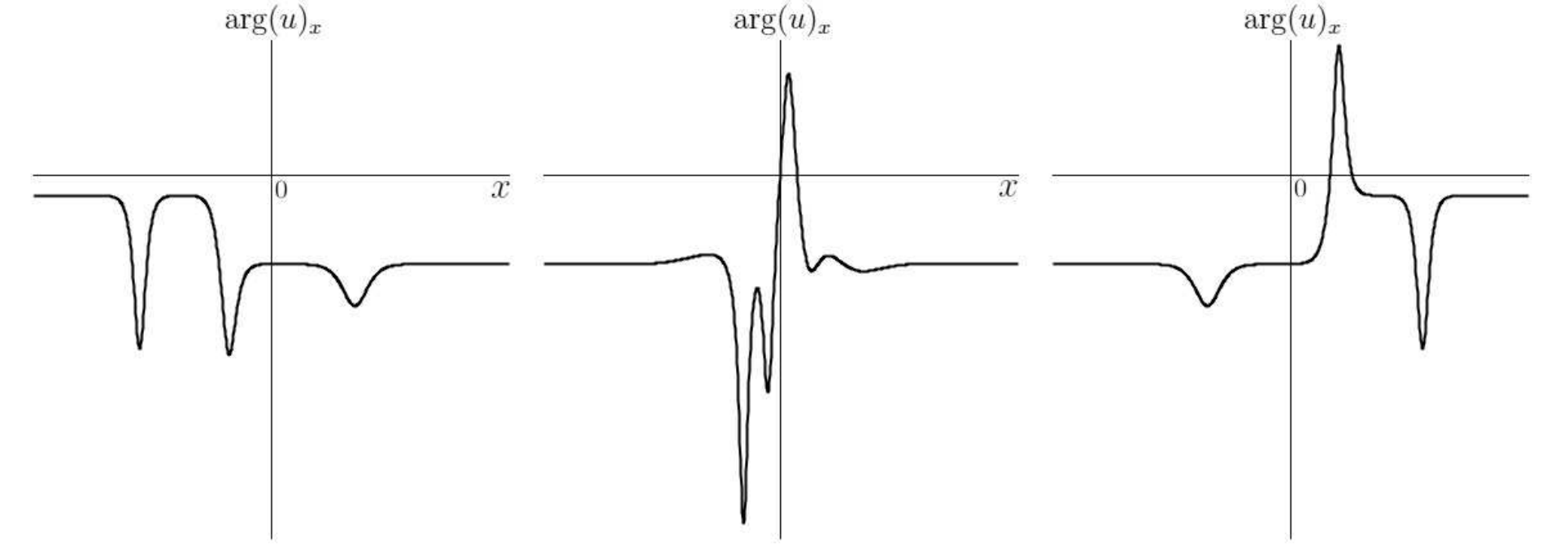}
\captionof{figure}{phase gradient}
\end{subfigure}
\caption{Sasa-Satsuma oscillatory $2$-soliton head-on
with $c_1=4$, $c_2=-2$, $\nu_1=2$, $\nu_2=5$, $\phi_1=0$, $\phi_2=\pi/2$ $(t_0=-1.5$)}
\label{headon_ss_c1=4_c2=-2_nu1=2_nu2=5_phi1=0_phi2=halfpi}
\end{figure}

As seen from 
\figref{right-overtake_hirota_c1=4_c2=2_nu1=2_nu2=5_phi1=0_phi2=halfpi}--\figref{headon_hirota_c1=4_c2=-2_nu1=2_nu2=5_phi1=0_phi2=halfpi}
for the Hirota equation \eqref{hmkdveqscaled}
and 
\figref{right-overtake_ss_c1=4_c2=2_nu1=2_nu2=5_phi1=0_phi2=halfpi}--\figref{headon_ss_c1=4_c2=-2_nu1=2_nu2=5_phi1=0_phi2=halfpi}
for the Sasa-Satsuma equation \eqref{ssmkdveqscaled}, 
oscillatory $2$-solitons describe collisions between two oscillatory waves
at $t=t_0,0,-t_0$. 
The collision is a right-overtake when $c_1>c_2\geq 0$, 
a left-overtake when $0\geq c_1>c_2$, 
and a head-on when $c_1>0>c_2$. 



In all collisions, 
the speed and the shape of both oscillatory waves are preserved. 
We will show analytically in a sequel paper \cite{AncWilMia} that 
a collision produces a shift in the asymptotic phase and position of each oscillatory wave.

\subsection{Oscillatory breathers}
Last we discuss a few aspects of the oscillatory breather solutions from Theorem~\ref{thm:oscilbreather}. 

An oscillatory breather \eqref{oscilbreathersoln}
is parameterized by a speed $c$, 
an envelope frequency $\nu\neq 0$ and phase $\phi$, 
and a modulation frequency $\nu_0$ and phase $\phi_0$, 
where the speed and frequencies satisfy 
the kinematic relation $(c/3)^3 +(\nu/2)^2 > 0$. 
Depending on the combined frequency $|\nu|+|\nu_0|$, 
the speed of a breather can be positive, negative, or zero. 
The additional parameter $\chi$ controls the size of the oscillations,
such that the oscillations disappear in the limit $|\chi|\gg 1$. 

From the expressions \eqref{hoscilbrX1}--\eqref{hoscilbrY}
for the Hirota oscillatory breather 
and \eqref{ssoscilbrX1}--\eqref{ssoscilbrY} 
for the Sasa-Satsuma oscillatory breather, 
we see that the amplitude $|u|=|\tilde f(\xi,\tau)|$ 
has exponential decay in the moving coordinate $|\xi|=|x-ct|$
and is periodic in the oscillation coordinate $\tau=\nu t+\phi$. 
In particular, these breathers are distinguished from oscillatory waves
by having an oscillating amplitude with a temporal period of $T=\pi/\nu$. 
We also see that the phase $\arg(u)$ of these breathers 
is time-periodic if either the frequency $\nu_0$ vanishes,
or the two frequencies $\nu_0\neq 0$ and $\nu\neq 0$ are commensurate, 
where the condition $\nu_0\neq 0$ distinguishes an oscillatory breather from 
an ordinary breather. 

\begin{figure}[H]
\begin{subfigure}[t]{0.45\textwidth}
\includegraphics[width=\textwidth]{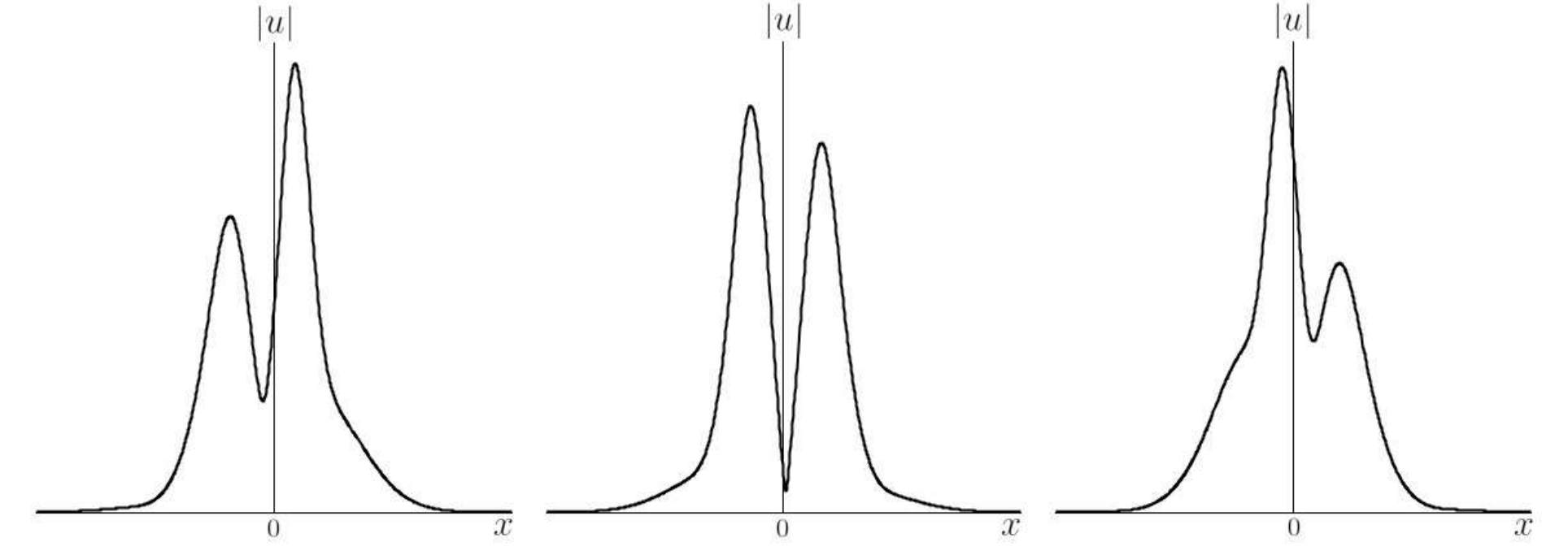}
\captionof{figure}{amplitude}
\end{subfigure}%
\begin{subfigure}[t]{0.45\textwidth}
\includegraphics[width=\textwidth]{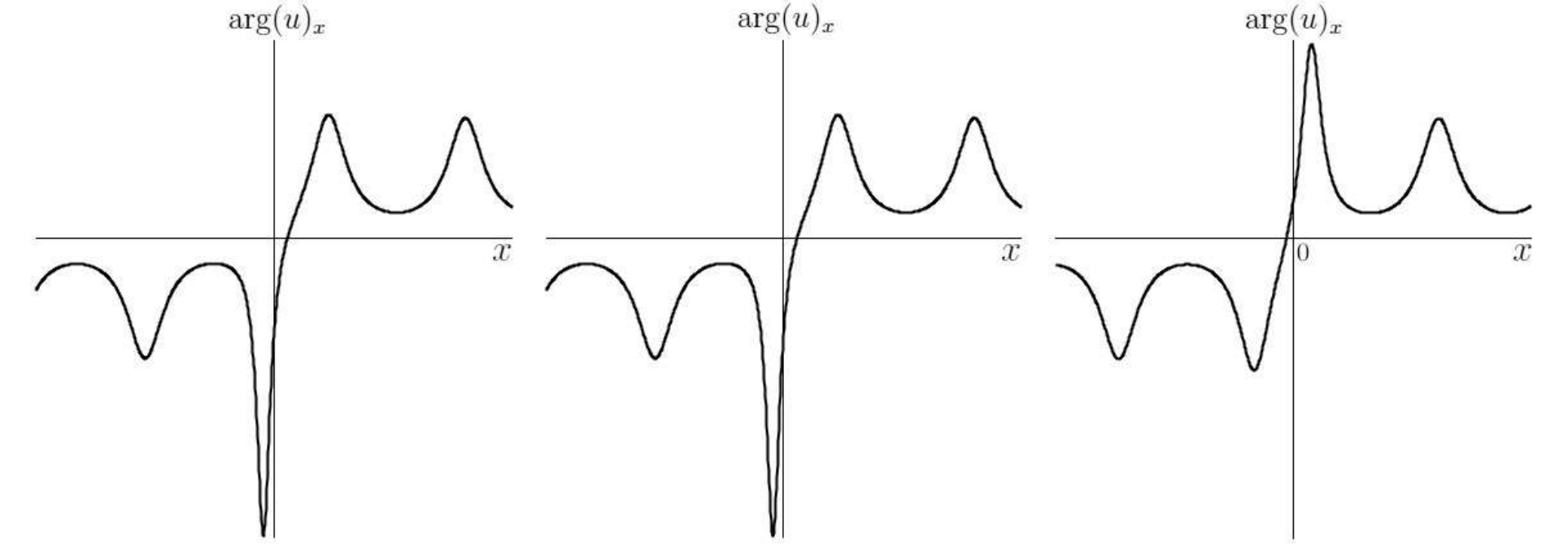}
\captionof{figure}{phase gradient}
\end{subfigure}
\caption{Hirota stationary breather 
with $\chi=0.5$, $c=0$, $\nu=2$, $\phi=0$, $\phi_0=0$ ($t_0=-2$)}
\label{breather_hirota_2mu=1_c=0_nu=2_phi0=0_phi=0}
\end{figure}

\begin{figure}[H]
\begin{subfigure}[t]{0.45\textwidth}
\includegraphics[width=\textwidth]{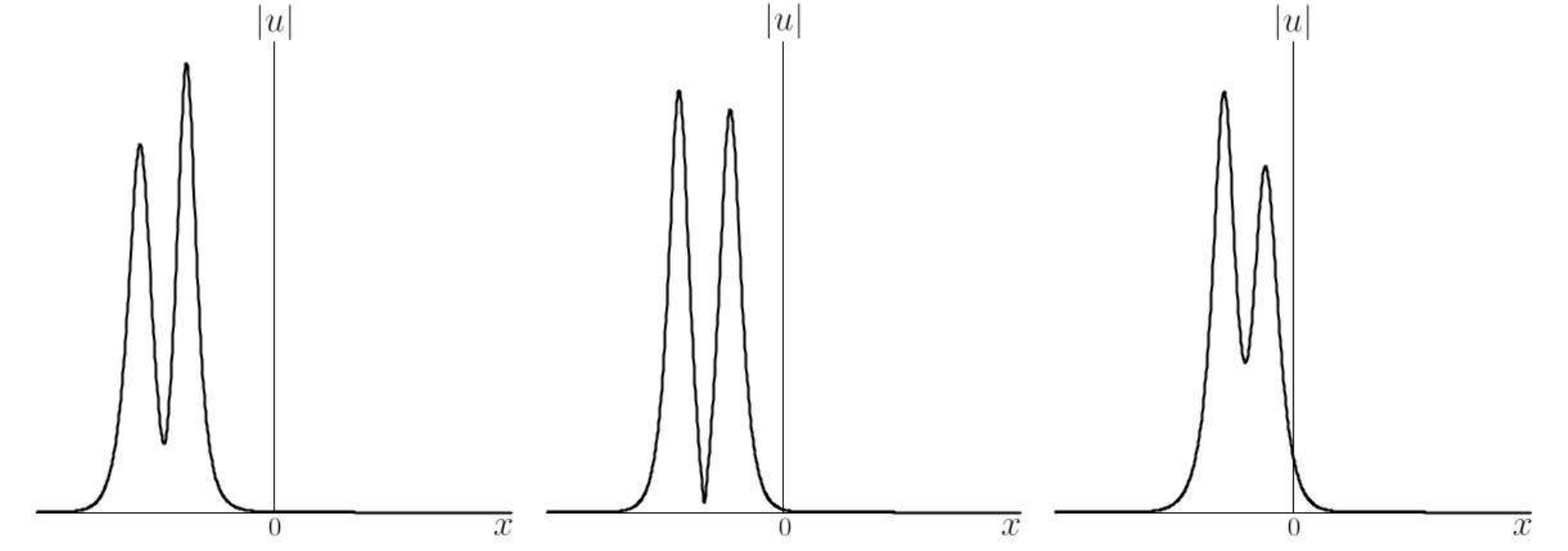}
\captionof{figure}{amplitude}
\end{subfigure}%
\begin{subfigure}[t]{0.45\textwidth}
\includegraphics[width=\textwidth]{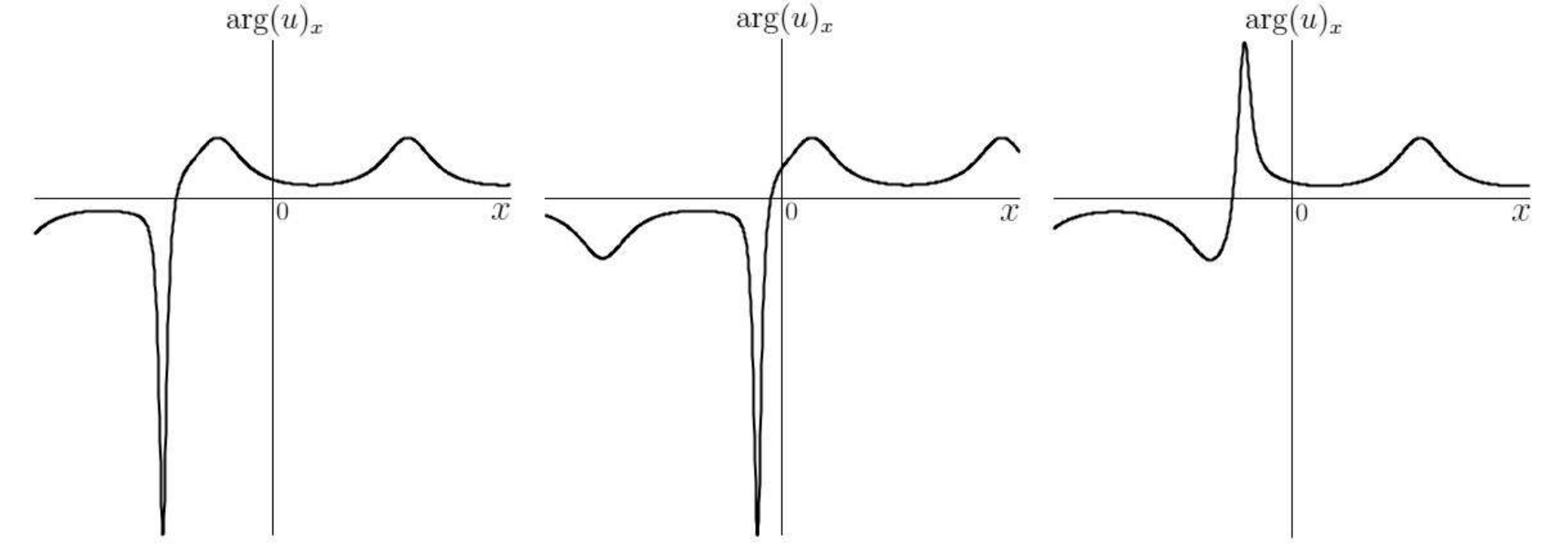}
\captionof{figure}{phase gradient}
\end{subfigure}
\caption{Hirota moving breather 
with $\chi=0.5$, $c=3$, $\nu=2$, $\phi=0$, $\phi_0=0$ ($t_0=-2$)}
\label{right-breather_hirota_2mu=1_c=3_nu=2_phi0=0_phi=0}
\end{figure}

\begin{figure}[H]
\begin{subfigure}[t]{0.45\textwidth}
\includegraphics[width=\textwidth]{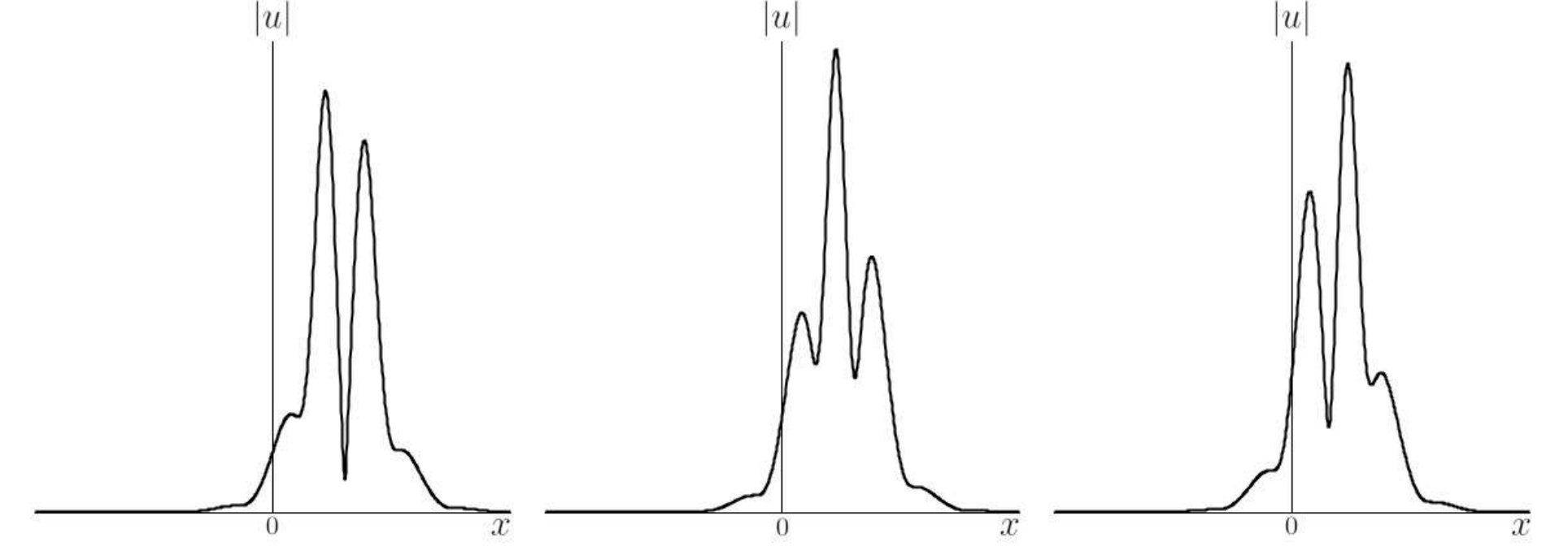}
\captionof{figure}{amplitude}
\end{subfigure}%
\begin{subfigure}[t]{0.45\textwidth}
\includegraphics[width=\textwidth]{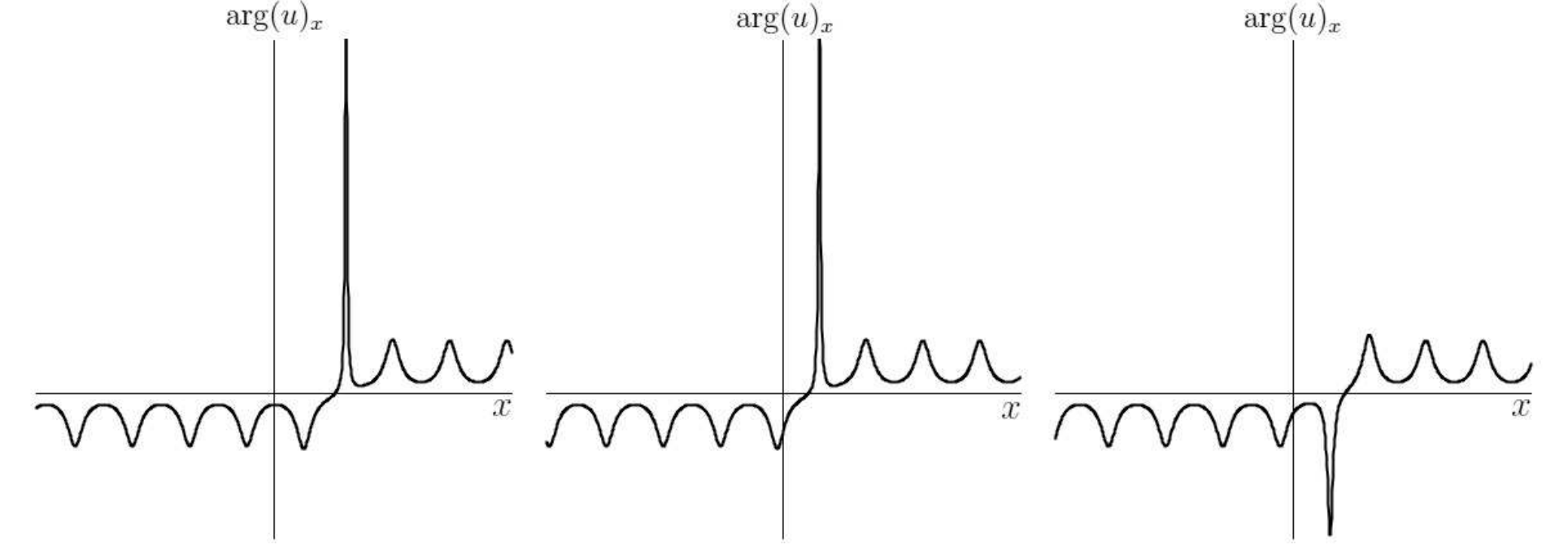}
\captionof{figure}{phase gradient}
\end{subfigure}
\caption{Hirota moving breather 
with $\chi=0.5$, $c=-2$, $\nu=3$, $\phi=0$, $\phi_0=0$ ($t_0=-2$)}
\label{left-breather_hirota_2mu=1_c=-2_nu=3_phi0=0_phi=0}
\end{figure}

\begin{figure}[H]
\begin{subfigure}[t]{0.45\textwidth}
\includegraphics[width=\textwidth]{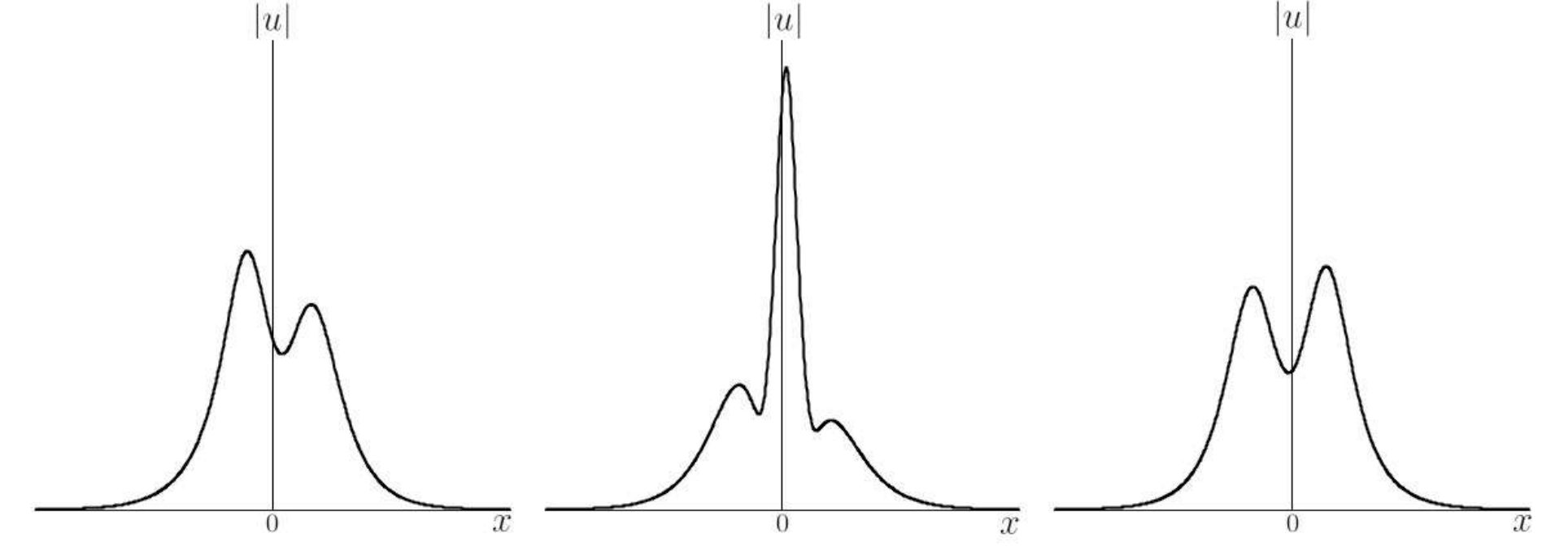}
\captionof{figure}{amplitude}
\end{subfigure}%
\begin{subfigure}[t]{0.45\textwidth}
\includegraphics[width=\textwidth]{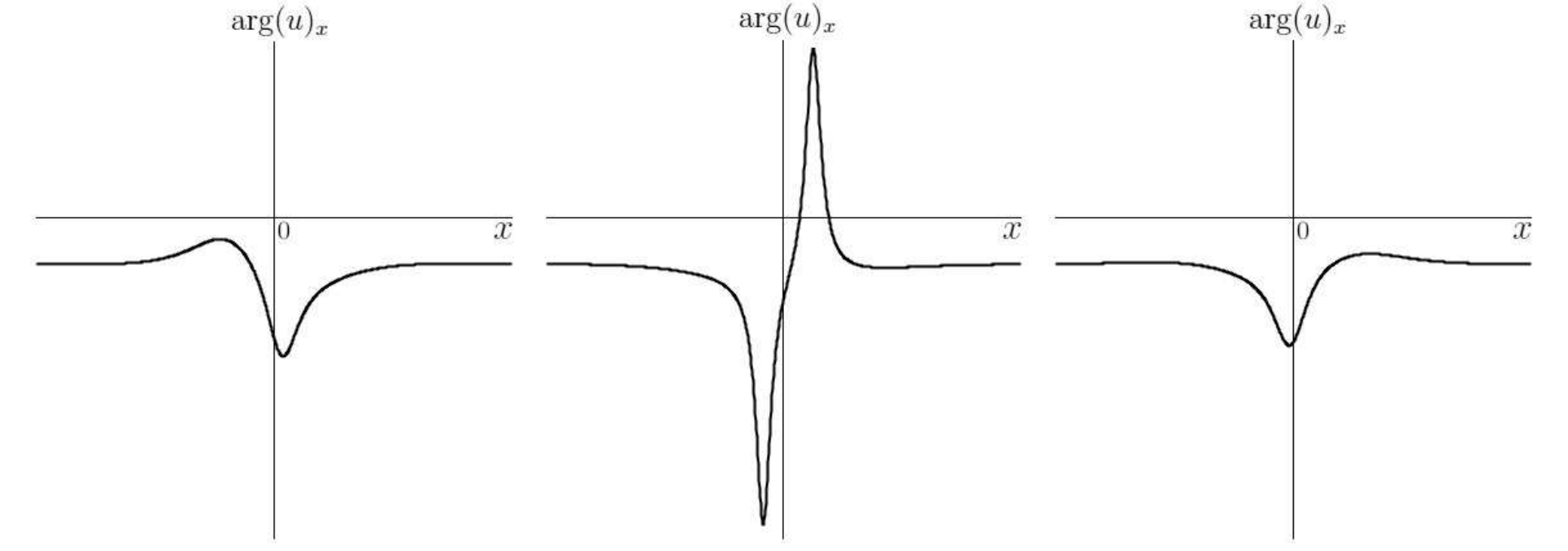}
\captionof{figure}{phase gradient}
\end{subfigure}
\caption{Hirota stationary oscillatory breather 
with $\chi=0$, $c=0$, $\nu=2$, $\phi=0$, $\nu_0=3$, $\phi_0=0$ ($t_0=-2$)}
\label{oscilbreather_hirota_c=0_nu1=5_nu2=1_phi1=0_phi2=0}
\end{figure}

\begin{figure}[H]
\begin{subfigure}[t]{0.45\textwidth}
\includegraphics[width=\textwidth]{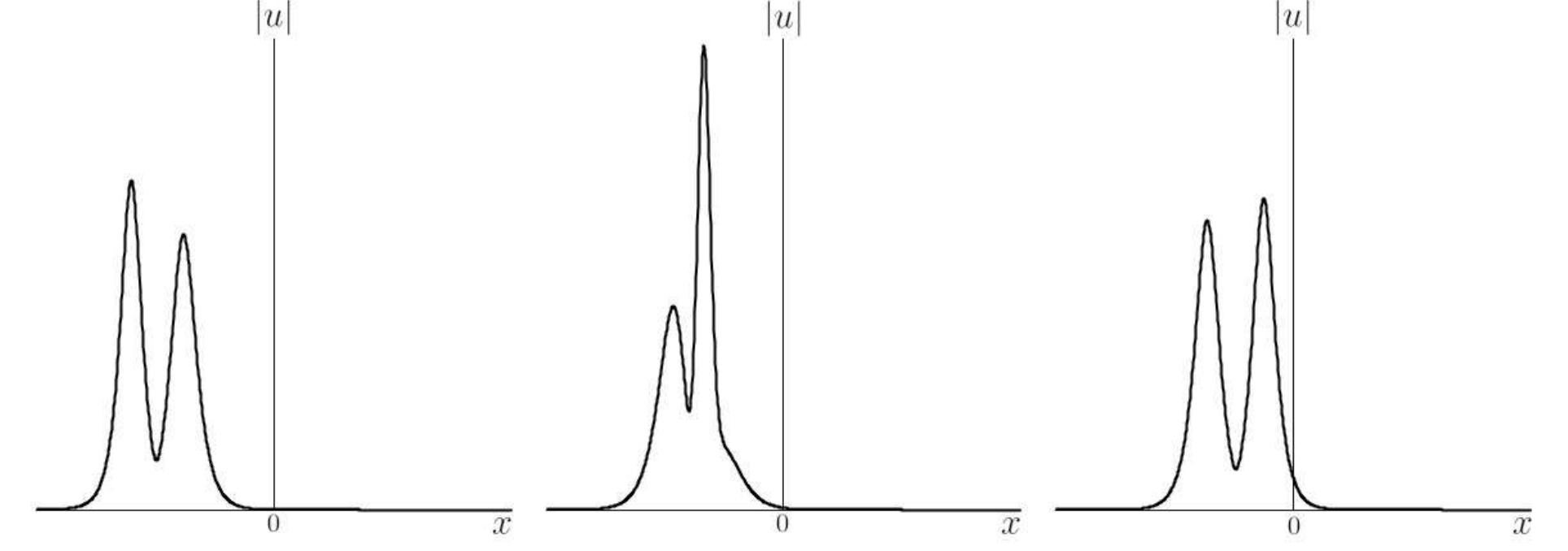}
\captionof{figure}{amplitude}
\end{subfigure}%
\begin{subfigure}[t]{0.45\textwidth}
\includegraphics[width=\textwidth]{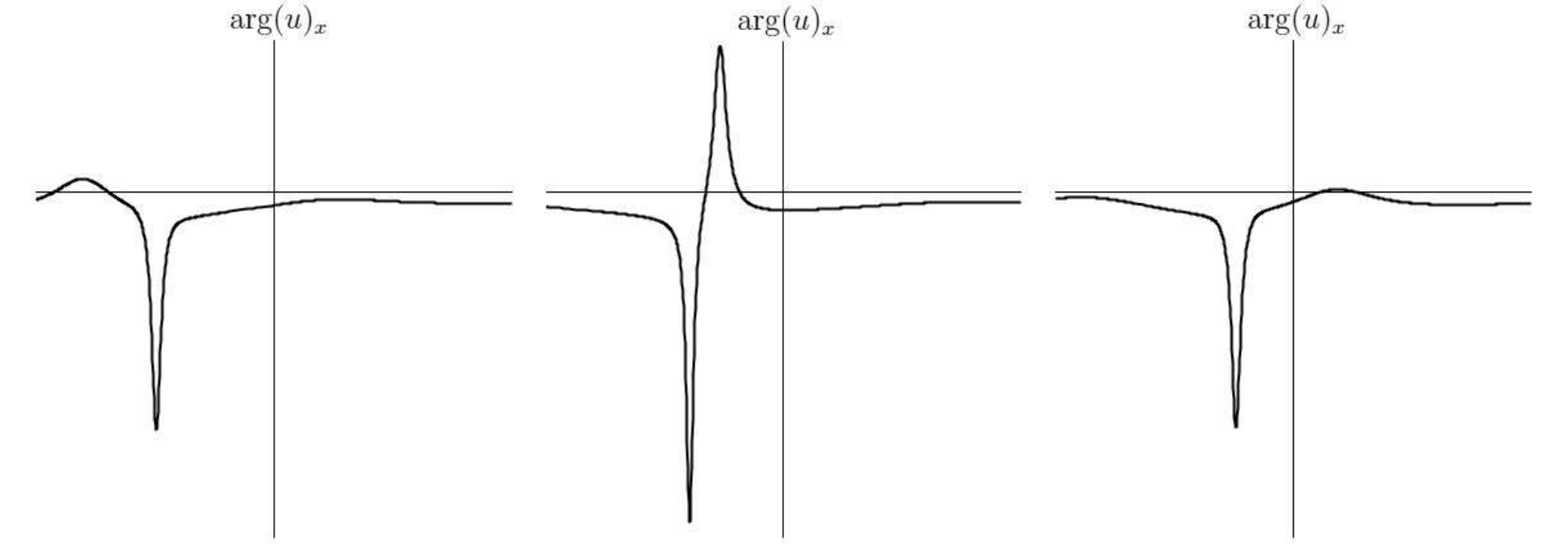}
\captionof{figure}{phase gradient}
\end{subfigure}
\caption{Hirota moving oscillatory breather 
with $\chi=0$, $c=3$, $\nu=2$, $\phi=0$, $\nu_0=3$, $\phi_0=0$ ($t_0=-2$)}
\label{right-oscilbreather_hirota_c=3_nu1=5_nu2=1_phi1=0_phi2=0}
\end{figure}

\begin{figure}[H]
\begin{subfigure}[t]{0.45\textwidth}
\includegraphics[width=\textwidth]{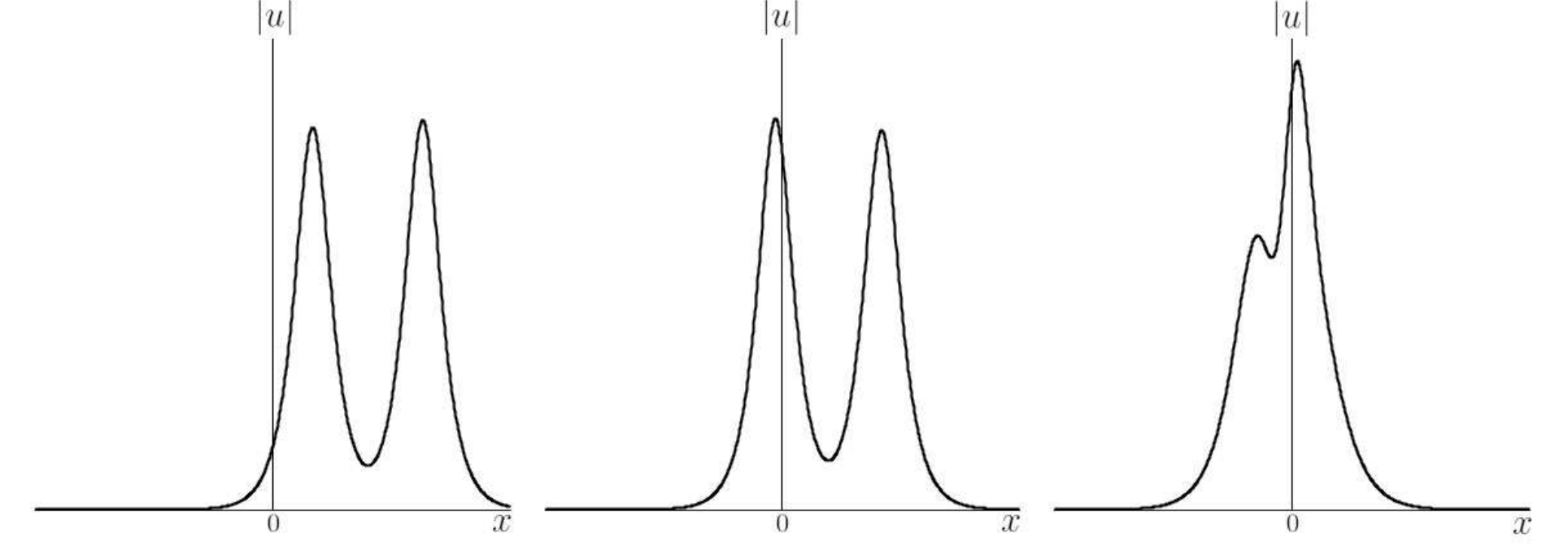}
\captionof{figure}{amplitude}
\end{subfigure}%
\begin{subfigure}[t]{0.45\textwidth}
\includegraphics[width=\textwidth]{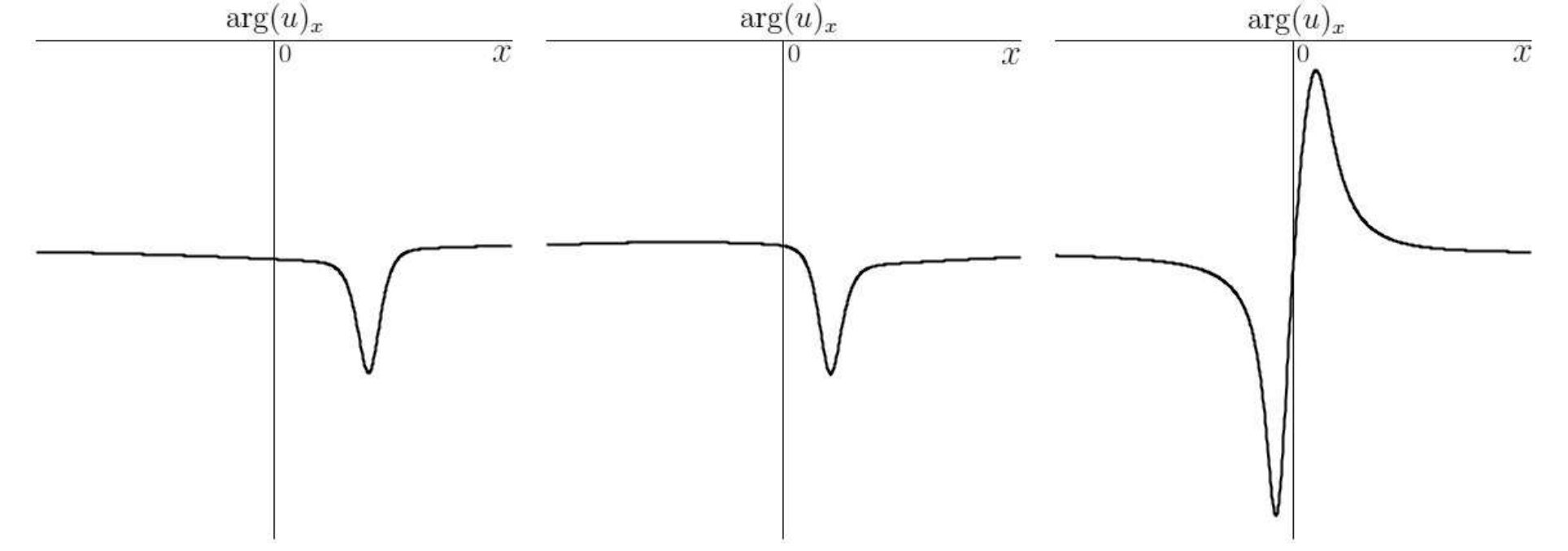}
\captionof{figure}{phase gradient}
\end{subfigure}
\caption{Hirota moving oscillatory breather 
with $\chi=0$, $c=-2$, $\nu=1$, $\phi=0$, $\nu_0=9$, $\phi_0=0$ ($t_0=-2$)}
\label{left-oscilbreather_phase_c=-2_nu1=10_nu2=8_phi1=0_phi2=0}
\end{figure}

\begin{figure}[H]
\begin{subfigure}[t]{0.45\textwidth}
\includegraphics[width=\textwidth]{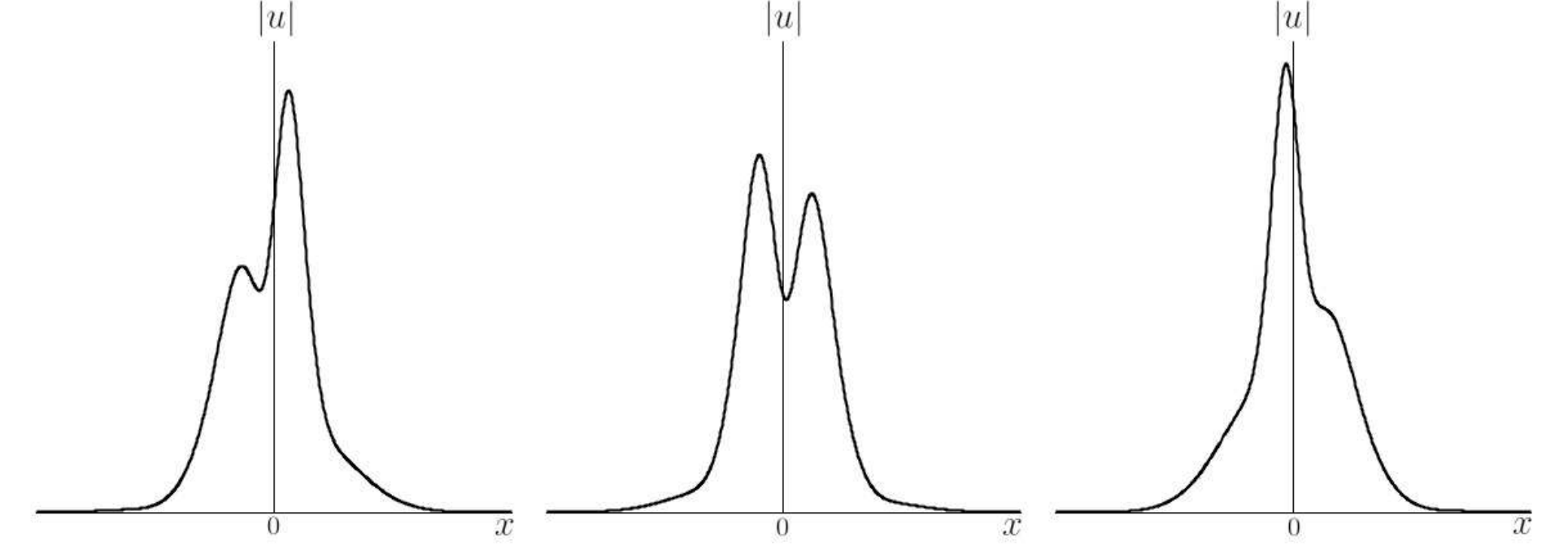}
\captionof{figure}{amplitude}
\end{subfigure}%
\begin{subfigure}[t]{0.45\textwidth}
\includegraphics[width=\textwidth]{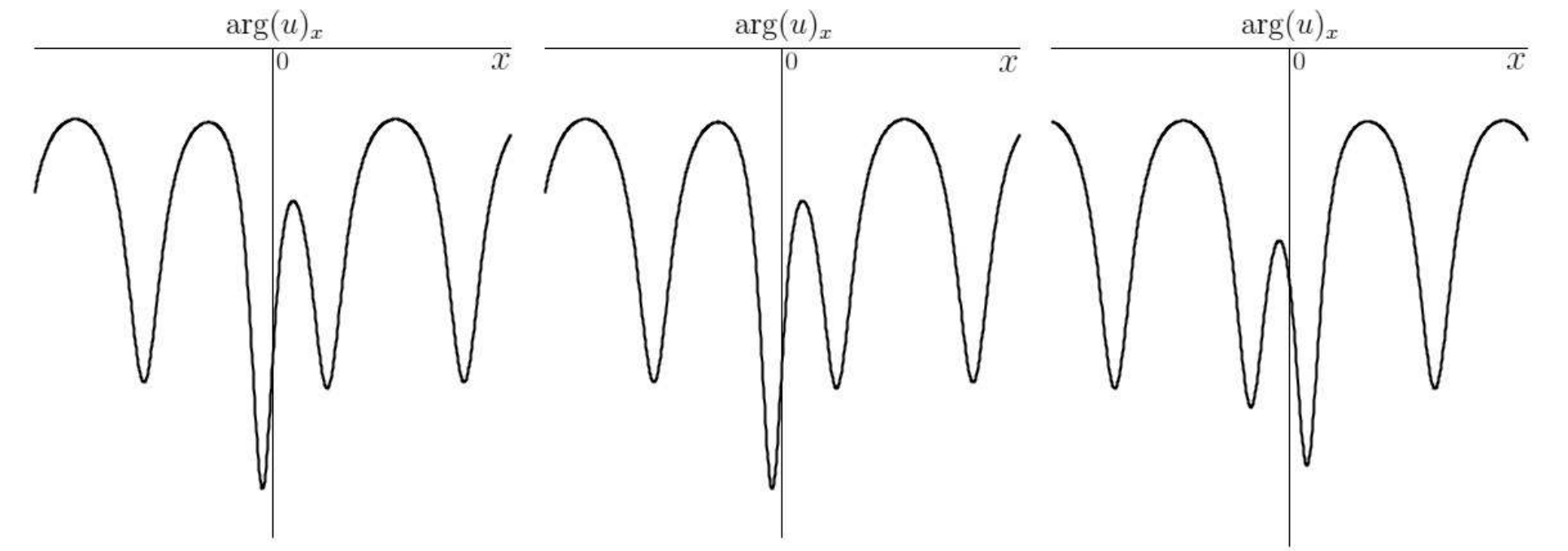}
\captionof{figure}{phase gradient}
\end{subfigure}
\caption{Sasa-Satsuma stationary breather 
with $\chi=0.5$, $c=0$, $\nu=2$, $\phi=0$, $\phi_0=0$ ($t_0=-2$)}
\label{breather_ss_2mu=1_c=0_nu=2_phi0=0_phi=0}
\end{figure}

\begin{figure}[H]
\begin{subfigure}[t]{0.45\textwidth}
\includegraphics[width=\textwidth]{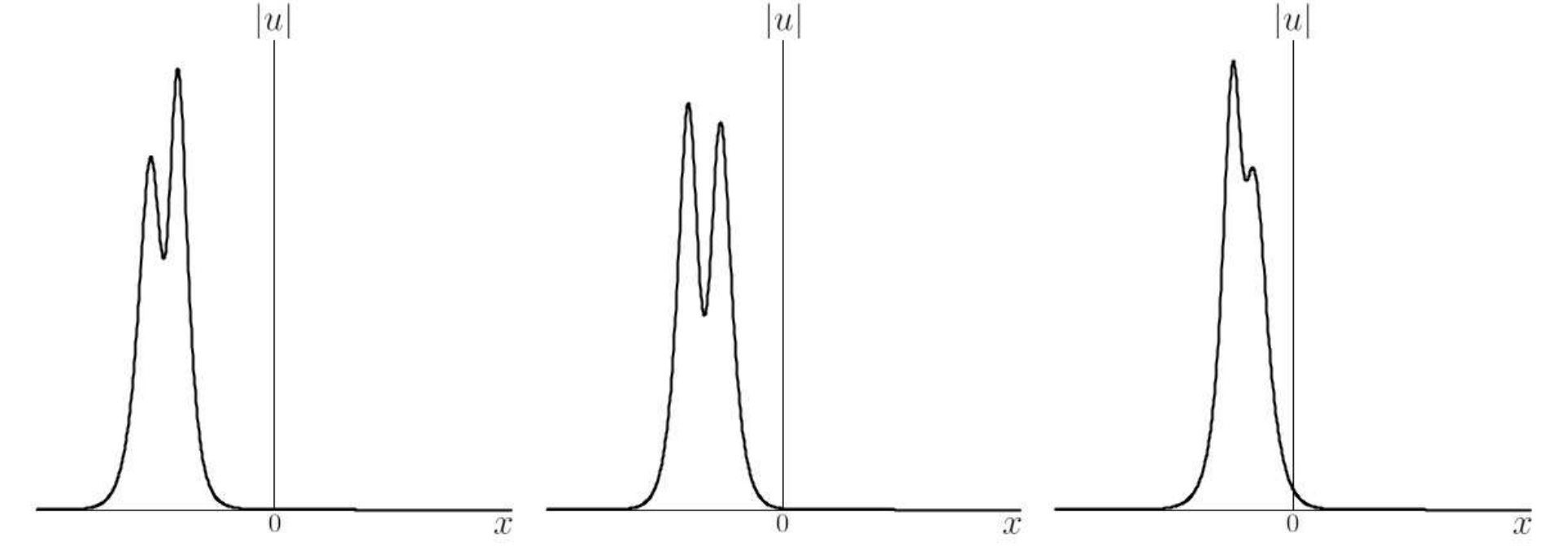}
\captionof{figure}{amplitude}
\end{subfigure}%
\begin{subfigure}[t]{0.45\textwidth}
\includegraphics[width=\textwidth]{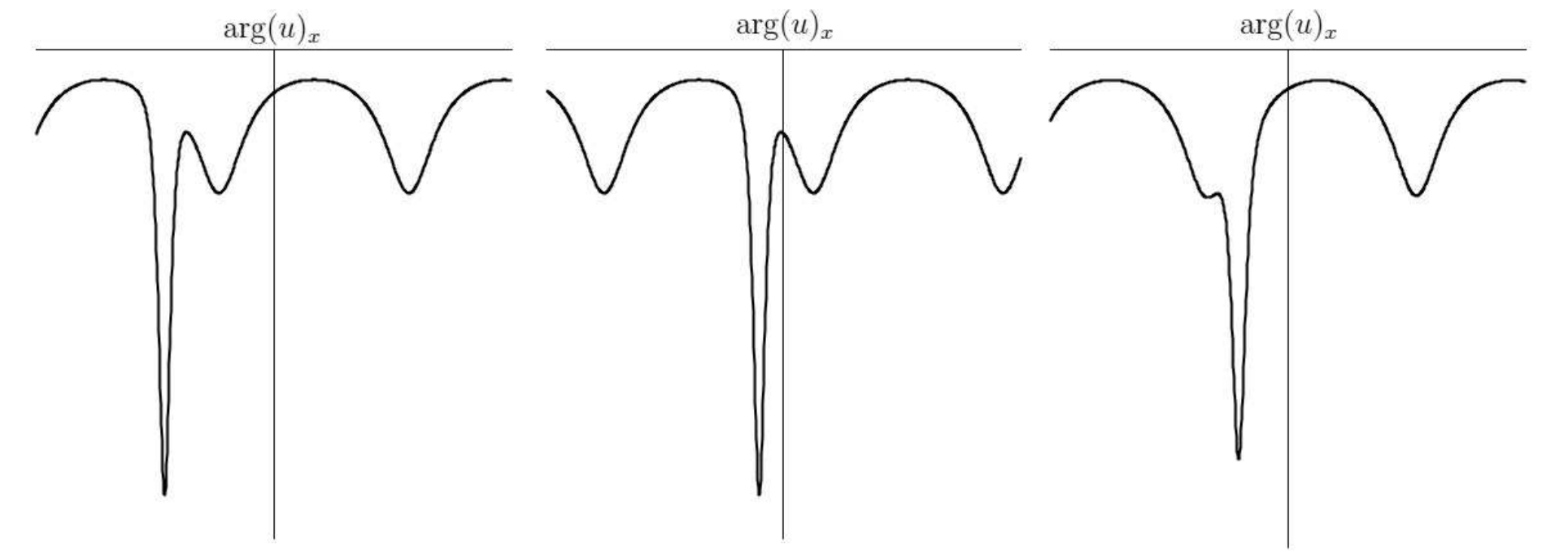}
\captionof{figure}{phase gradient}
\end{subfigure}
\caption{Sasa-Satsuma moving breather 
with $\chi=0.5$, $c=3$, $\nu=2$, $\phi=0$, $\phi_0=0$ ($t_0=-2$)}
\label{right-breather_ss_2mu=1_c=3_nu=2_phi0=0_phi=0}
\end{figure}

\begin{figure}[H]
\begin{subfigure}[t]{0.45\textwidth}
\includegraphics[width=\textwidth]{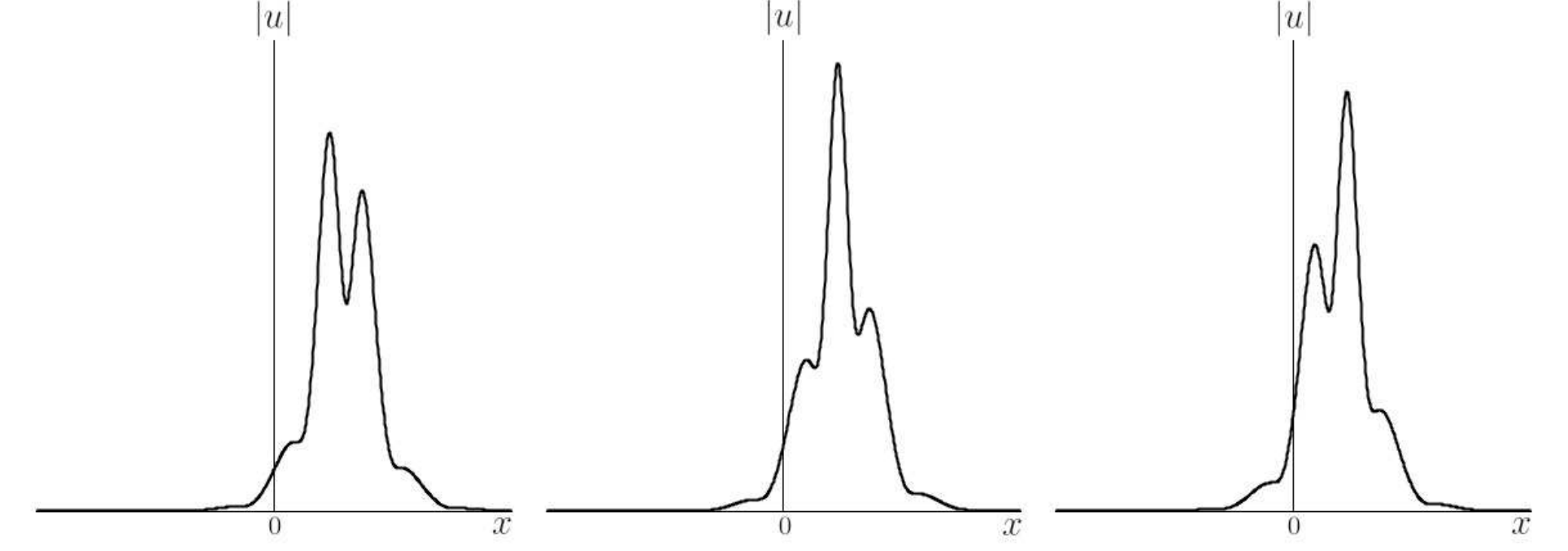}
\captionof{figure}{amplitude}
\end{subfigure}%
\begin{subfigure}[t]{0.45\textwidth}
\includegraphics[width=\textwidth]{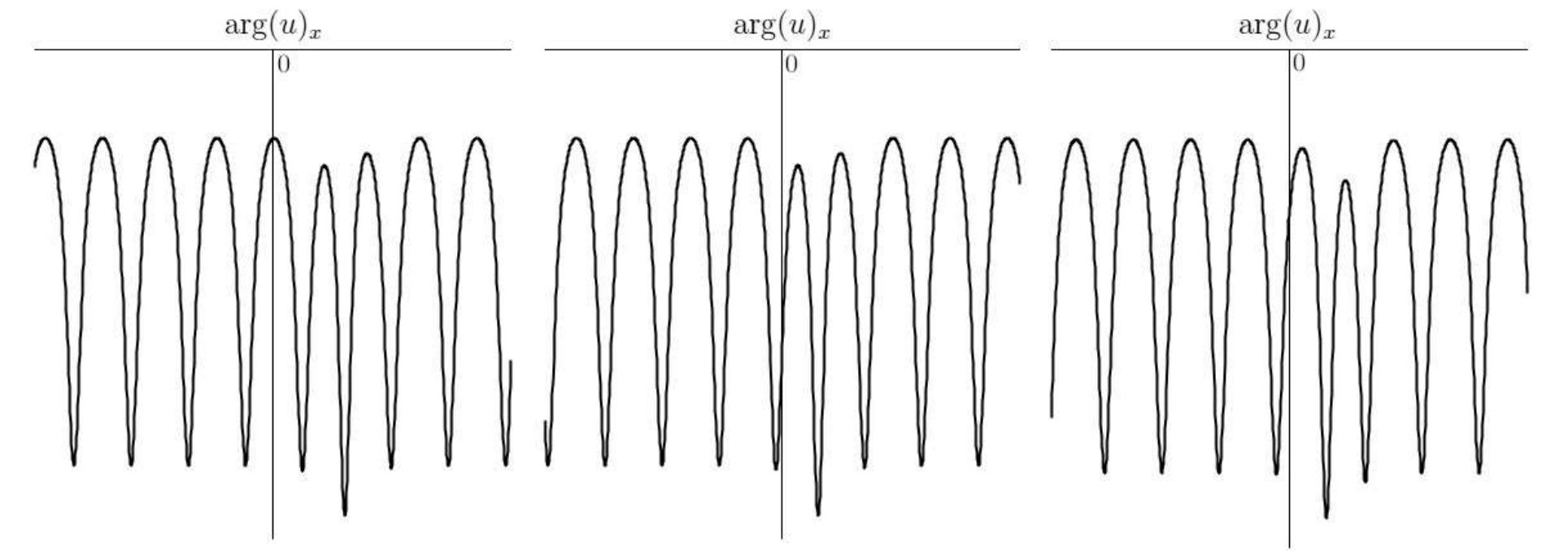}
\captionof{figure}{phase gradient}
\end{subfigure}
\caption{Sasa-Satsuma moving breather 
with $\chi=0.5$, $c=-2$, $\nu=3$, $\phi=0$, $\phi_0=0$ ($t_0=-2$)}
\label{left-breather_ss_2mu=1_c=-2_nu=3_phi0=0_phi=0}
\end{figure}

\begin{figure}[H]
\begin{subfigure}[t]{0.45\textwidth}
\includegraphics[width=\textwidth]{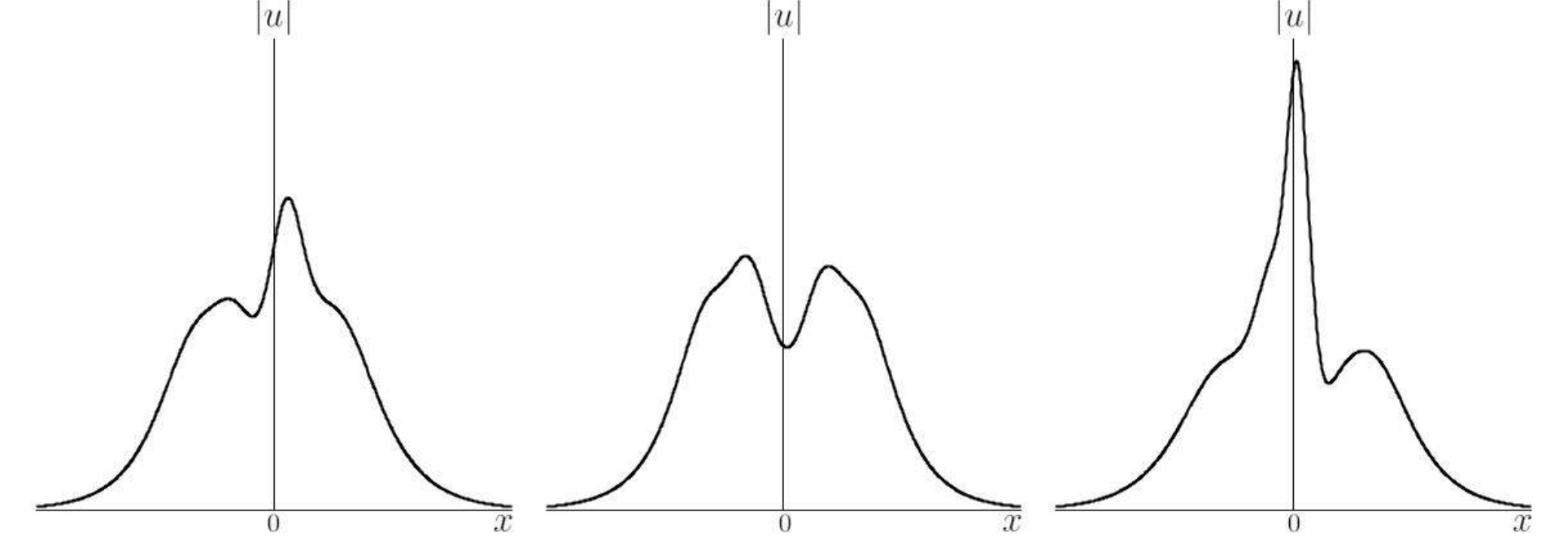}
\captionof{figure}{amplitude}
\end{subfigure}%
\begin{subfigure}[t]{0.45\textwidth}
\includegraphics[width=\textwidth]{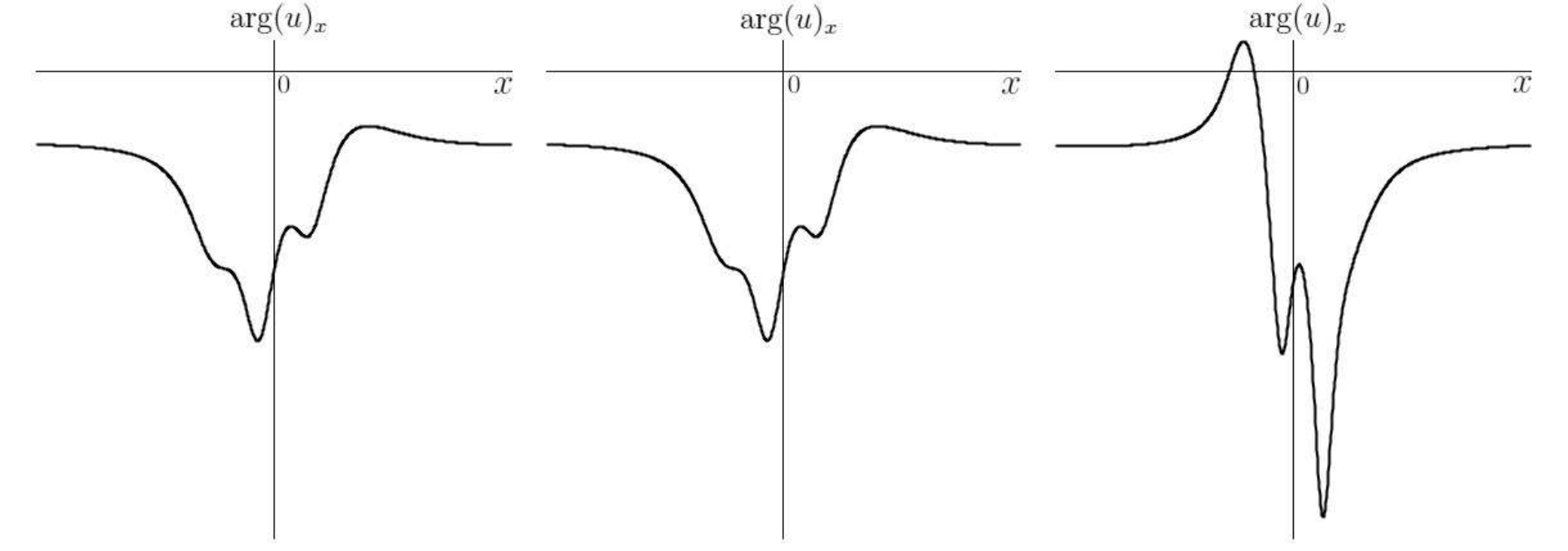}
\captionof{figure}{phase gradient}
\end{subfigure}
\caption{Sasa-Satsuma stationary oscillatory breather 
with $\chi=0$, $c=0$, $\nu=2$, $\phi=0$, $\nu_0=3$, $\phi_0=0$ ($t_0=-2$)}
\label{oscilbreather_ss_c=0_nu1=5_nu2=1_phi1=0_phi2=0}
\end{figure}

\begin{figure}[H]
\begin{subfigure}[t]{0.45\textwidth}
\includegraphics[width=\textwidth]{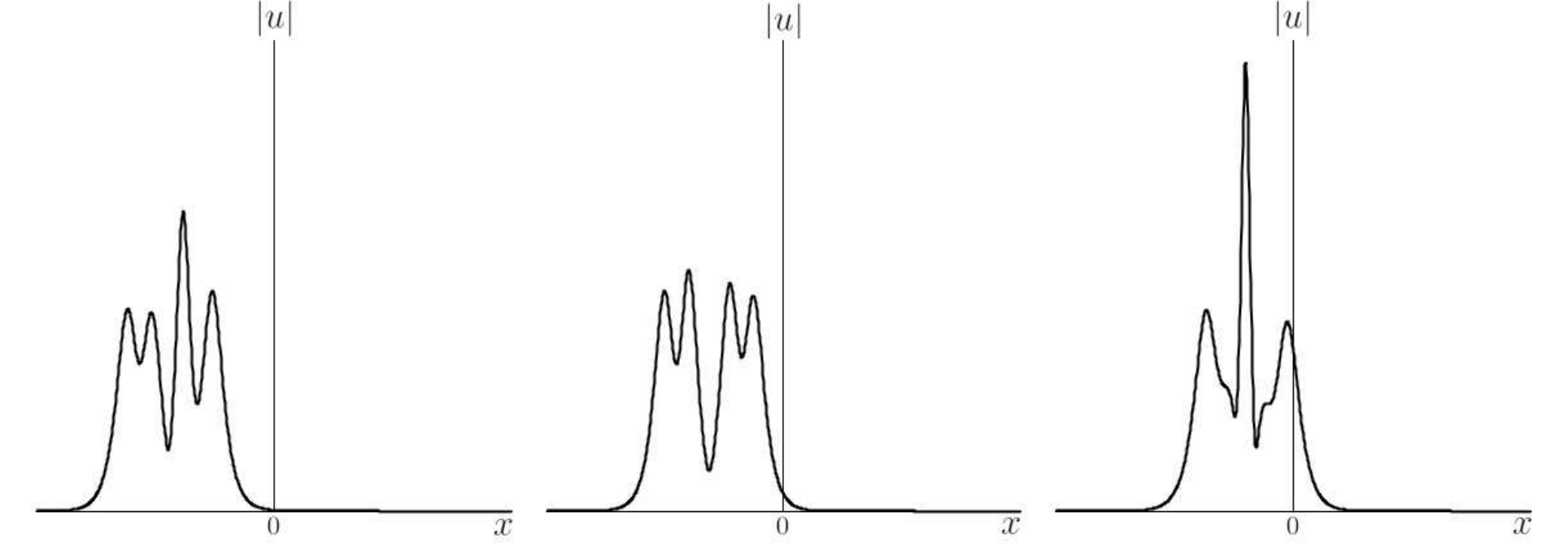}
\captionof{figure}{amplitude}
\end{subfigure}%
\begin{subfigure}[t]{0.45\textwidth}
\includegraphics[width=\textwidth]{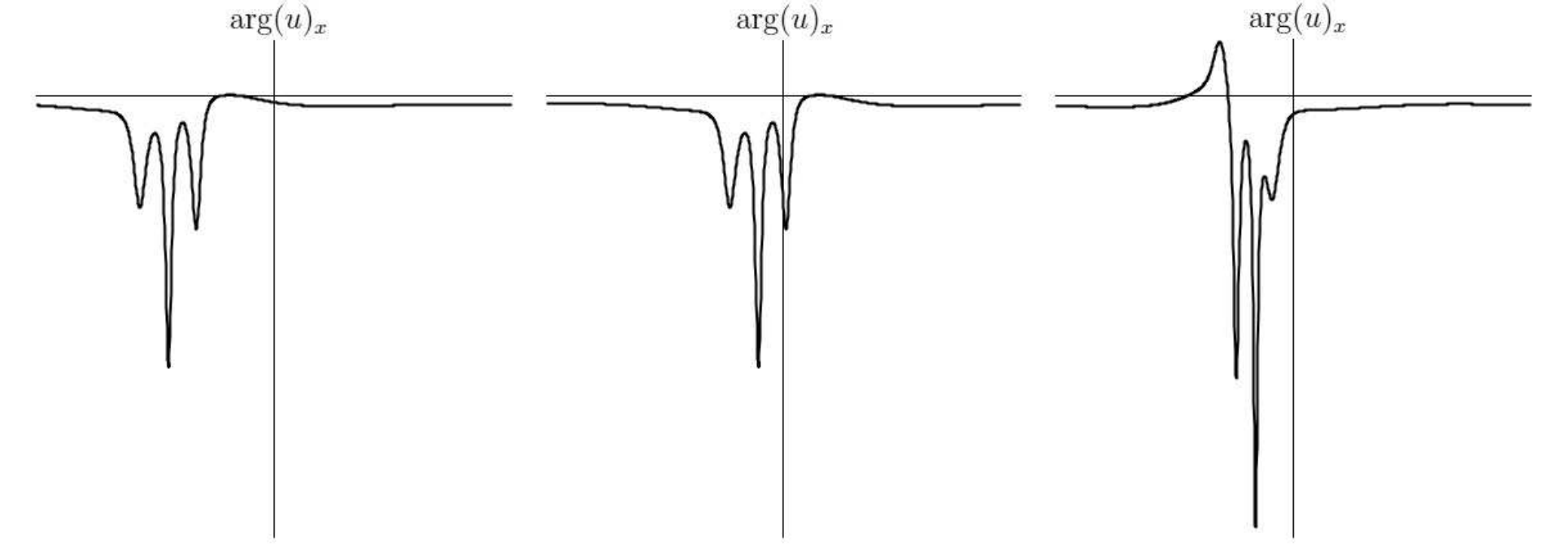}
\captionof{figure}{phase gradient}
\end{subfigure}
\caption{Sasa-Satsuma moving oscillatory breather 
with $\chi=0$, $c=3$, $\nu=2$, $\phi=0$, $\nu_0=3$, $\phi_0=0$ ($t_0=-2$)}
\label{right-oscilbreather_ss_c=3_nu1=5_nu2=1_phi1=0_phi2=0}
\end{figure}

\begin{figure}[H]
\begin{subfigure}[t]{0.45\textwidth}
\includegraphics[width=\textwidth]{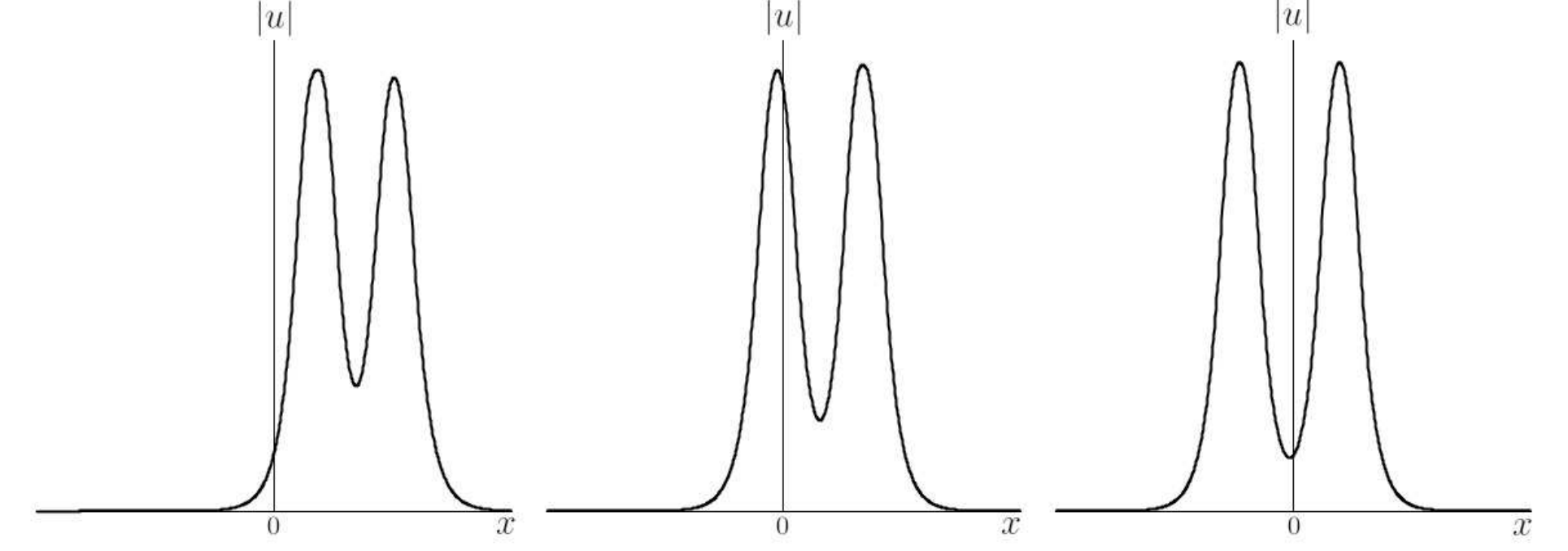}
\captionof{figure}{amplitude}
\end{subfigure}%
\begin{subfigure}[t]{0.45\textwidth}
\includegraphics[width=\textwidth]{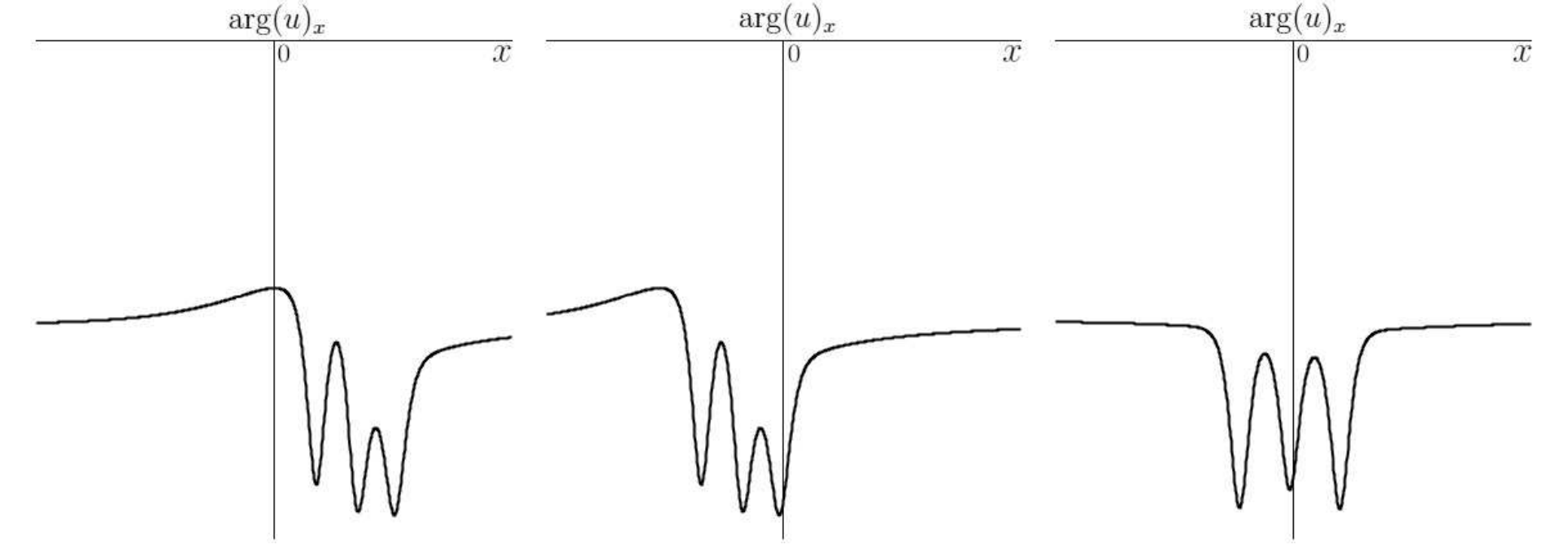}
\captionof{figure}{phase gradient}
\end{subfigure}
\caption{Sasa-Satsuma moving oscillatory breather 
with $\chi=0$, $c=-2$, $\nu=1$, $\phi=0$, $\nu_0=9$, $\phi_0=0$ ($t_0=-2$)}
\label{left-oscilbreather_ss_c=-2_nu1=10_nu2=8_phi1=0_phi2=0}
\end{figure}

The amplitude and phase gradient of the breathers solutions 
\eqref{hoscilbrX1}--\eqref{hoscilbrY} 
and \eqref{ssoscilbrX1}--\eqref{ssoscilbrY} 
are illustrated in 
\figref{breather_hirota_2mu=1_c=0_nu=2_phi0=0_phi=0}--\figref{left-oscilbreather_phase_c=-2_nu1=10_nu2=8_phi1=0_phi2=0}
for the Hirota equation \eqref{hmkdveqscaled}
and 
\figref{breather_ss_2mu=1_c=0_nu=2_phi0=0_phi=0}--\figref{left-oscilbreather_ss_c=-2_nu1=10_nu2=8_phi1=0_phi2=0}
for the Sasa-Satsuma equation \eqref{ssmkdveqscaled}. 
Each figure shows $t=t_0(=t_0+T),t_0+T/3,t_0+2T/3$. 





\section{Concluding remarks}
\label{conclude}

In the literature, 
harmonically modulated solitons \eqref{envel1soliton} and \eqref{envel2soliton}
are commonly called an {\em envelope soliton}. 
Strictly speaking, however, 
the factorization of such solitons \eqref{envel1soliton}
into a solitary wave part $f(kx+wt)$ 
and a harmonic wave part $\exp(i(\kappa x +\omega t))$ 
is well-defined only if the function $f$ is real-valued,
so that the modulation of the solitary wave envelope 
is fully contained in the phase $\arg(u(t,x)) = \kappa x +\omega t$,
as seen in the Hirota envelope soliton \eqref{henvel}. 
Otherwise, 
when the function $f$ is complex-valued, 
as happens in the Sasa-Satsuma harmonically modulated soliton \eqref{ssenvel} 
as well as in the harmonically modulated $2$-solitons 
for both the Hirota and Sasa-Satsuma equations, 
the solitary wave envelope is given by the modulus $|f(kx+wt)|$ 
while its modulation comes from the combined phase 
$\kappa x +\omega t +\arg(f(kx+wt))$. 
In general the only mathematically and physically meaningful way 
to decompose this phase is to write it as 
a travelling wave part $\arg(\tilde f(kx+wt))$ 
plus a temporal part $\nu t$,
corresponding to the oscillatory forms 
\eqref{oscil1soliton} and \eqref{oscil2soliton}. 
Thus the oscillatory parameterization that we have introduced in this paper 
for harmonically modulated solitons is 
mathematically clearer and physically simpler 
than the usual envelope parameterization 
\eqref{envel1soliton} and \eqref{envel2soliton}. 

In a sequel paper \cite{AncWilMia}, 
we will study the main features of the amplitude and phase of 
colliding oscillatory waves as described by 
the oscillatory $2$-soliton solutions \eqref{2soliton}--\eqref{ssoscilY} 
presented in Theorem~\ref{thm:2soliton} 
for the Hirota equation \eqref{hmkdveqscaled}
and the Sasa-Satsuma equation \eqref{ssmkdveqscaled}.

\section*{Acknowledgement}
S. Anco is supported by an NSERC research grant. 
The authors thank the referee for remarks which have helped to improve this paper. 
Nestor Tchegoum Ngatat is thanked for assistance 
in an early stage of this work. 

Email:
{\lowercase{
\scshape{sanco@brocku.ca}, 
\scshape{sattar\_ju@yahoo.com}, 
\scshape{markw@math.ubc.ca}
}}

\appendix\section{}\label{A}

We will first review the derivation of the harmonically modulated $1$-soliton solution 
from the split bilinear system \eqref{hdegs}
for the Hirota equation \eqref{hmkdveqscaled}. 
For $N$=1, the lowest degree terms in the ansatz \eqref{hsolitonansatz}
are given by 
\begin{equation}
G^{(1)}=A e^{\Theta}, 
\quad
F^{(2)}=B e^{\Theta+\bar\Theta}
 \end{equation}
with 
\begin{equation}
\Theta = \k x +\w t,
\quad
\bar\Theta = \bar\k x +\bar\w t
\end{equation}
where $\k,\w,A$ are complex constants and $B$ is a real constant. 
The two lowest degree equations \eqref{hdeg1} and \eqref{hdeg2} yield
\begin{gather}
\w=-\k^3 ,
\label{hwkrel}\\
B=\frac{4A\bar A}{(\k+\bar\k)^2} .
\label{Bsol}
\end{gather}
The inhomogeneous terms in the next equation \eqref{hdeg3} 
turn out to vanish, which determines $G^{(3)}=0$. 
Likewise, the next equation \eqref{hdeg4} determines $F^{(4)}=0$. 
Hence the ansatz \eqref{hsolitonansatz} terminates at degrees $1$ and $2$,
respectively. 
This yields the $1$-soliton solution
\begin{equation}\label{h1soliton}
u = \frac{A e^\Theta}{1+(|A|/\Re\k)^2 e^{\Theta+\bar\Theta}} ,
\quad
\Theta =\k(x-\k^2 t) .
\end{equation}
It can be written in the form of an harmonically modulated soliton \eqref{envel1soliton}, \eqref{wkrels}, \eqref{henvel} 
in the following way. 
By putting 
\begin{equation}\label{kwnotation}
\Re\k =k, \quad
\Im\k =\kappa,
\quad
\Re\w =w, \quad
\Im\w =\omega, 
\end{equation}
we see that the algebraic relations \eqref{wkrels} and \eqref{hwkrel} match. 
Next expressing
\begin{equation}\label{thetaAnotation}
\begin{gathered}
\exp(\Theta)= \exp(i\Im\Theta)\exp(\Re\Theta)
= \exp(i(\kappa x +\omega t)) \exp(kx+wt) , 
\\
A= |A|\exp(i\phi), 
\quad 
\phi = \arg A , 
\end{gathered}
\end{equation}
we see that the solution \eqref{h1soliton} has 
the general harmonically modulated form \eqref{envel1soliton} with 
\begin{equation}
f= |A| \exp(kx+wt)/(1+(|A|/k)^2 \exp(2kx+2wt)) . 
\end{equation}
Then writing 
\begin{equation}
|A/k| = e^{-ak}
\end{equation}
and using the identity 
$\sech\theta = 2\exp\theta/(1+\exp 2\theta)$
where
\begin{equation}
\theta=\Re\Theta-ak=k(x-a)+wt , 
\end{equation}
we find $f$ matches the Hirota envelope function \eqref{henvel}
up to a space translation $x\rightarrow x-a$. 
Hence we have 
\begin{equation}\label{henvel1soliton}
u(t,x) = \exp(i(\varphi+\kappa (x-a) +\omega t)) (|k|/2)\sech(k(x-a)+wt)
\end{equation}
after putting $\phi=\varphi -a\kappa$. 
This yields the rational cosh form shown in Proposition~\ref{prop:hrationalcosh1soliton}. 

We will next derive the explicit form of the harmonically modulated $2$-soliton solution. 
For $N=2$, the lowest degree terms in the ansatz \eqref{hsolitonansatz}
are given by 
\begin{equation}
G^{(1)} =A_1 e^{\Theta_1} + A_2 e^{\Theta_2}, 
\quad
F^{(2)} =B_1 e^{\Theta_1+\bar\Theta_1} + B_2 e^{\Theta_2+\bar\Theta_2} + C e^{\Theta_1+\bar\Theta_2} + \bar C e^{\Theta_2+\bar\Theta_1}
 \end{equation}
with 
\begin{equation}
\Theta_1 = \k_1 x +\w_1 t,
\quad
\Theta_2 = \k_2 x +\w_2 t,
\quad
\bar\Theta_1 = \bar\k_1 x +\bar\w_1 t, 
\quad
\bar\Theta_2 = \bar\k_2 x +\bar\w_2 t, 
\end{equation}
where $\k_1,\k_2,\w_1,\w_2,A_1,A_2,C$ are complex constants 
and $B_1,B_2$ are real constants. 
Similarly to the $N=1$ case, 
the two lowest degree equations \eqref{hdeg1} and \eqref{hdeg2} 
in the split bilinear system yield
\begin{gather}
\w_1=-\k_1^3, 
\quad
\w_2=-\k_2^3,
\label{hwkrels}\\ 
B_1=\frac{4A_1\bar A_1}{(\k_1+\bar\k_1)^2} ,
\quad
B_2=\frac{4A_2\bar A_2}{(\k_2+\bar\k_2)^2} ,
\quad
C=\frac{4 A_1\bar A_2}{(\k_1+\bar\k_2)^2} .
\label{BCsol}
\end{gather}
The inhomogeneous terms in the next two equations \eqref{hdeg3} and \eqref{hdeg4} 
no longer vanish. 
Instead, equation \eqref{hdeg3} now contains monomial terms 
$e^{\Theta_1+\Theta_2+\bar\Theta_2}$ and $e^{\Theta_2+\Theta_1+\bar\Theta_1}$,
while equation \eqref{hdeg4} contains a single monomial term
$e^{\Theta_1+\Theta_2+\bar\Theta_1+\bar\Theta_2}$. 
To balance these degree $3$ and $4$ terms,
the ansatz \eqref{hsolitonansatz} needs to contain
the corresponding monomial terms
\begin{equation}
G^{(3)} =D_1 e^{\Theta_2+\Theta_1+\bar\Theta_1} + D_2 e^{\Theta_1+\Theta_2+\bar\Theta_2} , 
\quad
F^{(4)} = E e^{\Theta_1+\Theta_2+\bar\Theta_1+\bar\Theta_2}
 \end{equation}
where $D_1,D_2$ are complex constants 
and $E$ is a real constant. 
Equations \eqref{hdeg3} and \eqref{hdeg4} then yield
\begin{equation}
D_1 =\frac{4A_1 A_2\bar A_1(\k_1-\k_2)^2}{(\bar\k_1+\k_2)^2(\k_1+\bar\k_1)^2}, 
\quad
D_2=\frac{4A_1A_2\bar A_2(\k_1-\k_2)^2}{(\k_1+\bar\k_2)^2(\k_2+\bar\k_2)^2}
\end{equation}
and
\begin{equation}
E=\frac{16 A_1A_2\bar A_1 \bar A_2(\k_1-\k_2)^2(\bar\k_1-\bar\k_2)^2}{(\k_1+\bar\k_1)^2(\k_1+\bar\k_2)^2(\bar\k_1+\k_2)^2(\k_2+\bar\k_2)^2} . 
\end{equation}

The next two higher degree equations in the split bilinear system \eqref{hdegs}
are given by 
\begin{subequations}
\begin{align}
\label{hdeg5}
& \begin{aligned}
D_t(G^{(5)},1) + D_x^3(G^{(5)},1) 
= & - D_t(G^{(3)},F^{(2)}) - D_x^3(G^{(3)},F^{(2)}) 
\\&\qquad 
- D_t(G^{(1)},F^{(4)}) - D_x^3(G^{(1)},F^{(4)}) 
\end{aligned}
\\
\label{hdeg6}
& D_x^2(F^{(6)},1) - 4 (G^{(5)}\bar G^{(1)} + G^{(1)}\bar G^{(5)}) 
= 4G^{(3)}\bar G^{(3)} - D_x^2(F^{(4)},F^{(2)}) . 
\end{align}
\end{subequations}
The inhomogeneous terms in these equations are found to vanish,
which determines $G^{(5)}=0$ and $F^{(6)}=0$. 
Hence the ansatz \eqref{hsolitonansatz} terminates at degrees $3$ and $4$,
respectively. 
This yields the $2$-soliton solution
\begin{equation}\label{h2soliton}
u = \frac{A_1 e^{\Theta_1}(1+V_1) + A_2 e^{\Theta_2}(1+ V_2)}{1+W}
\end{equation}
with 
\begin{gather}
V_1 = |A_2|^2 \Gamma_2 e^{2\Re\Theta_2},
\quad
V_2 = |A_1|^2 \Gamma_1 e^{2\Re\Theta_1}
\\
\begin{aligned}
W= & |A_1|^2 \Omega_1 e^{2\Re\Theta_1}+ |A_2|^2 \Omega_2 e^{2\Re\Theta_2}
+ |A_1|^2|A_2|^2 \Omega_1\Omega_2\Gamma^2 e^{2\Re(\Theta_1+\Theta_2)}
\\&\qquad
+ 8\Re\big(A_1\bar A_2 \Phi e^{i\Im(\Theta_1-\Theta_2)} \big) e^{\Re(\Theta_1+\Theta_2)}
\end{aligned}
\end{gather}
where
\begin{gather}
\Theta_1 =\k_1(x-\k_1^2 t) , 
\quad
\Theta_2 =\k_2(x-\k_2^2 t) , 
\\
\Gamma_1= \frac{D_1}{A_2|A_1|^2}
= \frac{(\k_1-\k_2)^2}{(\Re\k_1)^2(\bar\k_1+\k_2)^2} , 
\quad
\Gamma_2= \frac{D_2}{A_1|A_2|^2}
= \frac{(\k_1-\k_2)^2}{(\Re\k_2)^2(\k_1+\bar\k_2)^2} , 
\\
\Omega_1=\frac{B_1}{|A_1|^2}
=\frac{1}{(\Re\k_1)^2} ,
\quad
\Omega_2=\frac{B_2}{|A_2|^2}
=\frac{1}{(\Re\k_2)^2} , 
\\
\Phi =\frac{C}{4A_1\bar A_2}
= \frac{1}{(\k_1+\bar\k_2)^2} , 
\\
\Gamma =\frac{\sqrt{E}}{\sqrt{B_1B_2}}
= \frac{|\k_1-\k_2|^2}{|\k_1+\bar\k_2|^2} . 
\end{gather}

To write this solution \eqref{h2soliton}
in the form of an harmonically modulated soliton,
we proceed similarly to the $1$-soliton case. 
First we put 
\begin{equation}\label{kwReImparts}
\begin{aligned}
\Re\k_1 =k_1, \quad
\Im\k_1 =\kappa_1,
\quad
\Re\w_1 =w_1, \quad
\Im\w_1 =\omega_1, 
\\
\Re\k_2 =k_2, \quad
\Im\k_2 =\kappa_2,
\quad
\Re\w_2 =w_2, \quad
\Im\w_2 =\omega_2, 
\end{aligned}
\end{equation}
so thus the algebraic relations \eqref{wkrels} become
\begin{equation}\label{w1w2rels}
\begin{gathered}
w_1=-k_1(k_1^2-3\kappa_1^2) ,
\quad 
\omega_1= -\kappa_1(3k_1^2-\kappa_1^2) ,
\\
w_2=-k_2(k_2^2-3\kappa_2^2) ,
\quad 
\omega_2= -\kappa_2(3k_2^2-\kappa_2^2) . 
\end{gathered}
\end{equation}
Next we write 
\begin{gather}
\label{polarA1A2}
\begin{gathered}
A_1= |A_1|\exp(i\phi_1), 
\quad 
\phi_1 = \arg A_1,
\quad
A_2= |A_2|\exp(i\phi_2), 
\quad 
\phi_2 = \arg A_2,
\end{gathered}\\
\begin{gathered}
\Gamma_1=\Gamma\Omega_1 \exp(i2\gamma_1) ,
\quad
\gamma_1=\frac{\arg(\Gamma_1)}{2}
=\arg\Big(\frac{\k_1-\k_2}{\bar\k_1+\k_2}\Big) ,
\\
\Gamma_2=\Gamma\Omega_2 \exp(i2\gamma_2) ,
\quad
\gamma_2=\frac{\arg(\Gamma_2)}{2}
=\arg\Big(\frac{\k_2-\k_1}{\k_1+\bar\k_2}\Big) . 
\end{gathered}
\end{gather}
We now express 
\begin{equation}
|A_1|\sqrt{\Gamma\Omega_1} = e^{-a_1k_1},
\quad
|A_2|\sqrt{\Gamma\Omega_2} = e^{-a_2k_2}
\end{equation}
and 
\begin{gather}
\Re\Theta_1 -a_1k_1 =k_1(x-a_1)+w_1t = \theta_1 , 
\quad
\Im\Theta_1 =\kappa_1x+\omega_1t = \vartheta_1 , 
\\
\Re\Theta_2 -a_2k_2 =k_2(x-a_2)+w_2t = \theta_2 ,
\quad
\Im\Theta_2 =\kappa_2x+\omega_2t = \vartheta_2 . 
%\Im\Theta_1 - \Im\Theta_2 = \vartheta_1 - \vartheta_2 = \vartheta 
\end{gather}
The expressions in the numerator and denominator of 
the $2$-soliton solution \eqref{h2soliton} are then given by 
\begin{gather}
V_1 = e^{i2\gamma_2}e^{2\theta_2},
\quad
V_2= e^{i2\gamma_1}e^{2\theta_1}
\\
W = e^{2(\theta_1+\theta_2)} +\frac{1}{\Gamma} (e^{2\theta_1}+e^{2\theta_2})
- 2\frac{|\Gamma-1|}{\Gamma} \Re( e^{i(\phi_1-\phi_2-\gamma_1+\gamma_2)} e^{i(\vartheta_1-\vartheta_2)} ) e^{\theta_1+\theta_2}
\end{gather}
where we have used the identities 
\begin{gather}
\Phi/\bar\Phi
= (\Gamma_2\Omega_1)/(\Gamma_1\Omega_2)
\\
\Phi\bar\Phi
= (\Gamma - 1)^2\Omega_1 \Omega_2/16
\\
\arg(-\Phi) = \gamma_2 -\gamma_1, 
\quad
|\Phi| = |\Gamma - 1|\sqrt{\Omega_1 \Omega_2}/4 . 
\end{gather}
Hence we have 
\begin{equation}\label{h2solitonrationalcosh}
u = e^{i\phi_1} e^{i\vartheta_1}f_1 +e^{i\phi_2} e^{i\vartheta_2}f_2
\end{equation}
with 
\begin{equation}
f_1=X_1/Y,
\quad
f_2=X_2/Y 
\end{equation}
given by 
\begin{align}
& 
X_1= (1/\sqrt{\Omega_1\Gamma}) (1+V_1) e^{-\theta_2} 
= (2/\sqrt{\Omega_1\Gamma})\exp(i\gamma_2) \cosh(\theta_2 +i\gamma_2) 
\label{X1h}\\
& 
X_2= (1/\sqrt{\Omega_2\Gamma}) (1+V_2) e^{-\theta_1} 
= (2/\sqrt{\Omega_2\Gamma})\exp(i\gamma_1) \cosh(\theta_1 +i\gamma_1) 
\label{X2h}\\
&\begin{aligned}
Y & = (1+W) e^{-\theta_1-\theta_2} 
\\& 
= 
(2/\Gamma)\big( \cosh(\theta_1-\theta_2) + \Gamma\cosh(\theta_1+\theta_2) -|\Gamma-1|\cos(\vartheta_1-\vartheta_2 +\phi_1-\phi_2+\gamma_2-\gamma_1) \big) . 
\label{Yh}
\end{aligned}
\end{align}

As expressed in this form, 
the $2$-soliton solution \eqref{h2solitonrationalcosh}--\eqref{Yh}
closely resembles a harmonically modulated $2$-soliton \eqref{envel2soliton} except for 
the presence of the shifts $a_1,a_2$ on $x$ in $\theta_1,\theta_2$ 
and the appearance of $\vartheta_1,\vartheta_2$ in the functions $f_1,f_2$.
However, 
under the assumption $w_1/k_1\neq w_2/k_2$, 
this solution can be converted exactly into the form \eqref{envel2soliton}. 
First we apply a combined space-time translation 
\begin{equation}\label{spacetimetrans}
x\rightarrow x-x_0 ,
\quad
t\rightarrow t-t_0
\end{equation}
such that
\begin{equation}\label{shiftcond}
0=k_1(x_0+a_1)+w_1t_0= k_2(x_0+a_2)+w_2t_0
\end{equation}
whereby 
\begin{equation}
\theta_1\rightarrow k_1x+w_1t, 
\quad
\theta_2\rightarrow k_2x+w_2t
\end{equation}
absorbs the shifts $a_1,a_2$ on $x$.
This transformation \eqref{spacetimetrans}--\eqref{shiftcond} 
exists provided $k_1w_2\neq k_2w_1$. 
It induces a corresponding transformation
\begin{equation}
\vartheta_1\rightarrow \kappa_1x+\omega_1t -(\kappa_1x_0 +\omega_1t_0), 
\quad
\vartheta_2\rightarrow \kappa_2x+\omega_2t -(\kappa_2x_0 +\omega_2t_0)
\end{equation}
producing additional phase angles 
which can be absorbed by shifts 
\begin{equation}\label{phaseshift}
\phi_1\rightarrow \phi_1-\varphi_1,
\quad
\phi_2\rightarrow \phi_2-\varphi_2
\end{equation}
so that 
\begin{equation}
\vartheta_1+\phi_1 +\gamma_2 \rightarrow \kappa_1x+\omega_1t +\phi_1,
\quad
\vartheta_2+\phi_2 +\gamma_1 \rightarrow \kappa_2x+\omega_2t +\phi_2
\end{equation}
via $\varphi_1=\gamma_2 -(\kappa_1x_0 +\omega_1t_0)$, 
$\varphi_2=\gamma_1 -(\kappa_2x_0 +\omega_2t_0)$. 

Next we use the identity
\begin{equation}\label{Imthetaid}
(\kappa_1-\kappa_2)x+(\omega_1-\omega_2)t
= \mu_1(k_1x+w_1t) - \mu_2(k_2x+w_2t)
\end{equation}
with 
\begin{equation}\label{mus}
\mu_1= \frac{w_2(\kappa_1-\kappa_2) - k_2(\omega_1-\omega_2)}{k_1w_2-k_2w_1},
\quad
\mu_2= \frac{w_1(\kappa_1-\kappa_2) - k_1(\omega_1-\omega_2)}{k_1w_2-k_2w_1}
\end{equation}
which holds provided $k_1w_2\neq k_2w_1$. 
This leads to the rational cosh form presented in Proposition~\ref{prop:henvel2soliton}.

\section{}\label{B}

We will first summarize the derivation of the harmonically modulated $1$-soliton solution
from the split bilinear system \eqref{ssdegs}
for the Sasa-Satsuma equation \eqref{ssmkdveqscaled}. 
For $N$=1, the lowest degree terms in the ansatz \eqref{sssolitonansatz}
are given by 
\begin{equation}
G^{(1)}=A e^{\Theta} , 
\quad
F^{(2)}=B e^{\Theta+\bar\Theta} , 
\quad
H^{(2)}=C e^{\Theta+\bar\Theta} , 
 \end{equation}
with 
\begin{equation}
\Theta = \k x +\w t,
\quad
\bar\Theta = \bar\k x +\bar\w t
\end{equation}
where $\k,\w,A,C$ are complex constants and $B$ is a real constant. 
The two lowest degree equations \eqref{ssdeg1} and \eqref{ssdeg2} yield
exactly the results \eqref{hwkrel} and \eqref{Bsol} 
obtained for the Hirota solution,
while the next equation \eqref{ssH2} determines
\begin{equation}
C= 6A\bar A(\k-\bar\k) . 
\end{equation}
In the next two lowest degree equations \eqref{ssdeg3} and \eqref{ssdeg4},
the inhomogeneous terms are found to consist of the respective monomials
$e^{2\Theta+\bar\Theta}$ and $e^{2\Theta+2\bar\Theta}$. 
These degree $3$ and $4$ terms need to be balanced by 
having the ansatz \eqref{sssolitonansatz} for $G$ and $F$ 
contain the corresponding monomial terms 
\begin{equation}
G^{(3)} =D e^{2\Theta+\bar\Theta}, 
\quad
F^{(4)} =E e^{2\Theta+2\bar\Theta}
\end{equation}
where $D$ is a complex constant 
and $E$ is a real constant. 
Equations \eqref{ssdeg3} and \eqref{ssdeg4} then yield
\begin{equation}
D=\frac{A^2\bar A(\k-\bar\k)}{\k(\k+\bar\k)^2} , 
\quad
E=\frac{-A^2\bar A^2(\k-\bar\k)^2}{\k\bar\k(\k+\bar\k)^4} . 
\end{equation}
Next, the inhomogeneous terms in equation \eqref{ssH4} turn out to vanish, 
which determines $H^{(4)}=0$. 
Likewise, the next two higher degree equations \eqref{ssdeg4} and \eqref{ssdeg5}
determine $G^{(5)}=0$ and $F^{(6)}=0$. 
Hence the ansatz \eqref{sssolitonansatz} terminates at degrees 
$3$, $4$, and $2$, respectively. 
This yields the $1$-soliton solution
\begin{equation}\label{ss1soliton}
u = 
\frac{A e^\Theta\Big(1+\dfrac{|A|^2}{(\Re\k)^2} \dfrac{i\Im\k}{2\k} 
e^{\Theta+\bar\Theta}\Big)}
{1+\dfrac{|A|^2}{(\Re\k)^2} e^{\Theta+\bar\Theta}
+\dfrac{|A|^4}{(\Re\k)^4}\dfrac{(\Im\k)^2}{4|\k|^2} e^{2\Theta+2\bar\Theta}}, 
\quad
\Theta =\k(x-\k^2 t)
\end{equation}
which can be written in the form of a harmonically modulated soliton \eqref{envel1soliton}, \eqref{wkrels}, \eqref{ssenvel} 
similarly to the Hirota case. 
In particular, 
using expressions \eqref{kwnotation} and \eqref{thetaAnotation}, 
we see that the solution \eqref{ss1soliton} has 
the general harmonically modulated form \eqref{envel1soliton} with 
\begin{equation}\label{ssf}
f= \frac{|A|\exp(kx+wt)\big(1+i(|A|/k)^2\Lambda\exp(2kx+2wt)\big)}
{1+\exp(2kx+2wt) +(|A|/k)^4|\Lambda|^2 \exp(4kx+4wt)}
\end{equation}
where
\begin{equation}
\Lambda = \frac{i\kappa}{2(k+i\kappa)}= \frac{\kappa(\kappa+ik)}{2(k^2+\kappa^2)} . 
\end{equation}
After we simplify this function $f$ in terms of 
$|A/k| = e^{-ak}$ and $\theta=\Re\Theta-ak=k(x-a)+wt$, 
it matches the Sasa-Satsuma envelope function \eqref{ssenvel}
up to a space translation $x\rightarrow x-a$. 
Hence we have 
\begin{equation}\label{ssenvel1soliton}
\begin{aligned}
u(t,x) & = \exp(i\phi)\exp(i(\kappa x +\omega t)) 
\frac{|k|\exp(k(x-a)+wt)\big(1+\Lambda\exp(2(k(x-a)+wt))\big)}
{1+\exp(2(k(x-a)+wt)) +|\Lambda|^2 \exp(4(k(x-a)+wt))}
\\
%& = \frac{\exp(i(\varphi+\kappa (x-a) +\omega t)) |k|\big(1+\Lambda\exp(2(k(x-a)+wt))\big)}{2\cosh(2(k(x-a)+wt)) +|\Lambda|^2 \exp(3(k(x-a)+wt))}
& = \frac{\exp(i\varphi)\exp(i(\kappa (x-a) +\omega t)) |k|\big(k^2+\kappa^2+(\kappa/2)(\kappa+ik)\exp(2(k(x-a)+wt))\big)}{2(k^2+\kappa^2)\cosh(k(x-a)+wt) +(\kappa/2)^2 \exp(3(k(x-a)+wt))}
\end{aligned}
\end{equation}
where $\varphi=\phi +a\kappa$. 

The harmonically modulated $1$-soliton solution \eqref{ssenvel1soliton} 
can be converted into a rational $\cosh$ form. 
This was carried out in \cite{AncNgaWil} when $\kappa=0$. 
For $\kappa\neq0$, 
the form of the highest-degree monomial terms 
in the numerator and denominator of expression \eqref{ss1soliton}
motivates writing 
\begin{gather}
|A|\sqrt{|\Lambda|}/|k| = e^{-\tilde a k}
\\
\tilde\theta=\Re\Theta-\tilde a k=k(x-\tilde a)+wt
\end{gather}
so that $\tilde a$ is absorbed into a space translation $x\rightarrow x-\tilde a$.  
The envelope function \eqref{ssf} then becomes
\begin{equation}
f= (|k|/\sqrt{|\Lambda|})  e^{\tilde\theta}(1+e^{i\lambda}e^{2\tilde\theta})/(1+e^{4\tilde\theta}+(1/|\Lambda|)e^{2\tilde\theta})
\end{equation}
where
\begin{gather}
|\Lambda| = \frac{|\kappa|}{2\sqrt{k^2+\kappa^2}} , 
\\
\lambda = \arg(\kappa(\kappa+ik)) . 
\label{sslam}
\end{gather}
We next use the identity 
$1+\exp(4\tilde\theta)=2\exp(2\tilde\theta)\cosh(2\tilde\theta)$,
which yields
\begin{equation}
f= (|k|/\sqrt{|\Lambda|})e^{i\lambda/2} \cosh(\tilde\theta+i\lambda/2)/(\cosh(2\tilde\theta)+1/(2|\Lambda|)) . 
\end{equation}
By now absorbing $\lambda/2$ into the phase angle 
$\phi\rightarrow \varphi=\phi +\tilde a\kappa +\lambda/2$, 
we obtain
\begin{equation}
u(t,x) = 
\frac{\exp(i\varphi)\exp(i(\kappa (x-\tilde a) +\omega t)) (|k|/\sqrt{|\Lambda|})\cosh(k(x-\tilde a)+wt+i\lambda/2)}
{\cosh(2(k(x-\tilde a)+wt)) +1/(2|\Lambda|)}
\end{equation}
This yields the rational cosh form shown in Proposition~\ref{prop:ssrationalcosh1soliton}. 

We will next derive the explicit form of the envelope $2$-soliton solution. 
For $N$=2, the lowest degree terms in the ansatz \eqref{sssolitonansatz}
are given by 
\begin{gather}
G^{(1)} =A_1 e^{\Theta_1} + A_2 e^{\Theta_2} , 
\quad
F^{(2)} =B_1 e^{\Theta_1+\bar\Theta_1} + B_2 e^{\Theta_2+\bar\Theta_2} + C e^{\Theta_1+\bar\Theta_2} + \bar C e^{\Theta_2+\bar\Theta_1} ,
\\
H^{(2)} =D_1 e^{\Theta_1+\bar\Theta_1} + D_2 e^{\Theta_2+\bar\Theta_2} + E_1 e^{\Theta_1+\bar\Theta_2} + E_2 e^{\Theta_2+\bar\Theta_1} , 
\end{gather}
with 
\begin{equation}
\Theta_1 = \k_1 x +\w_1 t,
\quad
\Theta_2 = \k_2 x +\w_2 t,
\quad
\bar\Theta_1 = \bar\k_1 x +\bar\w_1 t, 
\quad
\bar\Theta_2 = \bar\k_2 x +\bar\w_2 t, 
\end{equation}
where $\k_1,\k_2,\w_1,\w_2,A_1,A_2,C,D_1,D_2,E_1,E_2$ are complex constants 
and $B_1,B_2$ are real constants. 
Similarly to the $N=1$ case, 
the two lowest degree equations \eqref{hdeg1} and \eqref{hdeg2} 
in the split bilinear system yield exactly the results 
\eqref{hwkrels} and \eqref{BCsol} obtained for the Hirota solution,
while the next equation \eqref{ssH2} determines
\begin{equation}
D_1= 6A_1\bar A_1(\k_1-\bar\k_1) ,
\quad
D_2= 6A_2\bar A_2(\k_2-\bar\k_2) ,
\quad
E_1= 6A_1\bar A_2(\k_1-\bar\k_2) = -\bar E_2 . 
\end{equation}
In contrast 
the next two lowest degree equations \eqref{ssdeg3} and \eqref{ssdeg4} 
now contain many more inhomogeneous monomial terms than in the Hirota case. 
As a consequence, 
the ansatz \eqref{sssolitonansatz} continues past degrees $3$ and $4$ 
for $G$ and $F$ 
and finally turns out to terminate at degree $7$ for $G$, degree $8$ for $F$, 
and degree $6$ for $H$. 
The higher degree equations that determine these terms are given by 
\begin{subequations}\label{sshigherdegs}
\begin{align}
\label{ssH6}
& H_6 = 6D_x(G^{(1)},\bar G^{(5)}) + 6D_x(G^{(5)},\bar G^{(1)}) + 6D_x(G^{(3)},\bar G^{(3)}) 
-F^{(4)} H^{(2)} -F^{(2)} H^{(4)}
\\
\label{ssdeg7}
& \begin{aligned}
D_t(G^{(7)},1) + D_x^3(G^{(7)},1) =& 
- D_t(G^{(5)},F^{(2)}) - D_x^3(G^{(5)},F^{(2)}) +G^{(5)} H^{(2)} 
\\&\qquad 
- D_t(G^{(3)},F^{(4)}) - D_x^3(G^{(3)},F^{(4)})+G^{(3)} H^{(4)} 
\\&\qquad 
- D_t(G^{(1)},F^{(6)}) - D_x^3(G^{(1)},F^{(6)}) +G^{(1)} H^{(2)}
\end{aligned}
\\
\label{ssdeg8}
& \begin{aligned}
D_x^2(F^{(8)},1) - 4 (G^{(7)}\bar G^{(1)} + G^{(1)}\bar G^{(7)}) =&
4(G^{(5)}\bar G^{(3)} + G^{(3)}\bar G^{(5)}) -\tfrac{1}{2} D_x^2(F^{(4)},F^{(4)}) 
\\&\qquad 
- D_x^2(F^{(6)},F^{(2)}) .
\end{aligned}
\end{align}
\end{subequations}

Omitting all details, 
we find that equations \eqref{ssH4}--\eqref{ssdeg6} and \eqref{ssH6}--\eqref{ssdeg8} 
in the split bilinear system lead to the following results. 
The higher degree terms in $G$ and $F$ consist of 
\begin{align}
\label{g3}
& G^{(3)}=
F_1 e^{2\Theta_1+\bar\Theta_1}+F_2 e^{2\Theta_2+\bar\Theta_2}
+F_3 e^{2\Theta_1+\bar\Theta_2}+F_4 e^{2\Theta_2+\bar\Theta_1} 
+G_1 e^{\Theta_1+\Theta_2+\bar\Theta_2}+G_2 e^{\Theta_2+\Theta_1+\bar\Theta_1}
\\
\label{g5}  
&\begin{aligned}
G^{(5)} & = 
H_1 e^{\Theta_1+2\Theta_2+2\bar\Theta_1}+H_2 e^{2\Theta_1+\Theta_2+2\bar\Theta_2}
+H_3 e^{\Theta_1+2\Theta_2+2\bar\Theta_2}+H_4 e^{2\Theta_1+\Theta_2+2\bar\Theta_1}
\\&\qquad 
+I_1 e^{2\Theta_1+\Theta_2+\bar\Theta_1+\bar\Theta_2}
+I_2 e^{\Theta_1+2\Theta_2+\bar\Theta_1+\bar\Theta_2} 
\end{aligned}
\\
\label{g7}
& G^{(7)}=
J_1 e^{2\Theta_1+2\Theta_2+\bar\Theta_1+2\bar\Theta_2}
+J_2 e^{2\Theta_1+2\Theta_2+\bar\Theta_2+2\bar\Theta_1}
\end{align}
and
\begin{align}
\label{f4}  
&\begin{aligned}
F^{(4)} & = 
K_1 e^{2\Theta_1+2\bar\Theta_1}+K_2 e^{2\Theta_2+2\bar\Theta_2}
+K_3 e^{2\Theta_1+2\bar\Theta_2}+\bar K_3 e^{2\Theta_2+2\bar\Theta_1}
+L_1 e^{\Theta_1+\Theta_2+2\bar\Theta_1}+\bar L_1 e^{2\Theta_1+\bar\Theta_1+\bar\Theta_2}
\\&\qquad 
+L_2 e^{\Theta_1+\Theta_2+2\bar\Theta_2}+\bar L_2 e^{2\Theta_2+\bar\Theta_1+\bar\Theta_2} 
+M e^{\Theta_1+\Theta_2+{\bar\Theta_1}+\bar\Theta_2}  
\end{aligned}
\\
\label{f6}  
& F^{(6)}=
N_1 e^{2\Theta_1+\Theta_2+2\bar\Theta_1+\bar\Theta_2}
+N_2 e^{\Theta_1+2\Theta_2+\bar\Theta_1+2\bar\Theta_2}    
+ O e^{2\Theta_1+\Theta_2+\bar\Theta_1+2\bar\Theta_2}+\bar O e^{\Theta_1+2\Theta_2+2\bar\Theta_1+\bar\Theta_2}
\\
\label{f8}
& F^{(8)}=P e^{2\Theta_1+2\Theta_2+2\bar\Theta_1+2\bar\Theta_2}      
\end{align}
where $F_1,F_2,F_3,F_4,G_1,G_2,H_1,H_2,H_3,H_4,I_1,I_2,J_1,J_2$ are complex constants 
given by 
\begin{gather}\label{SS-firstGterm}
F_1=\frac{A_1^2\bar A_1(\k_1-\bar\k_1)}{\k_1(\k_1+\bar\k_1)^2} , 
\quad
F_2=\frac{A_2^2\bar A_2(\k_2-\bar\k_2)}{\k_2(\k_2+\bar\k_2)^2} , 
\\
F_3=\frac{A_1^2\bar A_2(\k_1-\bar\k_2)}{\k_1(\k_1+\bar\k_2)^2} , 
\quad
F_4=\frac{A_2^2\bar A_1 (\k_2-\bar\k_1)}{\k_2(\k_2+\bar\k_1)^2} , 
\\
G_1=
\frac{2 A_1A_2\bar A_2((\k_1+\k_2)(\k_1-\k_2)^2+(\k_1-\bar\k_2)(\k_1+\bar\k_2)^2+(\k_2-\bar\k_2)(\k_2+\bar\k_2)^2)}{(\k_1+\k_2)(\k_1+\bar\k_2)^2(\k_2+\bar\k_2)^2} , 
\\
G_2=
\frac{2 A_1A_2\bar A_1((\k_1+\k_2)(\k_1-\k_2)^2+(\k_1-\bar\k_1)(\k_1+\bar\k_1)^2+(\k_2-\bar\k_1)(\k_2+\bar\k_1)^2)}{(\k_1+\k_2)(\k_1+\bar\k_1)^2(\k_2+\bar\k_1)^2} , 
\end{gather}
\begin{gather}
H_1=
\frac{-A_1 A_2^2{\bar A_1}^2(\k_1-\bar\k_1)(\k_1-\k_2)^2(\k_2-\bar\k_1)^2}
{\k_2\bar\k_1(\k_1+\k_2)(\k_1+\bar\k_1)^2(\k_2+\bar\k_1)^4} , 
\\
H_2=
\frac{-A_2 A_1^2{\bar A_2}^2(\k_2-\bar\k_2)(\k_1-\bar\k_2)^2(\k_1-\k_2)^2}
{\k_1\bar\k_2(\k_1+\k_2)(\k_2+\bar\k_2)^2(\k_1+\bar\k_2)^{4}} , 
\\
H_3=
\frac{-A_1A_2^2{\bar A_2}^2(\k_1-\bar\k_2)(\k_1-\k_2)^2(\k_2-\bar\k_2)^2}
{\k_2\bar\k_2(\k_1+\k_2)(\k_1+\bar\k_2)^2(\k_2+\bar\k_2)^4} , 
\\
H_4=
\frac{-A_2A_1^2{\bar A_1}^2(\k_2-\bar\k_1)(\k_1-\bar\k_1)^2(\k_1-\k_2)^2}
{\k_1\bar\k_1(\k_1+\k_2)(\k_2+\bar\k_1)^2(\k_1+\bar\k_1)^{4}} , 
\\
\begin{aligned}
I_1 = & 
-2A_1^2A_2\bar A_1\bar A_2(\k_1-\bar\k_1)(\k_1-\bar\k_2)(\k_1-\k_2)^2 \times
\\&\qquad
\frac{(\k_2-\bar\k_1)(\k_2+\bar\k_1)^2+(\k_2-\bar\k_2) (\k_2 +\bar\k_2)^2 -(\bar\k_1+\bar\k_2)(\bar\k_1-\bar\k_2)^2}
{\k_1(\k_1+\k_2)(\bar\k_1+\bar\k_2)(\k_1+\bar\k_1)^2(\k_1+\bar\k_2)^2 (\k_2+\bar\k_1)^2(\k_2+\bar\k_2)^2} , 
\end{aligned}
\\
\begin{aligned}
I_2 = & 
-2A_1A_2^2\bar A_1\bar A_2(\k_2-\bar\k_1)(\k_2-\bar\k_2)(\k_1-\k_2)^2  \times
\\&\qquad
\frac{(\k_1-\bar\k_1)(\k_1+\bar\k_1)^2+(\k_1-\bar\k_2)(\k_1 + \bar\k_2)^2 -(\bar\k_1+\bar\k_2)(\bar\k_1-\bar\k_2)^2 }
{\k_2(\k_1+\k_2)(\bar\k_1+\bar\k_2)(\k_1+\bar\k_1)^2(\k_1+\bar\k_2)^2 (\k_2+\bar\k_1)^2(\k_2+\bar\k_2)^2} , 
\end{aligned}
\end{gather}
\begin{gather}
J_1=
\frac{ A_1^2A_2^2\bar A_1{\bar A_2}^2(\k_1-\bar\k_1)(\k_2-\bar\k_1)(\k_1-\bar\k_2)^2(\k_2-\bar\k_2)^2(\bar\k_1-\bar\k_2)^2(\k_1-\k_2)^4}
{\k_1\k_2\bar\k_2(\bar\k_1+\bar\k_2)(\k_1+\k_2)^2(\k_1+\bar\k_1)^2(\k_2+\bar\k_1)^2(\k_1+\bar\k_2)^4(\k_2+\bar\k_2)^4} , 
\\
J_2=
\frac{ A_1^2A_2^2{\bar A_1}^2\bar A_2(\k_1-\bar\k_2)(\k_2-\bar\k_2)(\k_1-\bar\k_1)^2(\k_2-\bar\k_1)^2(\bar\k_1-\bar\k_2)^2(\k_1-\k_2)^4} 
{\k_1\k_2\bar\k_1(\bar\k_1+\bar\k_2)(\k_1+\k_2)^2(\k_1+\bar\k_2)^2(\k_2+\bar{\k_2})^2(\k_1+\bar\k_1)^4(\k_2+\bar{\k_1})^4} , 
\label{SS-lastGterm}
\end{gather}
and where $K_1,K_2,M,N_1,N_2,P$ are real constants 
and $K_3,L_1,L_2,O$ are complex constants
given by 
\begin{gather}\label{SS-firstFterm}
K_1=\frac{-A_1^2{\bar A_1}^2(\k_1-\bar\k_1)^2}{\k_1\bar\k_1(\k_1+\bar\k_1)^4} , 
\quad
K_2=\frac{-A_2^2{\bar A_2}^2(\k_2-\bar\k_2)^2}{\k_2\bar\k_2(\k_2+\bar\k_2)^4} , 
\\
K_3=\frac{- A_1^2{\bar A_2}^2(\k_1-\bar\k_2)^2}{\k_1\bar\k_2(\k_1+\bar\k_2)^4} , 
\\
L_1=\frac{-4A_1A_2{\bar A_1}^2(\k_1-\bar\k_1)(\k_2-\bar\k_1)}
{\bar\k_1(\k_1+\k_2)(\k_1+\bar\k_1)^2(\k_2+\bar\k_1)^2} , 
\\
L_2=\frac{-4A_1A_2{\bar A_2}^2(\k_1-\bar\k_2)(\k_2-\bar\k_2)}
{\bar\k_2(\k_1+\k_2)(\k_1+\bar\k_2)^2(\k_2+\bar\k_2)^2} , 
\\
M= 
\frac{8A_1\bar A_1A_2\bar A_2 (M_1-M_2-M_3)}
{(\k_1+\k_2)(\bar\k_1+\bar\k_2)(\k_1+\bar\k_1)^2(\k_1+\bar\k_2)^2
(\k_2+\bar\k_1)^2(\k_2+\bar\k_2)^2} , 
\\
M_1 = 
(\k_1+\k_2)(\k_1-\k_2)^2(\bar\k_1+\bar\k_2)(\bar\k_1-\bar\k_2)^2  , 
%= & (|\k_1|^2 +|\k_2|^2)\big( |\k_1^2-\k_2^2|^2 + |\k_1^2-{\bar\k_2}^2|^2 \big)
%=(\k_1\bar\k_1+\k_2\bar\k_2)\big(2(\k_1^2{\bar\k_1}^2+\k_2^2{\bar\k_2}^2)-(\k_1^2+{\bar\k_1}^2)(\k_2^2+{\bar\k_2}^2) \big)
%= \frac{1}{32}\big( (\k_1+\bar\k_1)^2+(\k_2+\bar\k_2)^2 -(\k_1-\bar\k_1)^2 -(\k_2-\bar\k_2)^2 \big)\times 
%\\&\qquad \big( ((\k_1+\bar\k_1)^2-(\k_2+\bar\k_2)^2)^2 +((\k_1-\bar\k_1)^2 -(\k_2-\bar\k_2)^2)^2 
%\\&\qquad -2((\k_1+\bar\k_1)^2+(\k_2+\bar\k_2)^2)((\k_1-\bar\k_1)^2 +(\k_2-\bar\k_2)^2) \big)
\\
M_2 = 
(\k_1+\bar\k_1)^2(\k_1-\bar\k_1)(\k_2+\bar\k_2)^2(\k_2-\bar\k_2) , 
%= -(\k_1\bar\k_2+\k_2\bar\k_1)\big( |\k_1^2-\k_2^2|^2 + (\k_1^2-{\bar\k_1}^2)(\k_2^2-{\bar\k_2}^2) \big)
%=(\k_1\bar\k_2+\k_2\bar\k_1)\big( 2(\k_1^2{\bar\k_2}^2+\k_2^2{\bar\k_1}^2)-(\k_1^2+{\bar\k_2}^2)(\k_2^2+{\bar\k_1}^2) \big) 
%\\& = \frac{1}{32}\big( (\k_1+\bar\k_2)^2+(\k_2+\bar\k_1)^2 -(\k_1-\bar\k_2)^2 -(\k_2-\bar\k_1)^2 \big)\times
%\\&\qquad \big( ((\k_1+\bar\k_2)^2-(\k_2+\bar\k_1)^2)^2 +((\k_1-\bar\k_2)^2 -(\k_2-\bar\k_1)^2)^2 
%\\&\qquad -2((\k_1+\bar\k_2)^2+(\k_2+\bar\k_1)^2)((\k_1-\bar\k_2)^2 +(\k_2-\bar\k_1)^2) \big)
\\
M_3 
= (\k_1+\bar\k_2)^2(\k_1-\bar\k_2)(\k_2+\bar\k_1)^2(\k_2-\bar\k_1) , 
% = -(\k_1\k_2+\bar\k_1\bar\k_2)\big( |\k_1^2-{\bar\k_2}^2|^2 - (\k_1^2-{\bar\k_1}^2)(\k_2^2-{\bar\k_2}^2) \big)
%= -(\k_1\k_2+\bar\k_1\bar\k_2)\big( 2(\k_1^2\k_2^2+{\bar\k_1}^2{\bar\k_2}^2)-(\k_1^2+\k_2^2)({\bar\k_1}^2+{\bar\k_2}^2) \big)
%\\& = \frac{1}{32}\big( (\k_1+\k_2)^2+(\bar\k_1+\bar\k_2)^2 -(\k_1-\k_2)^2 -(\bar\k_1-\bar\k_2)^2 \big)\times
%\\&\qquad \big( ((\k_1+\k_2)^2-(\bar\k_1+\bar\k_2)^2)^2 +((\k_1-\k_2)^2 -(\bar\k_1-\bar\k_2)^2)^2 
%\\&\qquad -2((\k_1+\k_2)^2+(\bar\k_1+\bar\k_2)^2)((\k_1-\k_2)^2 +(\bar\k_1-\bar\k_2)^2) \big)
\end{gather}
\begin{gather}
N_1=
\frac{4 A_1^2A_2{\bar A_1}^2\bar A_2(\k_1-\bar\k_2)(\k_2-\bar\k_1)(\k_1-\k_2)^2(\k_1-\bar\k_1)^2(\bar\k_1-\bar\k_2)^2}
{\k_1\bar\k_1(\k_1+\k_2)(\bar\k_1+\bar\k_2)(\k_1+\bar\k_2)^2(\k_2 +\bar\k_1)^2(\k_2 +\bar\k_2)^2(\k_1 +\bar\k_1)^4} , 
\\
N_2= 
\frac{4 A_1A_2^2\bar A_1{\bar A_2}^2(\k_1-\bar\k_2)(\k_2-\bar\k_1)(\k_1-\k_2)^2(\k_2-\bar\k_2)^2(\bar\k_1-\bar\k_2)^2}
{\k_2\bar\k_2(\k_1+\k_2)(\bar\k_1+\bar\k_2)(\k_1+\bar\k_1)^2(\k_1 +\bar\k_2)^2(\k_2 +\bar\k_1)^2(\k_2 +\bar\k_2)^4} , 
\\
O=
\frac{4 A_1^2A_2\bar A_1{\bar A_2}^2(\k_1-\bar\k_1)(\k_2-\bar\k_2)(\k_1-\k_2)^2(\k_1-\bar\k_2)^2(\bar\k_1-\bar\k_2)^2}
{\k_1\bar\k_2(\k_1+\k_2)(\bar\k_1+\bar\k_2)(\k_1+\bar\k_1)^2(\k_2 +\bar\k_1)^2(\k_2 +\bar\k_2)^2(\k_1 +\bar\k_2)^4} , 
\end{gather}
\begin{gather}
P=
\frac{ A_1^2A_2^2{\bar A_1}^2{\bar A_2}^2(\k_1-\bar\k_1)^2(\k_1-\bar\k_2)^2(\k_2-\bar\k_1)^2(\k_2-\bar\k_2)^2(\k_1-\k_2)^4(\bar\k_1-\bar\k_2)^4}
{\k_1\k_2\bar\k_1\bar\k_2(\k_1+\k_2)^2(\bar\k_1+\bar\k_2)^2(\k_1+\bar\k_1)^4(\k_1+\bar\k_2)^4(\k_2+\bar\k_1)^4(\k_2+\bar\k_2)^4} .
\label{SS-lastFterm}
\end{gather}
For completeness, we also list the higher degree terms in $H$:
\begin{align}
\label{h4}
& H^{(4)} =
Q_1 e^{\Theta_1+\Theta_2+2\bar\Theta_1}+Q_2 e^{\Theta_1+\Theta_2+2\bar\Theta_2}
+Q_3 e^{2\Theta_1+\bar\Theta_1+\bar\Theta_2}+Q_4 e^{2\Theta_2+\bar\Theta_1+\bar\Theta_2}+R e^{\Theta_1+\Theta_2+\bar\Theta_1+\bar\Theta_2}        
\\
\label{h6}
& H^{(6)}=
S_1 e^{2\Theta_1+\Theta_2+\bar\Theta_1+{\bar2\Theta}_2} +S_2 e^{\Theta_1+2\Theta_2+2\bar\Theta_1+\bar\Theta_2}
+S_3 e^{\Theta_1+2\Theta_2+\bar\Theta_1+2\bar\Theta_2}+S_4 e^{2\Theta_1+\Theta_2+2\bar\Theta_1+\bar\Theta_2}
\end{align}
where $Q_1,Q_2,Q_3,Q_4,R,S_1,S_2,S_3,S_4$ are complex constants 
given by 
\begin{equation}
Q_1=\frac{-6A_1A_2{\bar A_1}^2(\k_1-\bar\k_1)(\k_2-\bar\k_1)(\k_1-\k_2)^2}
{\bar\k_1(\k_1+\bar\k_1)^2(\k_2+\bar\k_1)^2}
=-\bar Q_3 , 
\end{equation}
\begin{equation}
Q_2=\frac{-6 A_1A_2{\bar A_2}^2(\k_1-\bar\k_2)(\k_2-\bar\k_2)(\k_1-\k_2)^2}
{\bar\k_2(\k_1+\bar\k_2)^2(\k_2+\bar\k_2)^2}
=-\bar Q_4 , 
\end{equation}
\begin{equation}
R= 
\frac{-12 A_1A_2\bar A_1\bar A_2 (R_1 + R_2 + R_3 -\bar R_3 + R_4 +\bar R_4)}
{(\k_1+\k_2)(\bar\k_1+\bar\k_2)(\k_1+\bar\k_1)^2(\k_1+\bar\k_2)^2(\k_2+\bar\k_1)^2(\k_2+\bar\k_2)^2} , 
\end{equation}
\begin{equation}
R_1 = 
-\k_1 \bar\k_1 (\k_1-\bar\k_1)(\k_1 +\bar\k_1)^2
\big( (\k_2^2-{\bar\k_2}^2)^2 + (\k_1^2- \k_2^2)({\bar\k_1}^2 -{\bar\k_2}^2) + (\k_1^2- {\bar\k_2}^2)({\bar\k_1}^2 -\k_2^2) \big) , 
\end{equation}
\begin{equation}
R_2 = 
-\k_2 \bar\k_2 (\k_2-\bar\k_2)(\k_2 +\bar\k_2)^2
\big( (\k_1^2-{\bar\k_1}^2)^2 + (\k_1^2- \k_2^2)({\bar\k_1}^2 -{\bar\k_2}^2) + (\k_1^2- {\bar\k_2}^2)({\bar\k_1}^2 -\k_2^2) \big) , 
\end{equation}
\begin{equation}
R_3 = 
\k_1 \k_2 (\k_1 +\k_2)(\k_1 -\k_2)^2
\big( ({\bar\k_1}^2-{\bar\k_2}^2)^2 + (\k_1^2- {\bar\k_1}^2)(\k_2^2 -{\bar\k_2}^2) + (\k_2^2-{\bar\k_1}^2)(\k_1^2-{\bar\k_2}^2) \big) , 
\end{equation}
\begin{equation}
R_4 = 
-\k_1 \bar\k_2 (\k_1-\bar\k_2)(\k_1 +\bar\k_2)^2
\big( ({\bar\k_1}^2-\k_2^2)^2 - (\k_1^2- {\bar\k_1}^2)(\k_2^2 -{\bar\k_2}^2) - ({\bar\k_1}^2- {\bar\k_2}^2)(\k_1^2 -\k_2^2) \big) , 
\end{equation}
\begin{equation}
S_1=
\frac{6 A_1^2A_2\bar A_1{\bar A_2}^2(\k_1-\bar\k_1)(\k_2-\bar\k_1)(\k_2-\bar\k_2)(\k_1-\k_2)^2(\k_1-\bar\k_2)^2(\bar\k_1-\bar\k_2)^2}
{\k_1\bar\k_2(\k_1+\k_2)(\bar\k_1+\bar\k_2)(\k_1+\bar\k_1)^2(\k_2 +\bar\k_2)^2(\k_1 +\bar\k_2)^4}
=-\bar S_2 , 
\end{equation}
\begin{equation}
S_3=
\frac{6 A_1A_2^2\bar A_1{\bar A_2}^2(\k_1-\bar\k_1)(\k_1-\bar\k_2)(\k_2-\bar\k_1)(\k_1-\k_2)^2(\k_2-\bar\k_2)^2(\bar\k_1-\bar\k_2)^2}
{\k_2\bar\k_2(\k_1+\k_2)(\bar\k_1+\bar\k_2)(\k_1+\bar\k_2)^2(\k_2 +\bar\k_1)^2(\k_2 +\bar\k_2)^4} , 
\end{equation}
\begin{equation}
S_4=
\frac{6 A_1^2A_2{\bar A_1}^2\bar A_2(\k_1-\bar\k_2)(\k_2-\bar\k_1)(\k_2-\bar\k_2)(\k_1-\k_2)^2(\k_1-\bar\k_1)^2(\bar\k_1-\bar\k_2)^2}
{\k_1\bar\k_1(\k_1+\k_2)(\bar\k_1+\bar{\k_2})(\k_1+\bar\k_2)^2(\k_2 +\bar\k_1)^2(\k_1 +\bar\k_1)^4} . 
\end{equation}

Expressions \eqref{SS-firstGterm}--\eqref{SS-lastFterm} 
yield the $2$-soliton solution
\begin{equation}\label{ss2soliton}
u = \frac{A_1e^{\Theta_1}(1+V_1) + A_2e^{\Theta_2}(1+V_2)}{1+W}
\end{equation}
with 
\begin{equation}
\begin{aligned}
V_1= &
|A_1|^2 \Lambda_1 e^{2\Re\Theta_1} 
+|A_2|^2(\Pi_1 +|A_2|^2|\Lambda_2|^2\Psi_2 e^{2\Re\Theta_2})e^{2\Re\Theta_2} 
\\&\qquad
+|A_1|^2 |A_2|^2\Lambda_1\Psi_2(\bar\Pi_1 +|A_2|^2|\Lambda_2|^2\Psi^2 e^{2\Re\Theta_2})e^{2\Re(\Theta_1+\Theta_2)}
\\&\qquad
+ A_1\bar A_2\Sigma_1(1 + |A_2|^2\bar\Lambda_2\Psi_2 e^{2\Re\Theta_2})e^{i\Im(\Theta_1-\Theta_2)} e^{\Re(\Theta_1+\Theta_2)} 
\end{aligned}
\end{equation}
\begin{equation}
\begin{aligned}
V_2= &
|A_2|^2 \Lambda_2 e^{2\Re\Theta_2} 
+|A_1|^2(\Pi_2 +|A_1|^2|\Lambda_1|^2\Psi_1 e^{2\Re\Theta_1})e^{2\Re\Theta_1} 
\\&\qquad
+ |A_1|^2 |A_2|^2\Lambda_2\Psi_1(\bar\Pi_2 +|A_1|^2|\Lambda_1|^2\Psi^2 e^{2\Re\Theta_1})e^{2\Re(\Theta_1+\Theta_2)}
\\&\qquad
+ A_2\bar A_1\Sigma_2(1 + |A_1|^2\bar\Lambda_1\Psi_1 e^{2\Re\Theta_1})e^{i\Im(\Theta_2 -\Theta_1)} e^{\Re(\Theta_1+\Theta_2)} 
\end{aligned}
\end{equation}
\begin{equation}
\begin{aligned}
W = & 
|A_1|^2(\Omega_1+|A_1|^2|\Lambda_1|^2 e^{2\Re\Theta_1})e^{2\Re\Theta_1}
+ |A_2|^2(\Omega_2+|A_2|^2|\Lambda_2|^2 e^{2\Re\Theta_2})e^{2\Re\Theta_2}
\\&\qquad
+|A_1|^2|A_2|^2\big(\Pi +|A_1|^2\Omega_2 |\Lambda_1|^2\Psi^2 e^{2\Re\Theta_1} +|A_2|^2\Omega_1 |\Lambda_2|^2\Psi^2 e^{2\Re\Theta_2}\big) e^{2\Re(\Theta_1+\Theta_2)} 
\\&\qquad
+|A_1|^4|A_2|^4|\Lambda_1|^2|\Lambda_2|^2\Psi^4 e^{4\Re(\Theta_1+\Theta_2)} 
%\\&\qquad
+2\Re\big((A_1\bar A_2)^2\Sigma_1\bar\Sigma_2 e^{i2\Im(\Theta_1-\Theta_2)}\big) e^{2\Re(\Theta_1+\Theta_2)} 
\\&\qquad
+ 8\Re\big( A_1\bar A_2\Phi(1 + |A_1|^2\Lambda_1\bar\Delta_1 e^{2\Re\Theta_1} + |A_2|^2\bar\Lambda_2\Delta_2 e^{2\Re\Theta_2}) e^{i\Im(\Theta_1-\Theta_2)} \big) e^{\Re(\Theta_1+\Theta_2)} 
\\&\qquad
+8|A_1|^2|A_2|^2\Re\big(A_1\bar A_2\bar\Phi\Lambda_1\bar\Psi_1 \bar\Lambda_2\Psi_2 e^{i\Im(\Theta_1-\Theta_2)}\big) e^{3\Re(\Theta_1+\Theta_2)}
\end{aligned}
\end{equation}
where
\begin{equation}
\Theta_1 =\k_1(x-\k_1^2 t) , 
\quad
\Theta_2 =\k_2(x-\k_2^2 t) , 
\end{equation}
\begin{equation}
\Omega_1= \frac{B_1}{|A_1|^2} 
= \frac{1}{(\Re\k_1)^2} ,
\quad
\Omega_2 = \frac{B_2}{|A_2|^2}
= \frac{1}{(\Re\k_2)^2} , 
\end{equation}
\begin{equation}
\Lambda_1= \frac{F_1}{A_1|A_1|^2} 
= \frac{i\Im\k_1}{2\k_1(\Re\k_1)^2} , 
\quad
\Lambda_2=\frac{F_2}{A_2|A_2|^2} 
= \frac{i\Im\k_2}{2\k_2(\Re\k_2)^2} , 
\end{equation}
\begin{equation}
\Sigma_1=\frac{F_3}{A_1^2\bar A_2}
= \frac{\k_1-\bar\k_2}{\k_1(\k_1+\bar\k_2)^2} , 
\quad
\Sigma_2=\frac{F_4}{A_2^2\bar A_1}
= \frac{\k_2-\bar\k_1}{\k_2(\k_2+\bar\k_1)^2} , 
\end{equation}
%\begin{equation}
%\Gamma_1 = \frac{H_1\bar C}{L_1F_4} = \frac{(\k_1-\k_2)^2}{(\k_2+\bar\k_1)^2}, 
%\quad
%\Gamma_2= \frac{H_2C}{L_2F_3 } = \frac{(\k_1-\k_2)^2}{(\k_1+\bar\k_2)^2}
%\end{equation}
\begin{equation}
\Delta_1= \frac{\bar A_1 L_1}{\bar F_1 \bar C}
= \frac{\k_2-\bar\k_1}{\k_1+\k_2} ,
\quad
\Delta_2 = \frac{\bar A_2 L_2}{\bar F_2 C}
= \frac{\k_1-\bar\k_2}{\k_1+\k_2} , 
\end{equation}
\begin{equation}
\Psi_1= \frac{\bar A_1 H_1}{\bar F_1 F_4} = \frac{|A_1|^2 H_4}{A_2|F_1|^2}
= \frac{(\k_2-\bar\k_1)(\k_1-\k_2)^2}{(\k_1+\k_2)(\k_2+\bar\k_1)^2} , 
%=\Delta_1\Gamma_1
\end{equation}
\begin{equation}
\Psi_2= \frac{\bar A_2 H_2}{\bar F_2 F_3} = \frac{|A_2|^2 H_3}{A_1|F_2|^2}
= \frac{(\k_1-\bar\k_2)(\k_1-\k_2)^2}{(\k_1+\k_2)(\k_1+\bar\k_2)^2} , 
%=\Delta_2\Gamma_2
\end{equation}
\begin{equation}
\Xi_1=  \frac{2\bar A_2 F_1L_2}{\bar F_2 F_3 B_1} 
= \frac{4i\Im\k_1}{\k_1+\k_2} , 
\quad
\Xi_2=  \frac{2\bar A_1 F_2L_1}{\bar F_1 F_4 B_2} 
= \frac{4i\Im\k_2}{\k_1+\k_2} , 
\end{equation}
\begin{equation}
\begin{aligned}
\Pi_1 &=\frac{G_1}{A_1|A_2|^2} = \frac{F_2\bar F_3 \bar I_1}{A_2\bar F_1 \bar H_2}
\\&
= \frac{(\k_1-\k_2)^2}{2(\Re\k_2)^2(\k_1+\bar\k_2)^2}
+ \frac{\k_1-\bar\k_2}{2(\Re\k_2)^2(\k_1+\k_2)} 
+ \frac{4i\Im \k_2}{(\k_1+\k_2)(\k_1+\bar\k_2)^2} , 
\end{aligned}
\end{equation}
\begin{equation}
\begin{aligned}
\Pi_2 &=\frac{G_2}{A_2|A_1|^2} = \frac{F_1\bar F_4 \bar I_2}{A_1\bar F_2 \bar H_1}
\\&
= \frac{(\k_1-\k_2)^2}{2(\Re\k_1)^2(\k_2+\bar\k_1)^2}
+\frac{\k_2-\bar\k_1}{2(\Re\k_1)^2(\k_1+\k_2)}
+\frac{4i\Im\k_1}{(\k_1+\k_2)(\k_2+\bar\k_1)^2} , 
\end{aligned}
\end{equation}
\begin{equation}
\Phi= \frac{C}{4A_1\bar A_2} 
= \frac{1}{(\k_1+\bar\k_2)^2} , 
\end{equation}
\begin{equation}
\begin{aligned}
\Psi^2 = |\Psi_1|^2=|\Psi_2|^2 
&= \frac{|A_1||A_2|\sqrt{P}}{|F_1||F_2|}
= \frac{|A_1|^2 N_1}{|F_1|^2 B_2}= \frac{|A_2|^2 N_2}{|F_2|^2 B_1} 
= \frac{A_1 J_1}{F_1 H_3}= \frac{A_2 J_2}{F_2 H_4} 
\\&
= \frac{|\k_1-\bar\k_2|^2|\k_1-\k_2|^4}{|\k_1+\k_2|^2|\k_1+\bar\k_2|^4} , 
%= (\Delta\Gamma)^2 
\end{aligned}
\end{equation}
\begin{equation}
\Delta = |\Delta_1|=|\Delta_2| 
= \frac{|A_1||L_1|}{|F_1||C|} = \frac{|A_2||L_2|}{|F_2||C|} 
= \frac{|\k_1-\bar\k_2|}{|\k_1+\k_2|} , 
\end{equation}
\begin{equation}
\Xi = 2\Re(\Xi_1\bar\Xi_2)
= \Re\bigg( \frac{A_1\bar A_2 L_2\bar L_1}{F_3\bar F_4} \bigg)\frac{8}{B_1B_2}
= \frac{32\Im\k_1\Im\k_2}{|\k_1+\k_2|^2} , 
\end{equation}
\begin{equation}
\begin{aligned}
\Pi &= \frac{M}{|A_1|^2|A_2|^2}
\\&
= \frac{|\k_1-\k_2|^4}{2(\Re\k_1\Re\k_2)^2|\k_1+\bar\k_2|^4}
+\frac{|\k_1-\bar\k_2|^2}{2(\Re\k_1\Re\k_2)^2|\k_1+\k_2|^2}
+\frac{32\Im\k_1\Im\k_2}{|\k_1+\k_2|^2|\k_1+\bar\k_2|^4} . 
% \\& = \frac{\Omega_1\Omega_2(\Delta^2+\Gamma^2)}{2} +\Xi|\Phi|^2 
\end{aligned}
\end{equation}

This solution \eqref{ss2soliton}
can be shown to reduce to the ordinary $2$-soliton solution \cite{GilHieNimOht, AncNgaWil}
for the Sasa-Satsuma equation when $\kappa_1=\kappa_2=0$. 
For $\kappa_1\neq 0$ and $\kappa_2\neq 0$, 
it can be written in the form of a harmonically modulated soliton
similarly to the Hirota case by use of the notation 
\eqref{kwReImparts}, \eqref{w1w2rels}, \eqref{polarA1A2}. 
To proceed, we first 
we observe that the highest-degree monomial terms 
in the numerator and denominator of expression \eqref{ss2soliton} consist of 
$|\Lambda_1|^2|\Lambda_2|^2\Psi^4 e^{4\Re\Theta_1}e^{4\Re\Theta_2}$, 
$|\Lambda_1|^2\Psi^2 \Lambda_2\Psi_1 e^{4\Re\Theta_1}e^{2\Re\Theta_2}$,
$|\Lambda_2|^2\Psi^2 \Lambda_1\Psi_2 e^{4\Re\Theta_2} e^{2\Re\Theta_1}$.
Based on the form of their coefficients, 
we write 
\begin{equation}
|A_1|\sqrt{|\Lambda_1|\Psi} = e^{-a_1k_1} ,
\quad
|A_2|\sqrt{|\Lambda_2|\Psi} = e^{-a_1k_2} , 
\end{equation}
and 
\begin{gather}
\Re\Theta_1 -a_1k_1 =k_1(x-a_1)+w_1t = \theta_1 ,
\quad
\Re\Theta_2 -a_2k_2 =k_2(x-a_2)+w_2t = \theta_2 , 
\\
\Im\Theta_1 =\kappa_1x +\omega_1t = \vartheta_1 ,
\quad
\Im\Theta_2 =\kappa_2x +\omega_2t = \vartheta_2 . 
\end{gather}
Then, in the denominator of the $2$-soliton solution \eqref{ss2soliton},
we have  
\begin{equation}
\begin{aligned}
W= & 
e^{4\theta_1 +4\theta_2}
+\frac{1}{\Psi^2}\Big( e^{4\theta_1} + e^{4\theta_2}\Big)
+\frac{\Omega_1}{\Psi|\Lambda_1|}\Big( 1 + e^{4\theta_2}\Big) e^{2\theta_1}
+ \frac{\Omega_2}{\Psi|\Lambda_2|}\Big( 1 + e^{4\theta_1}\Big) e^{2\theta_2}
\\&\quad
+\frac{1}{\Psi^2|\Lambda_1||\Lambda_2|}\Big(
\Pi +2\Re\big( \Sigma_1\bar\Sigma_2 e^{i2(\phi_1-\phi_2)} e^{i2(\vartheta_1-\vartheta_2)}\big) \Big)
e^{2\theta_1+2\theta_2}
\\&\quad
+\frac{8\Delta}{\Psi^2\sqrt{|\Lambda_1||\Lambda_2|}}
\Re\Big(\Phi e^{i(\phi_1-\phi_2)} e^{i(\vartheta_1-\vartheta_2)}
\Big(\frac{\bar\Lambda_2\Delta_2}{|\Lambda_2|\Delta} e^{2\theta_2} +\frac{\Lambda_1\bar\Delta_1}{|\Lambda_1|\Delta} e^{2\theta_1}\Big)\Big)
e^{\theta_1+\theta_2}
\\&\quad
+\frac{8}{\Psi\sqrt{|\Lambda_1||\Lambda_2|}}
\Re\Big(e^{i(\phi_1-\phi_2)} e^{i(\vartheta_1-\vartheta_2)}
\Big(\Phi+ \bar\Phi\frac{\bar\Lambda_2\Psi_2}{|\Lambda_2|\Psi}\frac{\Lambda_1\bar\Psi_1}{|\Lambda_1|\Psi} e^{2\theta_1+2\theta_2} \Big)\Big)
e^{\theta_1+\theta_2} . 
\end{aligned}
\end{equation}
In the numerator of the $2$-soliton solution \eqref{ss2soliton},
we have  
\begin{equation}
\begin{aligned}
V_1= & 
\frac{\Lambda_1\Psi_2}{|\Lambda_1|\Psi} 
e^{2\theta_1 + 4\theta_2}
+\frac{1}{\Psi}\Big( \frac{\Lambda_1}{|\Lambda_1|} e^{2\theta_1} + \frac{\Psi_2}{\Psi} 
e^{4\theta_2} \Big)
+\frac{1}{\Psi|\Lambda_2|}\Big( \Pi_1 + \frac{\Lambda_1\Psi_2}{|\Lambda_1|\Psi}\bar\Pi_1 e^{2\theta_1} \Big)
e^{2\theta_2}
\\&\quad
+\frac{\Sigma_1}{\Psi\sqrt{|\Lambda_1||\Lambda_2|}} \Big(1+\frac{\bar\Lambda_2\Psi_2}{|\Lambda_2|\Psi} e^{2\theta_2}\Big)
e^{i(\phi_1-\phi_2)} e^{i(\vartheta_1-\vartheta_2)}
e^{\theta_1 + \theta_2}
\end{aligned}
\end{equation}
and
\begin{equation}
\begin{aligned}
V_2= & 
\frac{\Lambda_2\Psi_1}{|\Lambda_2|\Psi} 
e^{2\theta_2 + 4\theta_1}
+\frac{1}{\Psi}\Big( \frac{\Lambda_2}{|\Lambda_2|} e^{2\theta_2} + \frac{\Psi_1}{\Psi} 
e^{4\theta_1} \Big)
+ \frac{1}{\Psi|\Lambda_1|}\Big( \Pi_2 +\frac{\Lambda_2\Psi_1}{|\Lambda_2|\Psi} \bar\Pi_2 e^{2\theta_2}\Big) 
e^{2\theta_1}
\\&\quad
+\frac{\Sigma_2}{\Psi\sqrt{|\Lambda_1||\Lambda_2|}}
\Big(1+\frac{\bar\Lambda_1\Psi_1}{|\Lambda_1|\Psi} e^{2\theta_1}\Big)
e^{i(\phi_2-\phi_1)} e^{i(\vartheta_2-\vartheta_1)}
e^{\theta_1 + \theta_2}
\end{aligned}
\end{equation}
where
\begin{gather}
\Pi_1 = \frac{\Omega_2}{2}\big(\frac{\Psi_2}{\Delta_2} +\Delta_2\big)+\Xi_2\Phi ,
\quad
\Pi_2 = \frac{\Omega_1}{2}\big(\frac{\Psi_1}{\Delta_1} +\Delta_1\big) +\Xi_1\bar\Phi , 
\\
\Pi = \Xi|\Phi|^2 + \frac{\Omega_1\Omega_2}{2}\Big(\frac{\Psi^2}{\Delta^2}+\Delta^2\Big) . 
\end{gather}
Next we write 
\begin{gather}
\lambda_1=\arg(\Lambda_1)=\arg(i\bar\k_1\Im\k_1) , 
\quad
\lambda_2=\arg(\Lambda_2)=\arg(i\bar\k_2\Im\k_2) , 
\\
\begin{aligned}
\delta_1=\arg(\Delta_1) = \arg\big((\bar\k_1+\bar\k_2)(\k_2-\bar\k_1)\big) , 
\\
\delta_2=\arg(\Delta_2)= \arg\big((\bar\k_1+\bar\k_2)(\k_1-\bar\k_2)\big) , 
\end{aligned}
\\
\begin{aligned}
\sigma_1=\arg(\Sigma_1) = \arg\big(\bar\k_1(\k_1-\bar\k_2)(\bar\k_1+\k_2)^2\big) ,
\\
\sigma_2=\arg(\Sigma_2)= \arg\big(\bar\k_2(\k_2-\bar\k_1)(\k_1+\bar\k_2)^2\big) , 
\end{aligned}
\\
\begin{aligned}
\psi_1=\arg(\Psi_1) =\arg\big((\bar\k_1+\bar\k_2)(\bar\k_2-\k_1)(\k_1+\bar\k_2)^2(\k_1-\k_2)^2\big) ,
\\
\psi_2=\arg(\Psi_2)=\arg\big((\bar\k_1+\bar\k_2)(\bar\k_1-\k_2)(\bar\k_1+\k_2)^2(\k_1-\k_2)^2\big) ,
\end{aligned}
\\
\begin{aligned}
\zeta_1=\arg(\Xi_1\bar\Phi) =\arg\big(i(\bar\k_1+\bar\k_2)(\k_1+\bar\k_2)^2\Im\k_1\big) ,
\\
\zeta_2=\arg(\Xi_2\Phi)=\arg\big(i(\bar\k_1+\bar\k_2)(\bar\k_1+\k_2)^2\Im\k_2\big) ,
\end{aligned}
\end{gather}
and use the identities 
\begin{gather}
\frac{\Phi}{\bar\Phi} = \frac{\Delta^2\bar\Psi_1 \Psi_2}{\Psi^2\bar\Delta_1 \Delta_2}, 
\quad
|\Phi| = -\Phi\Big(\frac{\Delta_2\Psi_1}{\Delta_1\Psi_2}\Big)^{1/2}
= -\bar\Phi\Big(\frac{\Delta_1\Psi_2}{\Delta_2\Psi_1}\Big)^{1/2} . 
\end{gather}

Hence the $2$-soliton solution takes the form 
\begin{equation}\label{ss2solitonrationalcosh}
u = 
e^{i\phi_1} e^{i\vartheta_1} f_1 + e^{i\phi_2} e^{i\vartheta_2} f_2
\end{equation}
with 
\begin{equation}
f_1=X_1/Y,
\quad
f_2=X_2/Y
\end{equation}
given by 
\begin{equation}\label{X1ss}
\begin{aligned}
X_1 & = (1/\sqrt{|\Lambda_1|\Psi}) (1+V_1) e^{-\theta_1-2\theta_2} 
\\& 
= \frac{2}{\sqrt{|\Lambda_1|\Psi^3}}\bigg(
\exp(i\tfrac{1}{2}(\lambda_1+\psi_2))\Big(
\Psi\cosh(\theta_1+2\theta_2+i\tfrac{1}{2}(\psi_2+\lambda_1))
\\&\qquad\qquad
+ \cosh(\theta_1-2\theta_2+i\tfrac{1}{2}(\lambda_1-\psi_2))
+\frac{|\Xi_2||\Phi|}{|\Lambda_2|}\cosh(\theta_1+i(\tfrac{1}{2}(\lambda_1+\psi_1)-\zeta_2))
\\&\qquad\qquad
+\frac{\Omega_2}{2|\Lambda_2|}\Big(
 \frac{\Psi}{\Delta}\cosh(\theta_1+i(\tfrac{1}{2}(\lambda_1-\psi_2)+\delta_2))
+\Delta\cosh(\theta_1+i(\tfrac{1}{2}(\lambda_1+\psi_2)-\delta_2)) \Big)
\Big)
\\&\qquad\qquad
+ \exp(i(\sigma_1+\tfrac{1}{2}(\psi_2-\lambda_2))) \Big(
\frac{|\Sigma_1|}{\sqrt{|\Lambda_1||\Lambda_2|}}
\exp(i(\vartheta_1-\vartheta_2+\phi_1-\phi_2))
\times
\\&\qquad\qquad\qquad
\cosh(\theta_2 +i\tfrac{1}{2}(\psi_2-\lambda_2)) \Big)
\bigg)
\end{aligned}
\end{equation}
\begin{equation}\label{X2ss}
\begin{aligned}
X_2 & = (1/\sqrt{|\Lambda_2|\Psi}) (1+V_2) e^{-\theta_2-2\theta_1}  
\\& 
= \frac{2}{\sqrt{|\Lambda_2|\Psi^3}}\bigg(
\exp(i\tfrac{1}{2}(\lambda_2+\psi_1))\Big(
\Psi\cosh(\theta_2+2\theta_1+i\tfrac{1}{2}(\psi_1+\lambda_2))
\\&\qquad\qquad
+ \cosh(\theta_2-2\theta_1+i\tfrac{1}{2}(\lambda_2-\psi_1))
+\frac{|\Xi_1||\Phi|}{|\Lambda_1|}\cosh(\theta_2+i(\tfrac{1}{2}(\lambda_2+\psi_2)-\zeta_1))
\\&\qquad\qquad
+\frac{\Omega_1}{2|\Lambda_1|}\Big(
\frac{\Psi}{\Delta}\cosh(\theta_2+i(\tfrac{1}{2}(\lambda_2-\psi_1)+\delta_1))
+\Delta\cosh(\theta_2+i(\tfrac{1}{2}(\lambda_2+\psi_1)-\delta_1)) \Big)
\Big)
\\&\qquad\qquad
+ \exp(i(\sigma_2+\tfrac{1}{2}(\psi_1-\lambda_1))) \Big( 
\frac{|\Sigma_2|}{\sqrt{|\Lambda_1||\Lambda_2|}} 
\exp(i(\vartheta_2-\vartheta_1+\phi_2-\phi_1))
\times
\\&\qquad\qquad\qquad
\cosh(\theta_1+i\tfrac{1}{2}(\psi_1-\lambda_1)) \Big)
\bigg)
\end{aligned}
\end{equation}
\begin{equation}\label{Yss}
\begin{aligned}
Y & = (1+W) e^{-2\theta_1-2\theta_2} 
\\& 
=\frac{2}{\Psi}\bigg( 
\frac{\Omega_1}{|\Lambda_1|}\cosh(2\theta_2)
+\frac{\Omega_2}{|\Lambda_2|}\cosh(2\theta_1)
%\\&\qquad
+ \Psi\cosh(2(\theta_1+\theta_2))
+ \frac{1}{\Psi}\cosh(2(\theta_1-\theta_2))
\\&\qquad
-8\frac{|\Phi|}{\sqrt{|\Lambda_1||\Lambda_2|}}\Re\Big(
\exp(i(\vartheta_1-\vartheta_2+\phi_1-\phi_2 +\tfrac{1}{2}(\lambda_1-\lambda_2+\psi_2-\psi_1)))
\times
\\&\qquad\qquad
\Big( \cosh(\theta_1+\theta_2+i\tfrac{1}{2}(\delta_2-\delta_1+\lambda_1-\lambda_2)) 
\\&\qquad\qquad
+\frac{\Delta}{\Psi} \cosh(\theta_1-\theta_2+i\tfrac{1}{2}(\lambda_1+\lambda_2-\delta_2-\delta_1)) 
\Big) \Big)
\\&\qquad
+\frac{1}{\Psi|\Lambda_1||\Lambda_2|}\Big(
|\Sigma_1||\Sigma_2| \cos(2(\vartheta_1-\vartheta_2)+2(\phi_1-\phi_2) +\sigma_1-\sigma_2) 
\\&\qquad\qquad
+\frac{\Xi|\Phi|^2}{2} 
+ \frac{\Omega_1\Omega_2}{4}\Big(\Delta^2+\frac{\Psi^2}{\Delta^2}\Big)
\Big)
\bigg) . 
\end{aligned}
\end{equation}

In this form, 
the $2$-soliton solution \eqref{ss2solitonrationalcosh}--\eqref{Yss}
closely resembles a harmonically modulated $2$-soliton \eqref{envel2soliton}, 
except that $a_1,a_2$ occur in $\theta_1,\theta_2$ 
while $\vartheta_1,\vartheta_2$ appear in the functions $f_1,f_2$. 
The same space-time translation \eqref{spacetimetrans}---\eqref{shiftcond} 
used in the Hirota case can be applied here to absorb $a_1,a_2$ on $x$,
and then the same identity \eqref{Imthetaid}
expressing $\vartheta_1-\vartheta_2$ 
as a linear combination of $\theta_1$ and $\theta_2$ can be used to convert
$f_1,f_2$ into the proper harmonically modulated form. 
Finally, the phase angles in the resulting expressions 
can be substantially simplified 
through phase shifts \eqref{phaseshift} given by 
\begin{equation}
\varphi_1=(\lambda_1+\psi_2)/2 ,
\quad
\varphi_2=(\lambda_2+\psi_1)/2 , 
\end{equation}
and through the angle identities
\begin{gather}
\zeta_1-\delta_2 -\rho_1
= \sigma_2 -\lambda_2+\rho_2 
= \arg(i(\bar\k_1-\k_2)(\k_1+\bar\k_2)^2) , 
\\
\zeta_2-\delta_1 -\rho_2
= \sigma_1-\lambda_1+\rho_1
= \arg(i(\bar\k_2-\k_1)(\bar\k_1+\k_2)^2) , 
\\
(\psi_1-\delta_1)/2= \arg((\k_1-\k_2)(\k_1+\bar\k_2)) , 
\\
(\psi_2-\delta_2)/2= \arg((\k_2-\k_1)(\bar\k_1+\k_2)) ,
\\
(\delta_1-\delta_2)/2= \rho+\arg(i(\bar\k_1-\k_2)) ,
\quad
(\delta_1+\delta_2)/2= \rho+\arg(i(\bar\k_1+\bar\k_2)) ,
\end{gather}
where
\begin{gather}
\rho_1=\arg(\Im\k_1) , 
\quad
\rho_2=\arg(\Im\k_2) ,
\\
\rho=\begin{cases}
0, & |\Re\k_1|\neq|\Re\k_2|\\
\arg(\Im\k_1+\Im\k_2), & |\Re\k_1|=|\Re\k_2|
\end{cases} .
\end{gather}
This leads to the rational cosh form presented in Proposition~\ref{prop:ssenvel2soliton}.

\end{document}